\documentclass[12pt]{article}
\usepackage[a4paper, margin=0.75in]{geometry}
\usepackage[utf8]{inputenc}         
\usepackage[english]{babel}
\usepackage{lmodern}
\usepackage{graphicx}

\usepackage[colorinlistoftodos]{todonotes}

\usepackage[colorlinks=true]{hyperref}

\usepackage[dvipsnames]{xcolor}

\usepackage[]{color-edits}
\addauthor{UF}{blue}
\newcommand{\ufc}[1]{{\UFcomment{#1}}}

\addauthor{VG}{red}
\newcommand{\vgc}[1]{{\VGcomment{#1}}}

\usepackage{amsthm}
\usepackage{amsmath}
\usepackage{amssymb}
\usepackage{amsfonts}
\usepackage{mathrsfs}
\usepackage{mathtools}
\usepackage{verbatim}
\usepackage{footnote}
\usepackage{lineno}
\usepackage{caption}
\usepackage{tikzsymbols}
\usepackage[capitalize, nameinlink]{cleveref}
\usepackage{tikz}
\usetikzlibrary{quantikz2}
\usepackage{nicematrix}
\usepackage{bm}

\usepackage{icomma}
\usepackage{enumitem}
\usepackage{array}
\usepackage{multirow}
\usepackage{setspace}
\usepackage{euscript}
\usepackage{indentfirst}
\usepackage{epigraph}
\usepackage{fancybox, fancyhdr}
\usepackage{titlesec}
\usepackage{dsfont}
\usepackage{csquotes}

\usepackage{algorithm}              
\usepackage{algorithmicx}           
\usepackage[noend]{algpseudocode}   
\usepackage{listings}

\floatname{algorithm}{Algorithm}

\crefname{starredthm}{\textbf{Theorem}}{}
\Crefname{starredthm}{\textbf{Theorem}}{}

\newtheorem{definition}{Definition}[section]

\newtheorem{assumption}{Assumption}
\newtheorem{theorem}{Theorem}[section]

\newtheorem{proposition}{Proposition}[section]
\newtheorem{conjecture}{Conjecture}

\newtheorem{corollary}{Corollary}[section]
\newtheorem{lemma}{Lemma}[section]
\newtheorem{observation}{Observation}[section]

\newtheorem{remark}{Remark}[section]

\crefname{conjecture}{\textbf{Conjecture}}{}
\crefname{lemma}{\textbf{Lemma}}{}
\crefname{claim}{\textbf{Claim}}{}
\crefname{theorem}{\textbf{Theorem}}{}
\crefname{corollary}{\textbf{Corollary}}{}
\crefname{observation}{\textbf{Observation}}{}
\crefname{proposition}{\textbf{Proposition}}{}
\crefname{assumption}{\textbf{Assumption}}{}

\newcommand{\Z}{\mathbb{Z}}

\usepackage{xparse}

\DeclarePairedDelimiterX\brackets[1]{(}{)}{
	
	#1
}
\DeclarePairedDelimiterX\squarebrackets[1]{[}{]}{
	
	#1
}
\DeclarePairedDelimiterX\figbrackets[1]{\{}{\}}{
	
	#1
}

\DeclarePairedDelimiterX\ntrs[1]{\langle}{\rangle}{
	
	#1
}
\DeclarePairedDelimiterX\kets[1]{\lvert}{\rangle}{
	
	#1
}
\DeclarePairedDelimiterX\bras[1]{\langle}{\rvert}{
	
	#1
}
\DeclarePairedDelimiterX\brakets[2]{\langle}{\rangle}{
	
	#1
    \delimsize\vert
    
    #2
}

\DeclarePairedDelimiterX\dprs[2]{\langle}{\rangle}{
	
	#1
    ,
    
    #2
}

\DeclarePairedDelimiterX\ketbras[2]{\lvert}{\rvert}{
	
	#1
    \delimsize\rangle\delimsize\langle
    
    #2
}
\DeclarePairedDelimiterX\qvals[3]{\langle}{\rangle}{
	
	#1
    \delimsize\vert
    #2
    \delimsize\vert
    
    #3
}
\DeclarePairedDelimiterX\qexps[2]{\langle}{\rangle}{
	
	#1
    \delimsize\vert
    #2
    \delimsize\vert
    
    #1
}

\let\P\undefined
\let\o\undefined
\let\O\undefined

\NewDocumentCommand{\P}{o g}{
	\IfNoValueTF{#1}{
		\IfNoValueTF{#2}{
			\mathop{\mathds{P}}
		}{
			\mathop{\mathds{P}}\squarebrackets*{#2}
		}
	}{
		\IfNoValueTF{#2}{
			\mathop{\mathds{P}}_{#1}
		}{
			\mathop{\mathds{P}}_{#1}\squarebrackets*{#2}
		}
	}
}

\NewDocumentCommand{\E}{o g}{
	\IfNoValueTF{#1}{
		\IfNoValueTF{#2}{
			\mathop{\mathds{E}}
		}{
			\mathop{\mathds{E}}\squarebrackets*{#2}
		}
	}{
		\IfNoValueTF{#2}{
			\mathop{\mathds{E}}_{#1}
		}{
			\mathop{\mathds{E}}_{#1}\squarebrackets*{#2}
		}
	}
}

\NewDocumentCommand{\D}{o g}{
	\IfNoValueTF{#1}{
		\IfNoValueTF{#2}{
			\mathop{\mathds{D}}
		}{
			\mathop{\mathds{D}}\squarebrackets*{#2}
		}
	}{
		\IfNoValueTF{#2}{
			\mathop{\mathds{D}}_{#1}
		}{
			\mathop{\mathds{D}}_{#1}\squarebrackets*{#2}
		}
	}
}

\NewDocumentCommand{\cov}{o g g}{
	\IfNoValueTF{#1}{
		\IfNoValueTF{#2}{
			\mathop{\mathrm{cov}}
		}{
			\IfNoValueTF{#3}{
				\mathop{\mathrm{cov}}
			}{
				\mathop{\mathrm{cov}}\brackets*{#2, #3}
			}
		}
	}{
		\IfNoValueTF{#2}{
			\mathop{\mathrm{cov}}_{#1}
		}{
			\IfNoValueTF{#3}{
				\mathop{\mathrm{cov}}_{#1}
			}{
				\mathop{\mathrm{cov}}_{#1}\brackets*{#2, #3}
			}
		}
	}
}

\NewDocumentCommand{\I}{g}{
	\IfNoValueTF{#1}{
		\mathop{\mathds{I}}
	}{
		\mathop{\mathds{I}}\figbrackets*{#1}
	}
}

\NewDocumentCommand{\V}{o g}{
	\IfNoValueTF{#1}{
		\IfNoValueTF{#2}{
			\mathop{\mathds{V}}
		}{
			\mathop{\mathds{V}}\squarebrackets*{#2}
		}
	}{
		\IfNoValueTF{#2}{
			\mathop{\mathds{V}}_{#1}
		}{
			\mathop{\mathds{V}}_{#1}\squarebrackets*{#2}
		}
	}
}

\NewDocumentCommand{\Tr}{o g}{
	\IfNoValueTF{#1}{
		\IfNoValueTF{#2}{
			\mathop{\mathrm{Tr}}
		}{
			\mathop{\mathrm{Tr}}\squarebrackets*{#2}
		}
	}{
		\IfNoValueTF{#2}{
			\mathop{\mathrm{Tr}}_{#1}
		}{
			\mathop{\mathrm{Tr}}_{#1}\squarebrackets*{#2}
		}
	}
}

\NewDocumentCommand{\o}{g}{\operatorname{\mathrm{o}}\IfNoValueTF{#1}{}{\brackets*{#1}}}
\NewDocumentCommand{\O}{g}{\operatorname{\mathrm{O}}\IfNoValueTF{#1}{}{\brackets*{#1}}}
\NewDocumentCommand{\Th}{g}{\operatorname{\Theta}\IfNoValueTF{#1}{}{\brackets*{#1}}}
\NewDocumentCommand{\om}{g}{\operatorname{\omega}\IfNoValueTF{#1}{}{\brackets*{#1}}}
\NewDocumentCommand{\Om}{g}{\operatorname{\Omega}\IfNoValueTF{#1}{}{\brackets*{#1}}}

\NewDocumentCommand{\ntr}{g}{\IfNoValueTF{#1}{}{\ntrs*{#1}}}
\let\ket\undefined
\NewDocumentCommand{\ket}{g}{\IfNoValueTF{#1}{}{\kets*{#1}}}
\let\bra\undefined
\NewDocumentCommand{\bra}{g}{\IfNoValueTF{#1}{}{\bras*{#1}}}
\NewDocumentCommand{\dpr}{g g}{\IfNoValueTF{#1}{}{\IfNoValueTF{#2}{}{\dprs*{#1}{#2}}}}
\let\braket\undefined
\NewDocumentCommand{\braket}{g g}{\IfNoValueTF{#1}{}{\IfNoValueTF{#2}{}{\brakets*{#1}{#2}}}}
\NewDocumentCommand{\ketbra}{g g}{\IfNoValueTF{#1}{}{\IfNoValueTF{#2}{}{\ketbras*{#1}{#2}}}}
\NewDocumentCommand{\qval}{g g g}{\IfNoValueTF{#1}{}{\IfNoValueTF{#2}{}{\IfNoValueTF{#3}{}{\qvals*{#1}{#2}{#3}}}}}
\NewDocumentCommand{\qexp}{g g}{\IfNoValueTF{#1}{}{\IfNoValueTF{#2}{}{\qexps*{#1}{#2}}}}

\newcommand{\argmin}{\mathop{\mathrm{argmin}}}

\newcommand{\mms}{\mathrm{MMS}}

\newcommand{\wx}{\tilde{x}}

\newcommand{\eps}{\varepsilon}

\newcommand{\One}{\mathds{1}}

\newcommand{\Ll}{\mathcal{L}}

\newcommand{\Bb}{\mathcal{B}}

\newcommand{\Pp}{\mathcal{P}}

\newcommand{\Xx}{\mathcal{X}}
\newcommand{\Mm}{\mathcal{M}}
\newcommand{\Tt}{\mathcal{T}}
\newcommand{\Rr}{\mathcal{R}}

\newcommand{\Ee}{\mathcal{E}}
\newcommand{\Hh}{\mathcal{H}}
\newcommand{\Ii}{\mathcal{I}}

\newcommand{\Nn}{\mathcal{N}}

\newcommand{\hx}{\hat{x}}
\newcommand{\hy}{\hat{y}}
\newcommand{\hz}{\hat{z}}

\newcommand{\ov}{\overline{v}}

\newcommand{\wX}{\widetilde{X}}

\newcommand{\hv}{\hat{v}}

\newcommand{\hX}{\hat{X}}

\newcommand{\oT}{\overline{T}}

\newcommand{\wXx}{\widetilde{\mathcal{X}}}

\newcommand{\Gg}{\mathcal{G}}

\title{Almost-MMS allocations with small agent redundancy}
\date{}
\author{Uriel Feige\thanks{Weizmann Institute of Science, Israel. {\tt uriel.feige@weizmann.ac.il}} \and Vadim Grinberg\thanks{Weizmann Institute of Science, Israel. {\tt vadim.grinberg@weizmann.ac.il}}}
\begin{document}

\maketitle

\begin{abstract}
For a fixed $\eps \ge 0$, let $\rho_{\eps, n}$ be the maximum value of $\rho_n$ such that, in every allocation instance with $m$ indivisible goods and $(1 + \eps)n$ agents, there is a subset of $n$ agents with the following property.
The allocation instance restricted to only this subset of $n$ agents has an allocation that gives each agent at least $\rho_n$-fraction of her $n$-maximin share (MMS).
 
When $\eps > 0$ we refer to $\eps n$ as the {\em agent redundancy}. 
The case $\eps = 0$ is the standard setting without agent redundancy, and we use $\rho_{0,n}$ and $\rho_n$ interchangeably.

For allocation instances with additive valuations, 
it is known that $\frac{7}{9} \le \rho_n < 1-\Omega(\frac{1}{\log^2 n})$ for every $n \ge 2$ (with stronger lower bounds for small $n$). 
It is an open question whether $\rho_n$ tends to~1 as $n$ grows. We show that with agent redundancy, the answer to this question is positive. That is, for every $\eps> 0$, $\rho_{\eps, n}$ tends to~1 as $n$ grows. A straightforward corollary of this result is that for every $c > 0$, if the number of items satisfies $m \le (2 - c)n$, then the answer is positive also for $\rho_n$ (with no agent redundancy).
Adapting our techniques to a related setting, we also improve a recent result of Ezra and Garbuz, and show that without agent redundancy, there always exists an allocation that gives $n - O(\log^2 n)$ agents their full MMS.
\end{abstract}

\section{Introduction}

We consider fair allocations of $m$ indivisible goods to $n$ agents.
Let $M$ denote the set of $m$ items to allocate, and let the set of agents be $N := [n]$.
Every agent $i \in \Nn$ has a \textit{valuation function} $v_i$ defined on the subsets of $M$.
We assume that $v_i$ satisfies $v_i(\varnothing) = 0$, and for every $S_1 \subseteq S_2$ it holds $v_i(S_1) \leq v_i(S_2)$, i.e $v_i$ is monotone and non-decreasing. 
In addition, we assume that every $v_i$ is \textbf{additive}: for every $S\subseteq M$ it holds $v_i(S) = \sum_{e \in S}v_i(e)$.

An \textit{allocation} of items $M$ to agents $N$ is a partition $(P_1, \ldots, P_n)$ of $M$ into $n$ disjoint sets, where agent $i$ ``receives'' the set $P_i$.
We would like this allocation to be \textit{fair}, for a certain definition of fairness.
In this work, we will focus on a share-based notion of fairness called the \textit{maximin share}~\cite{Budish11}, or MMS for short.
\begin{definition}
    Let $\Pp_n$ be the set of all partitions $(P_1, \ldots, P_n)$ of items $M$ into $n$ disjoint sets.
    For an agent $i$ with valuation $v_i$, her maximin share is
    \[
        \mms(M, v_i, n) = \max_{(P_1, \ldots, P_n)\in \Pp_n}\min_{j \in [n]}v_i(P_j).
    \]
    We will use $\mms(n)$ to denote the MMS-value with respect to partitions of $M$ into $n$ sets.
\end{definition}
A partition $(A_1, \ldots, A_n)$ of $M$ is called an \textit{MMS-allocation} if for every agent $i \in N$ it holds $v_i(A_i)\geq \mms(M, v_i, n)$.
That is, each $i$ values the bundle $A_i$ given to her at least as high as her MMS.
We will also consider approximate allocations: for a value $\rho \in (0, 1]$, we say that allocation $(A_1, \ldots, A_n)$ is an $\rho$-MMS allocation, if for every agent $i \in N$ it holds $v_i(A_i) \geq \rho\cdot \mms(M, v_i, n)$.
In this work, we will focus on the existence of exact / approximate MMS allocations for additive agents, in two closely related settings.

\subsection{Our results, put in context}

\label{sec:results}

There are known examples showing that with additive valuations, MMS allocations need not exist. The first such examples were given in~\cite{KPW18-1}. Later examples gave explicit upper bounds on $\rho$ for $\rho$-MMS allocations, with $\rho \le \frac{39}{40}$ when $n=3$~\cite{FST21}, and $\rho \le \frac{20}{21}$ when $n = 4$~\cite{EG26}. However, as $n$ grows, the known upper bounds on $\rho$ deteriorate and tend to~1. The motivating question for our current work is the following conjecture.

\begin{conjecture}
\label{conj:largen}
    For every $\rho < 1$, there exists some $n_\rho$ such that every allocation instance with additive valuations and $n \ge n_\rho$ agents has has a $\rho$-MMS allocation. 
\end{conjecture}

We currently do not have strong beliefs as to whether the conjecture is true. 

So far, attempts to refute the conjecture have not been successful. However, they did provide examples of allocation instances with successively stronger lower bounds on the possible value of $n_\rho$ as a function of $\rho$, or equivalently, allocation instances with no $\rho$-MMS allocation, with successively stronger lower bounds on the slackness $1 - \rho$ as a function of $n$. In~\cite{KPW18-1} the slackness was exponentially small in $1/n$, in~\cite{FST21}
it was improved to be polynomially small, and in~\cite{EG26} it was further improved to be polylogarithmically small. To refute the conjecture, the slackness needs to be a constant independent of $n$.

A recurring feature in all the above negative examples is that they  allow what we refer to as {\em agent redundancy}. That is, for each of them, for every $n$ and its corresponding $\rho_n$, in these examples one can have $d > n$ agents (for some $d$ that depends on $n$) in which no subset of $n$ agents has a $\rho_n$-$\mms(n)$ allocation. In~\cite{KPW18-1} and~\cite{FST21}, $d$ is nearly $2n$, whereas in~\cite{EG26} $d$ is even exponential in $n$. All negative examples for approximate MMS allocations with large $n$ that are known to us either hold in a setting with agent redundancy, or can be adapted to such a setting without significant change in the value of $\rho_n$. 

Our main positive result proves \cref{conj:largen}, but only in the setting with agent redundancy.

\begin{theorem}
 \label{thm:largen}
    For every $\eps > 0$ and $\rho < 1$, there exists some $n_{\eps,\rho}$ such that every $n \ge n_{\eps,\rho}$, in every allocation instance with additive valuations and $d \ge (1 + \eps)n$ agents, there is a subset of $n$ agents that have a $\rho$-$\mms(n)$ allocation.    
\end{theorem}

Note that in \cref{thm:largen}, every agent $i$ that is selected gets value $\rho \cdot \mms(M, v_i, n)$, and not only $\rho \cdot \mms(M, v_i, d)$. As $d \ge n$, $\mms(M, v_i, n) \ge \mms(M, v_i, d)$. Moreover, the ratio between the two might be unbounded, for every $d > n$ (including $d = n+1$).

A straightforward corollary of \cref{thm:largen} is that \cref{conj:largen} holds when the number of items $m$ is noticeably below $2n$.

\begin{corollary}
    \label{cor:largen}
    For every $\eps > 0$ and $\rho < 1$, there is some $n_{\eps,\rho}$ such that every allocation instance with additive valuations, $n \ge n_{\eps,\rho}$ agents and $m \le (2 - \eps)n$ items has a $\rho$-MMS allocation. 
\end{corollary}

We note that for $\rho = 1$, \cref{cor:largen} was  known to hold when $m \le n + \log n$~\cite{feige2026}, and known not to hold when $m \ge n + c\cdot \log n$ for some sufficiently large constant $c$~\cite{EG26}.

A closely related topic is a so called \textit{$1$-out-of-$d$ MMS} for $d > n$, also known as \textit{ordinal MMS}, first introduced in \cite{babaioff2021competitive} (though also considered earlier in \cite{Budish11} with $d = n + 1$).
A partition $(A_1, \ldots, A_n)$ of $M$ is called a \textit{$1$-out-of-$d$ MMS allocation}, if for every agent $i \in N$ it holds $v_i(A_i) \geq \mms(M, v_i, d)$.
In this setting, the main question is how large should $d$ be (as a function of $n$) to guarantee that exact $\mms(d)$ allocations exist.
A series of work leading to \cite{AGS24} established an upper bound of $d = 4\lceil n/3\rceil$, while a recent result by \cite{EG26-2} showed that only sublinear number of additional bundles is needed, with $d = n + O(\sqrt{n\log n})$.

We further reduce the required number of additional bundles to only polylogarithmic in $n$.
\begin{theorem}\label{thm:dmms}
    Any instance with $n$ agents admits an $\mms(d)$ allocation with $d = n + O(\log^2n)$.
\end{theorem}
Our \cref{thm:dmms} proof uses essentially the same techniques as our proof for \cref{thm:largen}.

We note that any results on 1-out-of-$d$ MMS allocations (i.e \cite{EG26-2} or \cref{thm:dmms}) have the following interpretation in the setting with agent redundancy, discussed earlier.
\begin{corollary}\label{thm:equiv}
    For every $\eps > 0$, there exists some $n_\eps$ such that for all $n \ge n_\eps$, in every allocation instance with additive valuations and $d \ge (1 + \eps)n$ agents, every subset of $n$ agents has an $\mms(d)$ allocation. 
\end{corollary}

In some aspects, \cref{thm:equiv} is stronger than  \cref{thm:largen}: it applies to every subset of size $n$ and not just some subsets of size $n$, and there is no need for $\rho < 1$. 
However, the big advantage of \cref{thm:largen} is that it refers to $\mms(n)$ whereas 
\cref{thm:equiv} refers to $\mms(d)$. 
This advantage allows us to infer \cref{cor:largen}. 

\subsection{Overview of our algorithm}

\label{sec:overview}

Here we explain the main steps of our algorithm that proves \cref{thm:largen}, the proof of \cref{thm:dmms} is easier and requires only a few of the ideas of \cref{thm:largen} (even though later we will first give a full proof of \cref{thm:dmms} and only then \cref{thm:largen}).
In the description, we slightly change notation.
The total number of agents is $n$ (this was $d$ in the theorem), and among them we wish to select $n$ agents (this was $d$ in the theorem), with $d = (1 - \eps)n$. 
We let $q$ be the smallest integer satisfying $\rho \le 1 - \frac{2}{q}$. 
Thus, achieving $(1 - \frac{2}{q})$-$\mms(d)$ suffices.

We make the following assumptions, standard in designing approximate MMS allocations.

\begin{assumption}\label{ass:wlog}
    Without loss of generality, in the setting of \cref{thm:largen} we can assume that
    \begin{itemize}
        \item for every agent $i \in N$ and item $e \in M$ it holds $v_i(e) \in [0, 1]$;
        \item for every $i \in N$, $d$-partition $\Bb_i:= (B_{i,1}, \ldots, B_{i,d})$ of $M$ witnessing $\mms(M, v_i, d)$ satisfies
        \[
            \forall j \in [d]:\; v_i(B_{i,j}) = 1;
        \]
        We refer to the bundles in this $d$-partition as the {\em MMS bundles} of agent $i$.
        \item (\cite{Bouveret2016}) items $M = \{e_1, \ldots, e_m\}$ are ordered so that
        for every agent $i \in N$ it holds that
        \[
            v_i(e_1) \geq v_i(e_2) \geq \ldots \geq v_i(e_m).
        \]
    \end{itemize}
\end{assumption}

Under \cref{ass:wlog}, every selected agent should be given a value of at least $1 - 2/q$.

{\bf Head and Tail items.}
For every agent $i$, we partition the items into two classes, the {\em head} items $\{e_1, \ldots, e_{h_i}\}$ that each has value at least $1/q$, and the {\em tail} items $\{e_{h_i + 1}, \ldots, m\}$ that each has value smaller than $1/q$. 
Recall that in approximating the MMS, we aim to lose at most $2/q$.
As we shall see, this loss is composed of two terms, each of value at most $1/q$. 
One term is due to tail items alone (technically, due to the gap between fractional and integral allocations of tail items). Another term is due to head items alone (technically, due to rounding down their values to a multiple of $1/q^2$). 

{\bf Only tail items.}
Consider first the case that all items are tail items for all agents. Namely, every item has value at most $1/q$. {This case can be handled by known techniques.}  Select $d$ arbitrary agents $\{1, \ldots, d\} \in N$, and give each selected agent $i$ all items $e_j$ with $j = k\cdot d + i$, where $k = \{0, 1, 2, \ldots\}$. Every agent $i$ gets a value of at least $v_i(M)/d - v_i(e_1) \ge 1 - 1/q$ (the inequality holds because of \cref{ass:wlog}).

{\bf Only head items.}
Consider now the case that all items are head items. Namely, every item has value at least $1/q$. This case is much more difficult to handle than that of tail items. (Later we will consider also the case of both head and tail items, but the explanation of the most innovative part of our algorithm becomes cleaner if we assume that there are no tail items.)

At a very high level, our plan has two steps. In step one, for some small $\eps'$, we select $(1 - \eps')d$ agents, and give each one of them an acceptable bundle. 
In step two, we gradually increase the set of selected agents to $d$. 
This is done in $\eps' n$ sub-steps, where in each sub-step we replace a small set of selected agents by a new set, larger by one. 
{These sub-steps are made possible by the fact that only a small fraction ($\eps'd$) of agents is missing, whereas there is a considerably larger set of agents from which they can be selected.  This allows the selection of new agents whose valuations are quite similar to each other. Having similar valuations allows for more efficient 
partitioning of items into MMS bundles (in analogy to the fact that MMS allocations exist if all agents have the same valuation function, but sometimes do not exist if valuations are different). Consequently, the number of added agents can be larger than the number of removed agents.}

The main difficulty in our proof is to show that the second step can be done successfully. 
The fact that the first step by itself can be done is an immediate consequence of earlier work, specifically the variation of \cref{thm:equiv} by \cite{EG26-2}. However, due to the need to enable the second step, we do not use \cref{thm:equiv}, and instead do the first step differently. Actually, already in that step we use the fact that we may give selected agents only $\rho$-MMS rather than their full MMS. We now provide more details.


{\bf Profiles and breakpoints.}
Round down the value of each item to the nearest multiple of $1/q^2$. Now, there are fewer than $q^2$ possible item values: $\{\frac{1}{q}, \frac{1}{q} + \frac{1}{q^2}, \ldots, 1\}$. For each agent, each such value defines a class of items that contains all items of this value. 
As the instance is ordered from highest value item to lowest, for each agent, the order of the items imposes an order on the classes, from highest value to lowest. As every MMS bundle contains at most $q$ head items, then after rounding down, its value remains at least $1 - q \cdot \frac{1}{q^2} = 1 - 1/q$.

With each MMS bundle we associate a {\em profile}, which {replaces each item with a corresponding value of its class.}
As there are at most $q$ items in a bundle and each belongs to one out of fewer than $q^2$ classes, the total number of possible profiles is at most $q^{2q}$, which is independent of $n$. Thus, for large $n$, many MMS bundles have the same profile.

Recall that items are ordered in order of decreasing value.
For every agent $i$ and every item $e_j$, we associate a {\em breakpoint} $b_{ij}$. This is the index $\ell$ of the least valuable item that satisfies $v_i(e_{\ell}) \ge v_i(e_j)$. Clearly, $\ell \ge j$. Note that if there are multiple items of the same value, then it can be that $\ell > j$. In our case, in which there are fewer than $q^2$ classes of items, for each agent there are fewer than $q^2$ breakpoint locations (for each non-empty class, the last index of an item in that class is a breakpoint). 

For each agent, for each class of items there is a single breakpoint location associated with all its items. 
With each profile of item classes, we can associate a profile of respective breakpoint locations, where the multiplicity of a breakpoint equals the multiplicity of the respective class in the profile.

Given $t \le n$ agents and a collection of MMS bundles, one for each agent, the profile of breakpoint locations for each of these bundles suffices in order to determine whether this collection of bundles is feasible. Basically, we need to find a matching between items and all breakpoints in the profiles (with multiplicities), where an item can be matched to a breakpoint if its index is not larger than the breakpoint location. A matching exists if Hall's condition holds. In our case, Hall's condition translates to the condition that for every $0 \le \ell \le m$, the number of breakpoints of value at most $\ell$ (we refer to this as the {\em load} on $\ell$) is not larger than $\ell$.

Importantly, even if the collection of selected MMS bundles was not feasible (that is, some items appear in more than one selected bundle), it may still happen that the collection of breakpoint profiles is feasible. This is because we need not give each agent the exact items appearing in her MMS bundle. We can replace each item $e_j$ by a different item $e_{\ell}$ of equal or higher value. As we are considering rounded values, it may well be that that we are allowed to make this replacement even if originally $v_i(e_{\ell}) < v_i(e_j)$.

We summarize the above discussion in the following lemma.

\begin{lemma}
    \label{thm:load}
    Let $A$ be a set of agents, with one MMS bundle selected for each agent. If for every $0 \le \ell \le m$, the load constraint is satisfied (namely, the number of breakpoints of the selected bundles of value at most $\ell$ does not exceed $\ell$), then there is an allocation giving each agent in $A$ her MMS value. (In our context of head items whose value was rounded down, the guarantee becomes $(1 - \frac{1}{q})$-MMS.) 
\end{lemma}

{\bf Satisfying nearly $d$ agents.} Before describing how we actually do the first step of our algorithm (for the case that there are only head items), it is instructive to see how one can do that step if there is no need to enable the future second step of our algorithm. The proof in our setting is considerably simpler than the proof for a variation of \cref{thm:equiv} given in~\cite{EG26-2}, and gives a much smaller value of $\eps$. We sketch our proof.

Take $d$ arbitrary agents. Of them, we will find a collection of $d - O(q^2 \log d)$ agents that satisfy the conditions of \cref{thm:load}. This will be done in $O(\log d)$ iterations, where in each iteration we will set up a linear program (LP, with empty objective), move to a basic feasible solution of the LP, and in the process reduce the number of fractional (non-integral) variables by a constant factor. In the process, some agents become integral (their fractional allocation changes into an integral allocation), until eventually all agents become integral. 
However, each iteration will violate the load constraints by an additive term of $O(q^2 )$. At the very end (after $\log d$ iterations), every load constrain is violated by at most $O(q^2  \log d)$. We fix this by discarding $O(q^2  \log d)$ agents, making all load constraints feasible. Then, \cref{thm:load} implies that all remaining $d - O(q^2  \log d)$ agents can be satisfied.
(Remark: in the rounding proof of \cref{appendix:rounding}, instead of discarding all the agents at the end, we discard $O(q^2)$ agents at every iteration, creating slackness to absorb potential violation when moving to a basic feasible solution).

In the first iteration, we start with the fractional allocation in which each agent selects a $\frac{1}{d}$-fraction
of each of her MMS bundles. Each agent contributes at most $d$ fractional variables (in fact, each agent has one variable for each profile rather than one for each MMS bundle, so the number is smaller). Thus we have a fractional solution with at most $d^2$ variables that satisfies all load constraints in a fractional sense. 

More generally, also at later iterations, we will have some $d'$ agents that did not yet become integral, and a fractional solution for these agents with $f' \ge 2d'$ non-zero variables.
Each variable with positive value corresponds to a profile, and the profile contributes at most $q$ breakpoint locations. So, overall, we have at most $qf'$ breakpoint locations. 

Now, instead of considering all load constraints, select $\frac{f' - d'}{2}$ item indices, referred to as {\em checkpoints}, and consider the loads only at these checkpoints. The checkpoints are selected so that they are ``evenly spaced", in the sense that between any two consecutive checkpoints, at most $O(q)$ of the $d'$ agents have a breakpoint.  Now move to a ``basic feasible solution" for the $\frac{f'+d'}{2}$ linear constraints (the load constraints for $\frac{f' - d'}{2}$ checkpoints, and one constraint for each agent, specifying that she gets one MMS bundle). In this basic feasible solution, there are at most $\frac{f' + d'}{2} \le \frac{3f'}{4}$ (because $f' \ge 2d'$) non-zero variables, reducing the number of fractional variables by a constant factor. As to the load constraints at non-checkpoint locations, they can be violated by at most $O(q^2)$: there are at most $O(q)$ agents contributing breakpoints after the closest earlier checkpoint (in which the load constraint was satisfied), and each such agent can contribute at most $q$ breakpoints. More details appear in \cref{appendix:rounding}.

As a consequence, we have the following proposition.

\begin{proposition}
    \label{prop:almost d}
    For every positive integer $q$, for every instance with $n$ agents with additive valuations in which for each agent every item has value at least $\frac{1}{q}$, there is a $(1 - \frac{1}{q})$-MMS allocation for $n - O(q^2 \log n)$ agents.
\end{proposition}

{\bf The first step, satisfying a set $A$ of $(1 - \eps')d$ agents.} In this step, for some small $\eps'$, we carefully select a set $A$ of $(1 - \eps')d$ agents, and for each selected agent, an associated MMS bundle. This is done in such a way that the collection of associated profiles is feasible. Moreover, this is done in a way that will enable the later performance of the second step of our algorithm. For this reason, we do not simply use \cref{prop:almost d}. Instead, we make a preliminary selection of a subset of agents and a subset of items, and apply \cref{prop:almost d} only on these subsets. We now explain properties of this preliminary selection. 

We will partition the set of agents into three sets: a large set $V$ with at least $(1 - \frac{\eps}{4})n$ agents, a discarded set $W$ of at most $\frac{\eps n}{8}$ agents, and a reserve set $U$ of at most $\frac{\eps n}{8}$ agents. $A$ will be a subset of $V$.

In addition, we partition the set of items into two sets, $E$ and $S$. Set $S$ of {\em saved items} is very small compared to $n$, of size $f(q,\eps)\log n$, where $f$ is a function that depends only on $q$ and $\eps$, but not on $n$. We will have that it suffices to use items of $E$ so as to satisfy all load constraints imposed by the agents in $A$. Items of $S$ are not used in the first step of our algorithm, but can be used in the second step.

Considering the sorted order of items in $M$, the items of $S$ form $O(q^2)$ consecutive intervals, where the first interval starts at $e_1$. Each interval by itself, and also the union of all intervals, is of length at most $f(q,\eps)\log n$. In addition, each interval of $S$ is immediately followed by an auxiliary interval of size somewhat larger than $|S|$. The items in these auxiliary intervals form the set $S'$, and remain in $E$. The significance of these auxiliary intervals is that no agent in $U \cup V$ has a breakpoint in them. That is, all agents that have a breakpoint in an auxiliary interval are moved into $W$. Our construction will be such that $|W| \le \frac{\eps n}{8}$. Moreover, for every two consecutive intervals of $S$ and the corresponding auxiliary intervals of $S'$, the number of agents that have breakpoints between the end point of the first $S'$ and the starting point of the second $S$ will be at most $\frac{n}{2}$. 
See \cref{sec:saved} for more details.

We sketch how the set $S$ is selected. Recall that each agent contributes fewer than $q^2$ breakpoints. So, the number of contributed breakpoints is at most $q^2n$. Consider their locations (several contributed breakpoints might fall on the same item), and mark $2q^2$ start locations (starting at $e_1$), such that between any two locations there are at most $\frac{n}{2}$ breakpoint contributions. These will be the start locations of intervals in $S$ (though intervals that overlap will be merged, so eventually we may have fewer that $2q^2$ intervals in $S$). Now, from each location, we start growing intervals in steps. The initial step size will be of $\Theta(q^4)\log n$ items, and each new step will increase the size of intervals by a factor of $q^{O(q^2)}$.  We stop the process when a certain step did not add more than $\frac{\eps n}{8}$ breakpoint contributions. Thus, the process stops after at most $\frac{8q^2}{\eps}$ iterations. The intervals added at the last iteration form $S'$, and the remaining parts of intervals form $S$. Note that $|S|$ might become as large as $q^{O(q^4/\eps)} \log n$, so we require $n$ to be exponential in a polynomial in $q$. Note also that any single interval of $S$ might contain breakpoints by more than $\frac{n}{2}$ agents, as it may be the result or merging several intervals. This potentially large number of breakpoints (combined with technical details of our algorithm) is the reason why we do not want to allow agents from $A$ to receive items from $S$.

Now, we explain why agents of $V$ can be satisfied by bundles from $E = M - S$. This is because for sufficiently large $n$, the size of $|S|$ is negligible compared to $n$ (and to $d$). Thus, only a negligible fraction of bundles in the MMS partition of an agent contain items from $S$. Starting with $n \ge d(1 + \eps)$ agents, the additional $\eps d$ agents that we have beyond the $d$ each comes with their own MMS partitions, so the number of total bundles that we have in MMS partitions is large enough to compensate for losing those MMS bundles that contain items from $S$. 

At the end of the first step, we will have an allocation of items from $E$ that satisfies all agents from $A$. This allocation will have the {\em large slackness property} that will be very useful for performing the second step. 

\begin{itemize}
    \item Agents from $A$ have no breakpoints in $S'$. Moreover, at every index $\ell$ that corresponds to an item of $E - S'$, the load imposed by the bundles selected by agents of $A$ is significantly smaller than the number of items of $E$ that precede $\ell$.
\end{itemize}

To get some intuition why we would have a large slackness property above, consider a uniform fractional allocation over the MMS partitions.
Observe that at most $q^2$ load constraints can be tight: for every agent, its slack at prefix $\ell$ can equal $0$ only if $\ell$ is a breakpoint, and the total number of breakpoints generated by MMS partitions is at most $q^2n$.
Therefore, for a random load constraint $\ell \in [m]$, the expected slackness of each agent is roughly $\frac{1}{n}\cdot\frac{m}{2q^2}$.
Indeed, between breakpoints the slack of an agent increases by one, so we can split $[m]$ into at most $q^2$ constant load intervals.
An interval with $t$ items contributes at least $\binom{t}{2}$ slack (each additional prefix makes one more item available).
So, the total slack across all prefixes is roughly $\frac{m^2}{2q^2}$, and at a random prefix $\ell \in [m]$ expected slackness of an agent w.r.t uniform allocation is around $\frac{1}{n}\cdot \frac{m}{2q^2}$.
It follows by the same reasoning that for every integer $s$, the number of load constraints with slackness smaller than $s$ is at most $2sq^2$.
So, the vast majority of constraints will have a substantial amount of slackness with respect to agents $A$ selected at the first step.


{\bf The second step, adding $\eps' d$ agents.} Recall the sets of agents $U, V, W$ and set of items $E, S$ from the first step. We start the second step with an allocation of items from $E$ to a set $A \subset V$ of agents, with $|A| = (1 - \eps')d$, where $\eps' d \le O(q^2 \log n)$. In the second step, we perform $\eps' d$ substeps. In each subset a certain number of agents will be discarded from the current $A$, and replaced by a new subset of agents from $U$, larger by~1. 

\begin{remark}
    As we also have agents remaining in $V - A$, it seems strange that we use $U$. However, the reason for distinguishing between $U$ and $V - A$ is due to existence of tail items, an aspect not present when there are only head items. When there are tail items, $U$ is the set of $\frac{\eps n}{8}$ agents with shortest tails.
\end{remark} 

We partition the agents in $U$ into equivalent classes. 
For this purpose, we first partition the set of items into classes, based on $S$. Each item of $S$ is in a class of its own. This gives $|S|$ classes. In addition, each consecutive interval that lies between two adjacent intervals of $S$ is also a class of its own. This gives $2q^2$ additional classes. As $|S| > 2q^2$, we will for simplicity treat $|S|$ as the number of classes of items.

The class of an agent is determined by the locations of its breakpoints. Each breakpoint can be in one of the classes of items, and thus we have at most $|S|^{q^2}$ classes of agents. For fixed $q$ and $\eps$, this term is polylogarithmic in $n$ (as $|S| \le q^{O(q^4/\eps) \log n}$). So, if $n$ is sufficiently large, at least one class of agents in $U$ contains many more agents than $O(q^2 \log n)$, the number of agents that we need to add to $A$. 
We let $U'$ be the largest class of agents within $U$, and ignore the rest of $U$.

Our way of analyzing the existence of allocations is based on locations of breakpoints and loads at selected checkpoints. The checkpoints that interest us in the second step are those at the beginning of each $S'$ interval.
With respect to these checkpoints, all agents in $U'$ are interchangeable. That is, for every profile of item values, if it is feasible for one agent in $U'$, then it is feasible for every agent in $U'$. 
(The same applies to breakpoints withing $S$, because we took these breakpoints into account when we defined the classes of items within $U$.)

It will be convenient for us to take an arbitrary agent $r \in U'$ as a {\em representative} of all agents. We refer to this agent as the {\em supplier}. The MMS partition of $r$ contains $d$ bundles. Consider now the following problem (for a given collection of checkpoints at which load constraints are required to hold): find a value $0 \le t \le |A|$, a set $A^-$ of $t$ agents from $A$ and a set $B^+$ of $t+1$ MMS bundles of $r$, such that replacing in $A$ the bundles of $A^-$ by the bundles of $B^+$ satisfies the load constraints at all checkpoints. This problem has a feasible solution, which is to choose $t = |A|$, and replace $A$ by $|A| + 1$ bundles from the MMS partition of $r$. However, we seek a feasible solution in which $t$ is small (we will need $t < |U'|$). Does such a solution necessarily exist?

The answer to this question in our setting is yes, because the number of checkpoints depends only on $q$. Showing that this is true makes use of a known lemma attributed to Dickson~\cite{Dickson13} (though apparently also proved by others). That lemma states that any collection of non-negative integer vectors of finite dimension has a finite set of minimal vectors (that do not dominate other vectors). We represent feasible solutions to our problem as integer vectors of dimension that depends only on $q$, in a system of vectors for which any minimal vector must have entries upper bounded by some (large) function of $q$. As our problem has a solution, it corresponds to some vector in this system. Consider any minimal vector dominated by our vector. We show that the minimal vector also corresponds to a feasible solution of our problem, with $t$ upper bounded by a function of $q$ (and hence smaller than $|U'|$, when $n$ is sufficiently large). 

The solution with $t < |U'|$ can be translated to a solution in which instead of taking the bundles of $B^+$ from the MMS partition of $r$, we select $t+1$ agents from $U'$, where for every $j$ the $j$th selected agent gets a bundle with the same profile as that of the $j$th bundle in $B^+$. As agents in $U'$ are interchangeable, the solution remains feasible. 

We remark that MMS bundles of agents from $U'$ may contain items from $S$. The combination of $S$ being small, having no breakpoints in $S'$ and the large slackness property for the initial set $A$ can be shown to imply that also at locations that are not among our selected checkpoints, the required load constraints hold. 
This finishes the description of head-only proof.

{\bf Mixture of head and tail items.} We have seen (an overview of) how to prove \cref{thm:largen} when either all items are head items, or all items are tail items. We now explain how to handle instances with combination of head items and tail items. We basically follow the approach outlined for the case of only head items, but with some adjustments. Note that the same item can be a head item for some agents, but a tail item for other agents, a fact that somewhat complicates the situation.

We shall use the known fact that fractional allocations can be rounded to integer allocations, where each agent receives a value that differs from that of her fractional allocation by the value of at most one item. Hence, for tail items (for which a single item has value at most $\frac{1}{q}$, and losing such a value can be tolerated), we may consider fractional allocations.

For each agent $i$, 
let $T_i$ denote the total value of the set of her tail items. 

For each bundle $P_j^i$ in the MMS partition of $i$, we consider head and tail items separately. For head items, we follow the approach outlined for the case in which there only are head items. That is, we round their values down to multiples of $\frac{1}{q^2}$, consider the profile that they induce and the respective breakpoints. Tail items are treated differently. We only keep track of the total value of tail items that an MMS bundle contains, but not their exact identity. This value is then translated into what fraction of $T_i$ the bundle needs to contain. Then, we think of this fraction as representing which fraction of each item in the tail of agent $i$ the bundle should contain.  

We distinguish between two cases, depending on whether the average value of $T_i$ is large or small. The breakpoint between these two cases is an average value of $g(q) \cdot \log n$, where $g$ is some specific function of $q$ and does not depend on $n$. 

If the average value of $T_i$ is large, we reduce the fraction that each agent desires to get from each tail item by a factor of $(1 - \frac{1}{q})$. (This affects the value of each MMS bundle by at most $\frac{1}{q}$, and hence introduced an additive error of at most $\frac{1}{q}$ in our final MMS approximation ratio.) This frees up a fraction of $\frac{1}{q}$ from each tail item. For those agents with relatively large $T_i$ (there must be such agents, because the average value of $T_i$ is not very small), the unused fractions of tail items can be used in order create additional bundles of value~1. This creates a substantial number of agents with an MMS partition into substantially more than $d$ parts. This can be used in order to make the first step of our algorithm produce a set of size $d$, without the need for the second step.

If the average value of $T_i$ over all $i$ is small, we modify the ``only head items" algorithm as follows. In the profiles associated with MMS bundles, we consider only head items. After discarding the set $W$ of agents, we choose the set $U$ to contain the $\frac{\eps n}{8}$ agents with fewest tail items.
For agents in $V$, we consider only MMS bundles for which the profile of head items has sufficient value, without need for tail items. As tail items contribute very little value on average, almost all agents in $V$ still maintain a partition into almost $d$ bundles. The analysis with respect to $V$ then proceeds in a way similar to the ``only head items" case, except that the resulting set $A$ of selected agents becomes slightly smaller. This has no significant effect of the proof. 


The main issue that needs to be handled in the proof is the second step of our algorithm. To use Dickson's lemma, we used a representative supplier agent $r$ from the set $U'$, and assumed that this agent had an MMS partition into $d$ parts. For agents in $U'$, this indeed holds, but the parts include also tail items. Hence, beyond handling the load on head items (which our use of Dickson's lemma does successfully), we need to also make sure that the load on tail items is feasible. This is accomplished by two clever choices that our algorithm makes. One is our choice of $U$, that implies that no agent from $A$ takes an item that is a tail item for an agent from $U'$. Thus, handling tail items for agents selected from $U'$ does not affect the allocation given to agents from $A$. This by itself does not suffice, the difficulty being that incompatibility between agents in $U'$ in terms of which is the starting item of the tail. The other clever choice is that of the supplier $r$. The supplier is selected from $U'$ as the agent that places the smallest demand on the tail, under a non-trivial definition of ``demand" that takes into account both value and length of tail. This ensures that the MMS bundles of the supplier chosen by the application of Dickson's lemma can indeed be replaced by bundles of other agents from $U'$, while satisfying load constraints for every items, even though agents in $U'$ may disagree on whether an item is a head item or a tail item.

\subsection{Related work}

The subject of fair division has been extensively studied in multiple different settings.
In our work, we focus specifically on indivisible goods and use MMS as our notion of fairness.
For work on divisible goods, see for example~\cite{Brams_Taylor_1996}.
For additional work on fair allocation of indivisible items, including envy-based notions like EF1 and EFX, see for example~\cite{AzLiMoWu22,  AmanatidisSurvey23}.
In addition, there has been a plenty of work done on fair allocations for valuations beyond additive, i.e submodular, XOS and subadditive classes from the complement free hierarchy \cite{LLN}, 
which is beyond the scope of this work.

For additive valuations and every $n \geq 3$, there are examples in which MMS allocations do not exist~\cite{KPW18-1}. {There has been a long line of work investigating the highest fraction of the MMS that can be guaranteed to each agent, and it is known to be between $\frac{7}{9}$~\cite{huang2025} and $\frac{20}{21}$~\cite{EG26}.}
For general $n$, \cite{FST21} showed an upper bound of $1 - \frac{1}{n^4}$, which was improved to $1-\Omega(\frac{1}{\log^2 n})$ by \cite{EG26}.

There has also been work on MMS allocation instances where the number of items $m$ is small relative to the number of agents $n$.
The most recent discoveries show that MMS allocations always exist when $m \leq n + \log n$ \cite{feige2026}, and may not exists when $m \leq n + c\log n$ for sufficiently large 
constant $c$ \cite{EG26}.
These works study exact MMS allocations, while our existence results consider approximate allocations in the regime $m < 2n$.

While we consider instances with large number of agents $n$, there has been additional results on instances with small $n$.
For additive valuations, for the case of $n = 3$ agents, it was shown in \cite{FeNo22} that $\frac{11}{12}$-MMS allocations exist, and for $n = 4$ agents, \cite{GhodsiHSSY22} proved an approximation ratio of $\frac{4}{5}$.

The concept of $1$-out-of-$d$ MMS allocations first appeared in \cite{Budish11}, and later was formally introduced in \cite{babaioff2021competitive}.
A series of works subsequently improved the upper bound on $d$, with notable milestones of $d = 4\lceil n/3\rceil$ by \cite{AGS24} and $d = n + O(\sqrt{n\log n})$ by \cite{EG26-2}.
As for the progress towards $d = n + 1$, \cite{schwer26} showed that $1$-out-of-$5$ allocations exist for $n = 4$ agents.

Regarding techniques, our algorithm utilizes the identical ordering reduction from \cite{Bouveret2016}.
When it comes to replacing MMS bundles with corresponding profiles of rounded item values, the idea is not new and has been utilized in various ways before.
In \cite{KulkarniMehtaTaki2021}, the authors compute MMS of a single agent by representing its MMS bundles via counts of its large items in rounded value classes, and completed selected bundle types using small items.
The construction somewhat resembling our head-tail profile framework was utilized in \cite{NR20} for welfare approximations.

In the recent work \cite{EG26-2}, most relevant to our topics, the authors also built their proofs using a sort of head and tail structure.
However, their definition of head profiles is vastly different and is based on counts of  items in certain buckets, and they control the ``loads'' of the buckets.
Our approach instead defines head profiles based on individual value of each item, and operates with prefix loads and value breakpoints.
This construction gives us much more flexibility, which is crucial to our main result.

Both of the works utilize linear programming rounding in order to obtain initial allocation for most of the agents.
While the rounding of \cite{EG26-2} is in some sense oblivious, our rounding algorithm is much more adaptive, which combined with our definition of head/tail profiles gives us the substantial improvement in the value of $d$ for $1$-out-of-$d$ allocations.

In order to compensate the head profiles with tail items, \cite{EG26-2} use faithful rounding method (see \cite{BEF22BoBW}) for the common tail.
Our tail completion approach is also inspired by faithful rounding, and develops its deterministic counterpart.
The main distinction in our case is that tails for different agents may have different lengths, and we need to control all of them simultaneously.

It is worth mentioning that the approach of \cite{EG26-2} was developed for a stronger notion of fairness called \textit{anyprice share} (APS), which considers fractional partitions of items.
For our results, it is crucial that we work with MMS partitions.
\subsection{Technical details}

We prove \cref{thm:dmms} first, in \cref{sec:dmms}.
The proof of \cref{thm:largen} is split between \cref{sec:large,sec:small,sec:supp}.
In \cref{sec:large}, we first consider mixed head-and-tail case when average value of tail $T_i$ is large, and also show how to complete the first step, i.e satisfying $(1 - \eps')d$ agents.
In \cref{sec:small}, we cover the first half of the second step (adding $\eps'd$ agents), and show how to construct set $S$ with corresponding checkpoints.
We complete the proof of \cref{thm:largen} in \cref{sec:supp}, where we describe how to use a supplier agent to add the required $\eps 'd$ agents, using Dickson's lemma.
In \cref{sec:corol}, we give a proof of \cref{cor:largen}, and mention some additional minor implications of our results.
As for appendices, \cref{appendix:etarounding,appendix:rounding} cover the proofs of LP-rounding algorithms that we used in respectively \cref{sec:dmms,sec:large}.
\cref{appendix:supplier} gives the full proof of the supplier profile selection guarantees via Dickson's lemma.
In \cref{appendix:constant} we give explicit constants and provide an upper bound on the claimed $n_{\eps,q}$ in \cref{thm:largen}.

\section{$1$-out-of-$d$ MMS for $d = n + O(\log^2 n)$}
\label{sec:dmms}
First we will prove the result on $1$-out-of-$d$ MMS (\cref{thm:dmms}).
Its proof is simpler, and will allow us to introduce the reader to the core techniques of our approach.
\begin{theorem}[\cref{thm:dmms} restated]\label{dmms}
    Consider a fair allocation instance with items $M$ and $n$ additive agents $N$ with valuations $\{v_i\}_{i \in N}$.
    There always exists an $\mms(d)$ allocation when
    \[
        d \geq n + 46(1 + \log n)^2,
    \]
\end{theorem}
In the setting of \cref{dmms}, we assume that \cref{ass:wlog} holds.
In particular, we associate $M$ with the set $[m]$, and for every $i \in N$ we simultaneously have
\[
    v_i(1) \geq v_i(2) \geq \ldots \geq v_i(m).
\]
In addition, each agent $i \in N$ has a $d$-partition $\Bb_i = \{B_{i,1}, \ldots, B_{i,d}\}$ of $M$, where $v_i(B_{i,j}) = 1$ for all $j \in [d]$.
Notably, $v_i(M) = d$ for every $i \in N$.

We will start with several straightforward observations.
\begin{lemma}\label{leftovers}
    Let $U\subseteq N$ and $E\subseteq M$ be such that $v_u(E) \geq 2|U| - 1$ for every $u \in U$.
    There exists an allocation of $E$ to $U$ such that every agent $u \in U$ receives a bundle of value at least $1$.
\end{lemma}
\begin{proof}
    If $|U| = 1$, give all of $E$ to a single agent in $U$.
    For $|U| > 1$, add goods in arbitrary order to an initially empty set $B$, until some agent $u \in U$ values $v_u(B) \geq 1$.
    Note that just before the last item was added to $B$, every agent valued $B$ less than $1$, and for every $u'\neq u$, $v_{u'}(B) < 2$.
    Hence, $v_{u'}(E - B) \geq 2|U| - 3 = 2(|U|-1) - 1$.
    The lemma follows by induction on $|U|$.
\end{proof}

If $d \geq 2n - 1$, we can take $U = N$ and $E = M$ and apply \cref{leftovers}.
For every agent $i \in N$, $v_i(E) = v_i(M) = d \geq 2n - 1 = 2|U| - 1$.
From now on, assume $n + 1 \leq d \leq 2n - 1$.

\cref{leftovers} suggests the following.
Suppose that we allocated all but a small fraction of agents (say, $n - O(\log^2 n)$), using some set of items $F\subseteq M$.
Then, as long as the total value of $F$ is not too large, the remaining agents can get their MMS from only the remaining items.

\begin{corollary}\label{leftoveralloc}
    Let $F\subseteq M$ satisfy $v_i(F) \leq n + 1$ for every $i \in N$.
    For any set $U\subseteq N$ of $|U|\leq (d - n)/2$ agents, there exists an allocation of $M- F$ to $U$ such that every agent $u \in U$ receives a bundle of value at least $1$.
\end{corollary}
\begin{proof}
    If $U = \varnothing$, there is nothing to do.
    Otherwise, as $v_u(M) = d$ for every $u \in U$,
    \[
    v_u(M - F)  = v_u(M) - v_u(F) \geq d - n - 1 \geq 2|U| - 1.
    \]
    This is exactly the condition of \cref{leftovers} with sets $E = M - F$ and $U$.
\end{proof}

We will now focus on finding a set $F\subseteq M$ of value $v_i(F)\leq n + 1$ for all $i \in N$ such that, for some set $N'\subset N$ of $n - O(\log^2n)$ agents, all $N'$ can get their MMS allocation from $F$ only.

By assumption, we associate $M$ with $[m]$, and for every agent $i \in N$ we simultaneously have
\[
    v_i(1) \geq v_i(2) \geq \ldots \geq v_i(m).
\]
We observe that, if one wants to bound the value of items $F\subseteq M$, it suffices to bound the total number of items from $F$ in every \textit{prefix} of $M = \{1, \ldots, m\}$.
\begin{observation}\label{prefixbound}
    Let $F\subseteq M$, and suppose that for every $0 \leq \ell \leq m$ it holds $|F\cap [\ell]| \leq \eta \ell$ for some value $\eta \in (0, 1]$.
    Then for every agent $i \in N$ it holds $v_i(F)\leq \eta d$.

    In particular, when $\eta = (n + 1)/d$, such set $F$ satisfies $v_i(F)\leq n + 1$.
\end{observation}
\begin{proof}
    Observe that for every $0 \leq \ell \leq m$, $\ell\in F$ if and only if $|F\cap [\ell]| - |F\cap [\ell - 1]| = 1$.
    Hence,
    \begin{multline*}
        v_i(F) = \sum_{\ell = 1}^mv_i(\ell)\cdot \big(|F\cap [\ell]| - |F\cap [\ell - 1]|\big) = \sum_{\ell = 1}^m\big(v_i(\ell) - v_i(\ell + 1)\big)\cdot |F\cap[\ell]|
        \\
         \leq \eta\sum_{\ell = 1}^m\big(v_i(\ell) - v_i(\ell + 1)\big)\cdot \ell = \eta\sum_{\ell = 1}^mv_i(\ell) =  \eta d.
    \end{multline*}
    We used a conventions $[0] = \varnothing$ and $v_i(m + 1) = 0$.
\end{proof}

We will refer to $|F\cap [\ell]|$ as a \textit{$\ell$-prefix load} of the set $F$.
Our approach to finding a set $F$ with bounded prefix loads is to, instead of working with actual items $M$ and allocating them to bundles, to consider \textit{profiles of values} of items from $M$.
We give more details below.

\subsection{Demand profiles}

\begin{definition}
    A \textbf{demand} is a positive value $\alpha \in (0, 1]$.
    We say that item $e \in M$ \textbf{satisfies }the demand $\alpha$ for agent $i \in N$ if $v_i(e) \geq \alpha$.
\end{definition}
\begin{definition}
    A \textbf{demand profile} is a finite multiset $\pi = \{\alpha_1, \ldots, \alpha_{|\pi|}\}$ of demands.
    A subset $B\subseteq M$ \textbf{satisfies} the demand profile $\pi$ for agent $i \in N$ if each demand $\alpha \in \pi$ can be assigned a distinct item $e\in B$ that satisfies it.
    That is, there exists $\{e_1, \ldots, e_{|\pi|}\}\subseteq B$ such that \[
        v_i(e_1) \geq \alpha_1,\ldots,  v_i(e_{|\pi|})\geq \alpha_{|\pi|}.
    \]
    A (guaranteed) \textbf{value of a profile} $\pi$ is 
    \[
        v(\pi) := \sum_{\alpha \in \pi}\alpha.
    \]
    If agent $i$ is assigned profile $\pi$, satisfying $\pi$ with some bundle $B\subseteq M$ guarantees $v_i(B) \geq v(\pi)$.
\end{definition}

Instead of measuring a prefix load of item set $F\subseteq M$, we will measure a prefix load of a \textit{demand profile} $\pi$.
Given some profile $\pi$, we want to know for each agent $i$ how many items from each prefix $\{1, \ldots, \ell\}$ agent $i$ needs to satisfy $\pi$.
We will do so using a notion of \textit{breakpoints}.
\begin{definition}\label{breakpoint}
    For $i \in N$ and demand $\alpha$, an $\alpha$-\textbf{breakpoint} for agent $i$ is the number $m_i(\alpha)$ of items in $M=[m]$ with value at least $\alpha$.
    That is, for every $e \leq m_i(\alpha)$ it holds $v_i(e) \geq \alpha$.
\end{definition}
By definition, agent $i$ can satisfy a demand $\alpha$ only by the items $\{1, \ldots, m_i(\alpha)\}$.
Hence, if $m_i(\alpha) \leq \ell$, then agent $i$ must use an item in prefix $\{1, \ldots, \ell\}$ to satisfy $\alpha$.
Therefore, it is natural to define the \textit{$\ell$-prefix load of a profile} $\pi$ as a total number of breakpoints of $\pi$ before $\ell$.
\begin{definition}\label{prefixload}
    Let $\pi = \{\alpha_1, \ldots, \alpha_{|\pi|}\}$ be a demand profile.
    For an integer $0 \leq \ell \leq m$ and agent $i \in N$, denote by $\Pp_i(\pi, \ell)$ the total number of demands in $\pi$ with breakpoints at most $\ell$:
    \[
        \Pp_i(\pi, \ell) = |\{\alpha \in \pi : m_i(\alpha) \leq \ell\}|.
    \]
    That is, $\Pp_i(\pi, \ell)$ counts how many demands in $\pi$ can be satisfied only by the first $\ell$ items.
    We refer to $\Pp_i(\pi, \ell)$ as an \textbf{$\ell$-prefix load} of $\pi$ for $i$.
\end{definition}
We extend the definition of prefix load to a collection of profiles.
\begin{definition}
    Let $I\subseteq N$, let $\Pi_I := \{\pi_i\}_{i \in I}$ be a collection of demand profiles, where $\pi_i$ is assigned to agent $i \in I$.
    For $0 \leq \ell \leq m$, the \textbf{total $\ell$-prefix load} of $\Pi_I$ is
    \[
        \Pp(\Pi_I, \ell) := \sum_{i \in I}\Pp_i(\pi_i, \ell),
    \]
    the total number of demands across all agents in $I$ that can be satisfied only by the first $\ell$ items.
\end{definition}

The following lemma shows that finding an allocation of $M$ to some agents $I\subseteq N$ is equivalent to finding a feasible collection of appropriate demand profiles $\Pi_I$.
\begin{lemma}\label{loadequiv}
    Let $I\subseteq N$, let $\pi_i = \{\alpha_{i,1}, \ldots, \alpha_{i,|\pi_i|}\}$ be some demand profile of agent $i \in I$, and let $\Pi_I := \{\pi_i\}_{i \in I}$.
    For any set $F\subseteq M$ of items, we can simultaneously satisfy all profiles of $\Pi_I$ using distinct items of $F$ if and only if 
    \[
        \Pp(\Pi_I, \ell) \leq |F\cap [\ell]|,\qquad\qquad\forall 0 \leq \ell \leq m.
    \]
    Furthermore, if $K := \sum_{i \in I}|\pi_i|$, it suffices to use only the first $K$ items of $F$.
    In particular, when $F = M$, we can satisfy all profiles of $\Pi_I$ using only items $\{1, \ldots, K\}$ as long as
    \[
        \Pp(\Pi_I, \ell) \leq \ell,\qquad\qquad\forall 0 \leq \ell \leq m.
    \]
\end{lemma}
\begin{proof}
    Suppose that we can satisfy all profiles of $\Pi_I$ using distinct items of $F$.
    By definition,
    \[
        \Pp(\Pi_I, \ell) = \sum_{i \in I}\Pp_i(\pi_i, \ell) =\sum_{i \in I}|\{\alpha \in \pi_i : m_i(\alpha) \leq \ell\}|.
    \]
    Consider all $\alpha \in \Pi_I$ with breakpoints at most $\ell$ (for corresponding agents), each of these values must be satisfied by an item in $\{1,\ldots, \ell\}$.
    Every such demand gets a distinct item in $F$ satisfying it, so there are at most $|F\cap [\ell]|$ demands with breakpoints at most $\ell$.
    Thus, $\Pp(\Pi_I, \ell) \leq |F\cap [\ell]|$.

    Now, suppose that $\Pp(\Pi_I, \ell) \leq |F\cap [\ell]|$ for every $0 \leq \ell \leq m$.
    Taking $\ell = 0$ we obtain $\Pp(\Pi_I, 0) \leq |F\cap[0]| = 0$, so no demand in the assigned profiles has breakpoint $0$.
    Consequently, every demand has at least one eligible item.
    
    List all individual demands $\alpha \in \Pi_I$ as $(i, \alpha)$ where $\alpha\in \pi_i$, including repetitions, and sort them according to breakpoint values $m_i(\alpha)$.
    We get a list of $K = \sum_{i \in I}|\pi_i| =\Pp(\Pi_I, m)$ pairs
    \[
        (i_1, \alpha_1), \ldots, (i_K, \alpha_K), \qquad \qquad 1\leq m_{i_1}(\alpha_1) \leq \ldots \leq m_{i_K}(\alpha_K)\leq m.
    \]
   Note that $K = \Pp(\Pi_I, m) \leq |F\cap [m]|$ implies $K \leq |F|$.

    Denote $m_k := m_{i_k}(\alpha_k)$, $m_1 \leq \ldots \leq m_K$.
    By definition, for each $k$ there are at least $k$ pairs $(i, \alpha)$ with breakpoint $m_i(\alpha) \leq m_k$.
    This means that there are at least $k$ different demands that want items among the first $m_k$, so $\Pp(\Pi_I, m_k) \geq k$.
    
    By assumption, $\Pp(\Pi_I, m_k) \leq |F\cap [m_k]|$.
    It follows that, if $F= \{f_1 < \ldots < f_{|F|}\}$, then $f_k \leq m_k$.
    Take this $m_k = m_{i_k}(\alpha_k)$ and allocate item $f_k$ to agent $i_k$ to satisfy $\alpha_k$.
    As $m_k \geq f_k$, it must be that $v_{i_k}(f_k) \geq \alpha_k$.
    We do so for every $1 \leq k \leq K$.
    Since we will always use distinct items, and $K \leq |F|$, we will use only the first $K$ items of $F$.
    The lemma follows.
\end{proof}

Recall that in \cref{prefixbound}, however, we seek for a set $F\subseteq M$ such that prefix load of $F$ is $|F\cap [\ell]|\leq \eta \ell$, while \cref{loadequiv} guarantees only a trivial bound of $\ell$.
A slight modification of the proof of \cref{loadequiv} allows us to get the desired set $F\subseteq M$, provided that the given collection of profiles $\Pi_I$ satisfies a stronger prefix load condition.

\begin{corollary}[Refinement of \cref{loadequiv}]\label{etaload}
    Let $I\subseteq N$, let $\pi_i$ be some demand profile of agent $i \in I$, and let $\Pi_I := \{\pi_i\}_{i \in I}$.
    Let $\eta \in (0, 1]$, and suppose that for every $0 \leq \ell \leq m$ it holds
    \[
        \Pp(\Pi_I, \ell) \leq \eta\ell.
    \]
    Then, all profiles of $\Pi_I$ can be simultaneously satisfied using distinct items of $F\subseteq M$ such that
    \[
        |F\cap [\ell]|\leq \eta \ell,\qquad \forall 0 \leq \ell \leq m.
    \]
\end{corollary}
\begin{proof}
    List all demands $\alpha \in \Pi_I$ as pairs $(i, \alpha)$ where $\alpha\in \pi_i$, including repetitions, and sort them according to breakpoint values $m_i(\alpha)$.
    We obtain a sorted list of $K = \sum_{i \in I}|\pi_i| =\Pp(\Pi_I, m)$ pairs
    \[
        (i_1, \alpha_1), \ldots, (i_K, \alpha_K), \qquad \qquad 1\leq m_{i_1}(\alpha_1) \leq \ldots \leq m_{i_K}(\alpha_K)\leq m.
    \]
    Denote $m_k := m_{i_k}(\alpha_k)$, $m_1 \leq \ldots \leq m_K$.
    Following the proof of \cref{loadequiv}, we have for every $k \in [K]$ that $k \leq \Pp(\Pi_I, m_k) \leq\eta m_k$.
    Take this $m_k = m_{i_k}(\alpha_k)$, and allocate item $f_k := \lceil k/\eta\rceil$ to agent $i_k$.
    Note that $1 \leq f_k \leq m_k \leq m$, so $f_k \in M$ and satisfies $\alpha_k$.
    Since $\eta \leq 1$, $f_{k + 1} \geq f_k + 1$, hence such allocation will use distinct items for every demand.

    Let $F = \{f_k\}_{k \in [K]}$ be the set of allocated items.
    We immediately obtain for every $0\leq \ell \leq m$:
    \[
        |F\cap [\ell]| = |\{ k \in [K] : \lceil k/\eta\rceil \leq \ell\}| = |\{k \in [K] : k \leq \eta\ell\}| =  \min\{K, \lfloor \eta \ell\rfloor\} \leq \eta \ell.
    \]
\end{proof}

So, to find a set $F\subseteq M$ with bounded prefix loads, it suffices to find a collection $\Pi_I$ of demand profiles of some agents $I \subseteq N$ with bounded prefix loads.
A natural starting point would be the profiles of $i \in N$ induced by its MMS bundles $\Bb_i = \{B_{i,1}, \ldots, B_{i, d}\}$.
That is, some $\pi_{i,j} = \{v_i(e) : e \in B_{i, j}\text{ and }v_i(e) > 0\}$.
The issue here is that the total number of demands in $\pi_{i,j}$ can be as large as $m$.
For this reason, we will split each profile $\pi$ into two: a \textit{head} profile with bounded number of demands, and a \textit{tail} profile.

\subsection{Head and tail profiles}

As per \cref{etaload}, we would like to find a set $I\subseteq N$ of agents and assign to them some value-$1$ profiles with bounded total prefix load.
If $|I|$ is sufficiently large, \cref{leftoveralloc} guarantees that unassigned agents will be able to receive a bundle of value at least $1$.

Instead of finding a collection of full profiles of value $1$ for some subset $I$ of agents, at once, we will split the task and separately find a collection $\Pi_I$ of \textit{head profiles} $\pi_i \in \Pi_I$ and a collection $\Psi_I$ of \textit{tail profiles} $\psi_i \in \Psi_I$.
Agent $i \in I$ in this case receives a full profile $\{\pi_i, \psi_i\}$.

\begin{definition}
    For $h \in [m]$, we call $\{1, \ldots, h\}$ \textbf{head items}, and $\{h + 1, \ldots, m\}$ \textbf{tail items}.
    For agent $i \in N$, the \textbf{tail value} of $i$ is
    \[
        T_i = \sum_{e > h}v_i(e).
    \]
\end{definition}
\begin{definition}
    For an agent $i \in N$, we call a demand profile $\pi$ a \textbf{head profile} for $i$, if every $\alpha \in \pi$ satisfies $\alpha = v_i(e)$ for some head item $e \leq h$.
    
    Similarly, we call a demand profile $\psi$ a \textbf{tail profile} for $i$, if every $\beta \in \psi$ satisfies $\beta = v_i(e)$ for some tail item $e > h$.
\end{definition}

We will also need two additional conventions.
\begin{definition}
    An \textbf{empty profile} is $\varnothing$.
    If agent $i \in N$ is assigned an empty head profile $\pi = \varnothing$ or an empty tail profile $\psi = \varnothing$, it means that he may receive no items from head or tail.
    It always holds $\Pp_i(\varnothing, \ell) = 0$ for all $0 \leq \ell \leq m$.
\end{definition}
We allow an agent $i \in N$ to be one of $I$ assigned ones, but get an empty profile $\pi_i = \varnothing$ or an empty tail profile $\psi_i = \varnothing$.

We will also define a ``no-profile'' $\bot$, which corresponds to an option of not choosing $i \in N$ into the final set of agents $I$.
\begin{definition}
    A \textbf{no-profile} is $\bot$.
    If agent $i \in N$ is assigned $\bot$, it means that we do not select this agent, and he receives no items at all.
    It always holds $\Pp_i(\bot, \ell) = 0$ for all $0 \leq \ell \leq m$.
\end{definition}

Our approach is to first find a sufficiently large set $I\subseteq N$ of agents and assign to them \textit{only head profiles} $\Pi_I = \{\pi_i\}_{i \in I}$, that use only items of $[h]$.
All the other $N - I$ agents will get $\bot$, and will get a bundle from \cref{leftoveralloc} later.
After each $i \in I$ gets a head profile $\pi_i$, we will find a corresponding tail profile $\psi_i$ so that $v(\pi_i) + v(\psi_i) \geq 1$.
\cref{etaload} then guarantees that $i$ will get a bundle of value at least $1$.

Since profiles $\Pi_I$ are not using all items of $M$, many profiles $\pi_i \in \Pi_I$ may have $v(\pi_i) < 1$.
We introduce a notion of profile deficit, to measure much value $i \in I$ misses from $\pi_i$.

\begin{definition}
    For a head profile $\pi \neq \bot$, its \textbf{deficit} is
    \[
        \Delta(\pi) := (1 - v(\pi))_+.
    \]
    Here $(\cdot)_+ := \max\{0, \cdot\}$.
\end{definition}

Suppose that agent $i \in N$ gets assigned a head profile $\pi_i\neq \bot$.
If $\Delta(\pi_i) > 0$, then satisfying $\pi_i$ with a subset of head items $[h]$ does not guarantee that $i$ gets a bundle of value $1$.
So, we will need to use some items from the tail $\{h + 1, \ldots, m\}$ to compensate for the missing value.

We do so using an approach inspired by faithful rounding method, which also appeared in \cite{BEF22BoBW,EG26-2}.
Suppose that, in addition to head items satisfying $\pi_i$, agent $i$ could take the same $\frac{\Delta(\pi_i)}{T_i}$-fraction from every tail item $e > h$.
Then, $i$ is guaranteed to receive a value at least
\[
    v(\pi_i) + \frac{\Delta(\pi_i)}{T_i}\sum_{e > h}v_i(e) = v(\pi_i) + \Delta(\pi_i) \geq 1.
\]

In essence, value $\frac{\Delta(\pi_i)}{T_i}$ defines a ``fractional allocation'' of the tail items to agent $i$.
While it is known  how to round a fractional allocation by sacrificing at most one item, we would like to give each agent total value $1$ exactly, so giving up items is not an option here.
We will provide sufficient conditions on the prefix load of head profiles $\pi_i$ that allows ``rounding'' this fractional allocation without any loss.

\begin{definition}
    For agent $i \in N$ and head profile $\pi$, let
    \[
        \tau_i(\pi) := \begin{cases}
            \Delta(\pi)/T_i,& \pi\neq \bot\text{ and }T_i > 0;\\
            0,& \text{otherwise}.
        \end{cases}
    \]
\end{definition}
Consider some prefix $\ell > h$, where we potentially may use tail items.
Taking the same $\tau_i(\pi_i)$-fraction from every tail item $e > h$ can be viewed as generating only a $\tau_i(\pi_i)$-fraction of a demand per item taken.
So, for a prefix $\ell > h$, the total additional load  agent $i$ generated via this process, i.e the amount of added demands, is ``essentially'' $\tau_i(\pi_i)(\ell - h)$.
This gives an idea on how much slackness the head profiles will need to allow completing the tail profiles.
\begin{lemma}[Tail completion]\label{etatailcomplete}
    Let $I\subseteq N$, let $\pi_i$ be a head profile assigned to agent $i \in I$, and let $\Pi_I = \{\pi_i\}_{i \in I}$.
    Assume that $T_i = 0$ implies $\Delta(\pi_i) = 0$ for every $i \in I$.
    For $\eta \in (0, 1]$, suppose that for every $0 \leq \ell \leq h$ we have $\Pp(\Pi_I, \ell) \leq \eta \ell$, and for every $h + 1 \leq \ell \leq m$ we have
    \[
         \Pp(\Pi_I, \ell) \leq \eta\ell -(\ell - h)\sum_{i \in I}\tau_i(\pi_i) - |I|.
    \]
    There exists a collection $\Psi_I := \{\psi_i\}_{i \in I}$ of tail profiles, where agent $i$ is assigned $\psi_i$, such that
    \[
        \forall 0 \leq \ell \leq m:\; \Pp(\Pi_I, \ell) + \Pp(\Psi_I, \ell) \leq \eta \ell,
    \]
    and for every $i \in I$ it holds $v(\pi_i) + v(\psi_i) \geq 1$.
\end{lemma}
\begin{proof}
    Consider $i \in I$, and denote $\tau_i := \tau_i(\pi_i)$.
    Construct the tail profile $\psi_i$ as follows.
    We break the tail $\{h + 1, \ldots, m\}$ into segments of length roughly $ 1/\tau_i$, and in each segment add the value of the first item to $\psi_i$.
    That is, for every $t = 1, \ldots, \lceil \tau_i\cdot  (m - h)\rceil$ add $v_i(e_{i,t})$ to $\psi_i$, where
    \[
        e_{i,t} = h + 1 + \left\lfloor \frac{t - 1}{\tau_i}\right\rfloor.
    \]
    The intuition behind this construction is straightforward: after taking the first item of every segment of length $1/\tau_i$, the total value of selected items is at least as large as if we took a $\tau_i$-fraction of every tail item.
    If $\tau_i = 0$, then we simply put $\psi_i = \varnothing$.
    
    We can describe this construction equivalently as follows.
    For $1 \leq k \leq m - h$, let $\beta_{i,k} := v_i(h + k)$, i.e the value of $k$-th tail item for agent $i$.
    Then, for every $k \in [m - h]$, we add exactly
    \[
        \lceil \tau_i \cdot k \rceil - \lceil \tau_i\cdot (k - 1)\rceil 
    \]
    copies of $\beta_{i,k}$ to the profile $\psi_i$.
    To see this, observe that
    \[
        e_{i,t} \leq h + k \iff   \left\lfloor \frac{t - 1}{\tau_i}\right\rfloor \leq k - 1 \iff t - 1 < \tau_i\cdot k \iff t \leq \lceil \tau_i\cdot k\rceil.
    \]
    So, the first $k$ tail items contain exactly $\lceil \tau_i\cdot k\rceil$ selected items from $\psi_i$, and the number of copies of tail item $k$ is exactly $\lceil \tau_i \cdot k \rceil - \lceil \tau_i\cdot (k - 1)\rceil $.
    Note that both constructions allow $\tau_i > 1$, in which case agent $i$ adds to $\psi_i$ multiple copies of the same demand.
    In our actual applications of this lemma, however, we will have $\tau_i \leq 1$, so each item will add at most one demand copy to $\psi_i$.

    It is easy to see that such choice of $\psi_i$ covers the entire deficit of agent $i$:
    \begin{multline*}
        v(\psi_i) = \sum_{k = 1}^{m - h}\beta_{i,k}\cdot \big(\lceil \tau_i \cdot k \rceil - \lceil \tau_i\cdot (k - 1)\rceil \big)
        =\sum_{k = 1}^{m - h}\lceil \tau_i\cdot k\rceil\cdot (\beta_{i,k} - \beta_{i, k + 1})\\ \geq \tau_i\sum_{k = 1}^{m - h}k\cdot (\beta_{i,k} - \beta_{i, k + 1}) = \tau_i\sum_{k = 1}^{m - h}\beta_{i,k} = \tau_iT_i = \Delta(\pi_i).
    \end{multline*}
    Since $\Delta(\pi_i)= (1 - v(\pi_i))_+$, it immediately follows that $v(\pi_i) + \psi(\pi_i) \geq 1$.

    It remains to bound the additional prefix load generated by $\Psi_I = \{\psi_i\}_{i \in I}$.
    Note that for every $i \in I$, the breakpoint of $\beta_{i,k} \in \psi_i$ is at least $h + k$.
    Hence, for every $0 \leq \ell \leq h$ we have 
    \[
        \Pp_i(\psi_i, \ell) = \sum_{\beta \in \psi_i}\One\{m_i(\beta) \leq \ell\} = 0,
    \]
    for every $i \in I$, and thus $\Pp(\Psi_I, \ell) = 0$.
    For $\ell > h$, as mentioned earlier the number of selected items from $\psi_i$ within the first $k = \ell - h$ tail items is exactly $\lceil \tau_i\cdot (\ell - h)\rceil\leq \tau_i\cdot (\ell - h) + 1$.
    So, the additional load generated by tail profiles $\Psi_I$ at prefix $\ell > h$ is at most
    \[
        \Pp(\Psi_I, \ell) \leq \sum_{i \in I}\big(\tau_i\cdot (\ell - h) + 1) = (\ell - h)\sum_{i \in I}\tau_i + |I|.
    \]
    The lemma follows.
\end{proof}

Thus, in order to get a set $I\subseteq N$ of agents with a collection of \textit{full profiles} satisfying \cref{etaload}, it suffices to find $I\subseteq N$ and a collection of \textit{only head profiles} $\Pi_I = \{\pi_i\}_{i \in I}$ satisfying \cref{etatailcomplete}.
We explain how to choose a sufficiently large set $I\subseteq N$ (of size roughly $n - O(\log^2 n)$) and corresponding profiles $\Pi_I$ with these properties in the following subsection.

\subsection{Fractional head profiles}
\label{sec:fracpro}

We will first define a collection of head profiles that agents $i \in N$ will choose from.
\begin{definition}
    For every agent $i \in N$, let $\Bb_i = \{B_{i,1},\ldots, B_{i,d}\}$ denote the $d$-partition of $M$ witnessing $\mms(M, v_i, d)$, for every $B_{i,j} \in \Bb_i$ it holds $v_i(B_{i,j}) = 1$.
    For every $j \in [d]$, define 
    \[
        \pi_{i,j} := \left\{v_i(e) :\; e \in B_{i,j}\cap [h], v_i(e) > 0\right\}.
    \]
    Let $\Pi := \{\varnothing, \bot\} + \bigcup_{i \in N}\{\pi_{i,j} : j \in [d]\}$.
\end{definition}

Instead of considering these profiles directly, for every agent $i \in N$ we will define a \textit{distribution} over profiles $\Pi$.
We will refer to these distributions as \textit{fractional profiles}.
Our goal then will be first to define a collection of fractional profiles that satisfy the prefix load condition of \cref{etatailcomplete}, and then round it to an integral collection while preserving every prefix constraint.

\begin{definition}
    Let $\Pi$ be a collection of head profiles, containing $\varnothing$ and $\bot$.
    A \textbf{fractional head profile} of agent $i \in N$ is a pair $(x_i, \Pi)$, where $x_i(\pi) \geq 0$ for every $\pi \in \Pi$ and $\sum_{\pi \in \Pi}x_i(\pi) = 1$.
\end{definition}

We naturally extend the definition of (head) prefix load to fractional profiles.

\begin{definition}
    For agent $i \in N$ with fractional head profile $(x_i, \Pi)$, his \textbf{fractional $\ell$-prefix head load} of $\Pi$ for $0 \leq \ell \leq m$ is 
    \[
        \Hh_i(x_i, \Pi, \ell) := \sum_{\pi\in \Pi}x_i(\pi)\cdot \Pp_i(\pi, \ell).
    \]
    Assume that every agent $i \in N$ has fractional head profile $(x_i, \Pi)$ for the shared set of profiles $\Pi$, and let $x = \{x_i\}_{i \in N}$.
    For $0 \leq \ell \leq m$, the \textbf{total fractional $\ell$-prefix head load} of $\Pi$ for $x$ is 
    \[
        \Hh(x, \Pi, \ell) := \sum_{i \in N}\Hh_i(x_i, \Pi, \ell) = \sum_{i \in N}\sum_{\pi\in \Pi}x_i(\pi)\cdot \Pp_i(\pi, \ell).
    \]
\end{definition}

In addition to the fractional head load, we will also consider the fractional \textit{tail load}.
As all agents share a common tail $\{h + 1, \ldots, m\}$, instead of defining the fractional tail load for every prefix $\ell$, we omit the multiplier $(\ell - h)$ from the definition and consider one common term.
\begin{definition}
    For agent $i \in N$ with fractional head profile $(x_i, \Pi)$, his \textbf{fractional tail load} of $\Pi$ is
    \[
        \tau_i(x_i, \Pi) := \sum_{\pi \in \Pi}x_i(\pi)\cdot \tau_i(\pi).
    \]
    Assume that every agent $i \in N$ has fractional head profile $(x_i, \Pi)$ for the shared set of profiles $\Pi$, and let $x = \{x_i\}_{i \in N}$.
    The \textbf{total fractional tail load} of $\Pi$ for $x$ is 
    \[
        \tau(x, \Pi) := \sum_{i \in N}\tau_i(x_i, \Pi) = \sum_{i \in N}\sum_{\pi\in \Pi}x_i(\pi)\cdot \tau_i(\pi).
    \]
\end{definition}
Finally, we will be interested in the expected number of agents $i \in N$ that are assigned a non-$\bot$ profile according with respect to fractional profiles $(x_i, \Pi)$.
\begin{definition}
    Assume that every agent $i \in N$ has fractional head profile $(x_i, \Pi)$ for the shared set of profiles $\Pi$, and let $x = \{x_i\}_{i \in N}$.
    The \textbf{fractional number of assigned profiles} with respect to $(x, \Pi)$ is
    \[
        \Ii(x, \Pi) := \sum_{i \in N}\sum_{\pi\neq\bot}x_i(\pi).
    \]
\end{definition}

For every agent $i \in N$, our starting fractional head profile $(x_i, \Pi)$ will simply be a uniform distribution over agent $i$'s MMS profiles.
\begin{definition}
    Let $\Pi := \{\varnothing, \bot\} + \bigcup_{i \in N}\{\pi_{i,j} : j \in [d]\}$.
    For every agent $i \in N$, consider the following fractional head profile $(x_i^*, \Pi)$: for every $\pi \in \Pi$ let
    \[
        x_i^*(\pi) = \frac{|\{j \in [d]: \pi_{i,j} = \pi\}|}{d}.
    \]
    We will refer to $(x_i^*, \Pi)$ as a \textbf{fractional MMS head profile} of agent $i$.
\end{definition}

We will now show that the MMS profiles $\pi_{i,j}$ of agents $i \in N$ satisfy the conditions of \cref{etatailcomplete} fractionally with respect to profiles $(x^*_i, \Pi)$.

\begin{lemma}\label{etafracload}
    Consider the fractional MMS head profiles $(x^*, \Pi)$ of agents $i \in N$.
    It holds that
    \[
        \forall 0 \leq\ell\leq m:\; \Hh(x^*, \Pi, \ell) \leq \frac{n}{d}\min\{\ell, h\};\quad\quad \tau(x^*, \Pi) \leq \frac{n}{d},\quad\quad \Ii(x^*, \Pi) = n.
    \]
    As a corollary, for $\eta = (n + 1)/d$ and $h = \min\{nd, m\}$ we have for every $0 \leq \ell \leq h$ that $\Hh(x^*, \Pi, \ell) \leq \eta \ell$, and for every $h + 1 \leq \ell \leq m$ that
    \[
        \Hh(x^*, \Pi, \ell) \leq \eta \ell - (\ell - h)\cdot \tau(x^*, \Pi) - \Ii(x^*, \Pi).
    \]
\end{lemma}
\begin{proof}
    We first observe that for every agent $i \in N$ and every $0 \leq \ell \leq m$ it holds that
    \[
        \sum_{j = 1}^d\Pp_i(\pi_{i,j}, \ell) \leq \min\{\ell, h\}.
    \]
    By definition $\Pp_i(\pi_{i,j}, \ell)$ counts how many demands in $\pi_{i,j}$ can be satisfied only by the first $\ell$ items, and the sum above counts the total number of such demands for agent $i$.
    Note that every demand in $\pi_{i,j}$ comes from a unique item of $B_{i,j}$, and MMS allocation $\Bb_i = \{B_{i,1},\ldots, B_{i,d}\}$ cannot use more than $\ell$ items in every $\ell$-prefix.
    Hence, are at most $\ell$ such demands across all profiles $\{\pi_{i,1},\ldots, \pi_{i,d}\}$.
    Since demands in every $\pi_{i,j}$ correspond only to head items in $[h]$, there cannot be more than $h$ total demands for $i$.
    Combining two bounds gives $\sum_{j = 1}^d\Pp_i(\pi_{i,j}, \ell) \leq \min(\ell, h)$.

    It immediately follows that
    \[
        \Hh(x^*, \Pi, \ell) = \frac{1}{d}\sum_{i \in N}\sum_{j = 1}^d\Pp_i(\pi_{i,j}, \ell) \leq \frac{n}{d}\min\{\ell, h\}.
    \]
    Next, for every $i \in N$ and $j \in [d]$ we have \[
        \Delta(\pi_{i,j}) = 1 - v(\pi_{i,j}) = v(B_{i,j} \cap \{h + 1, \ldots, m\}),
    \]
    so by additivity it holds $\sum_{j = 1}^d\Delta(\pi_{i,j}) = T_i$ and
    \[
         \tau(x^*, \Pi) = \frac{1}{d}\sum_{\substack{i \in N\\T_i > 0}}\sum_{j = 1}^d\frac{\Delta(\pi_{i,j})}{T_i} \leq \frac{n}{d}.
    \]
    The equality $\Ii(x^*, \Pi) = n$ follows from $\sum_{\pi\neq \bot}x_i^*(\pi_{i,j}) = 1$ for every $i \in N$.

    Now, put $\eta = (n + 1)/d$ and $h = \min\{nd, m\}$.
    If $h = m$, there is no tail, and $\Hh(x^*, \Pi, \ell)\leq \eta \ell$ for all $0 \leq \ell \leq m$ is automatic.
    If $h = nd < m$, then for every $\ell > h$ it holds $n < \ell/d$.
    For this $\ell$,
    \begin{multline*}
        \Hh(x^*, \Pi, \ell) + (\ell - h)\cdot \tau(x^*, \Pi) + \Ii(x^*, \Pi) \leq \frac{n}{d}\cdot h + (\ell - h)\cdot \frac{n}{d} + n \\
        < \frac{n}{d}\cdot \ell + \frac{\ell}{d} = \frac{n + 1}{d}\cdot \ell = \eta \ell.
    \end{multline*}
    The lemma follows.
\end{proof}

The path forward is now clear: set $h = \min\{nd, m\}$ and round the fractional MMS profiles $(x^*, \Pi)$ to some integral collection of profiles $\Pi_I = \{\pi_i\}_{i \in I}$ for sufficiently large set of agents $I\subseteq N$.
Specifically, to prove \cref{dmms} via \cref{leftoveralloc}, we will need $|I| \geq n - O(\log^2 n)$.
So, starting from $\Ii(x^*, \Pi) = n$, we are allowed to lose at most $O(\log^2 n)$ agents in the rounding.

Specifically, we establish the following theorem.
\begin{theorem}[Head rounding]\label{etarounding}
    Let $\Pi$ be a shared set of head profiles, such that $|\pi|\leq h$ for some integer $h \geq 1$.
    For every $i \in N$, let $(x_i, \Pi)$ be a fractional profile, denote $x = \{x_i\}_{i \in N}$.

    There exists a collection $y = \{y_i\}_{i \in N}$ of integer head profiles $(y_i, \Pi)$, such that for every $i, \pi$ we have $y_i(\pi) \in \{0, 1\}$ and $y_i(\pi) = 1$ for $\pi \neq \bot$ only if $x_i(\pi) > 0$, and
    \[\begin{aligned}
        & && \Hh(y, \Pi, \ell) \leq \Hh(x, \Pi, \ell),  & \forall 0 \leq \ell \leq m,\\
        & && \tau(y, \Pi) \leq\tau(x, \Pi),\\
        & && \Ii(y, \Pi) \leq \Ii(x, \Pi),\\
        &  && \Ii(y, \Pi)\geq \Ii(x, \Pi) - \rho(h, X), 
    \end{aligned}\]
    where $X = |\{(i, \pi) :\; 0 < x_i(\pi) < 1\}|$ and
    \[
        \rho(h, X) := 8 + 4\log \max\{1, X\}\cdot (1 + \ln h + \ln \max\{1, X\}).
    \]
\end{theorem}
We prove \cref{etarounding} in \cref{appendix:etarounding} (\cref{etaroundingproof}), and use it to prove \cref{dmms}.
\begin{proof}[Proof of \cref{dmms}]
    Take $h = \min\{nd, m\}$ and construct MMS head profiles $\pi_{i,j}$ for every $i \in N$, $j \in [d]$, and let $\Pi := \{\varnothing, \bot\} + \bigcup_{i \in N}\{\pi_{i,j} : j \in [d]\}$.
    Consider the fractional MMS head profiles $(x^*, \Pi)$.
    By definition of $\pi_{i,j}$, we automatically have for every $\pi \in \Pi$ that $|\pi|\leq h$.
    
    Note that the total number of strictly fractional entries in $x^*$ is at most $X^*\leq nd$.
    Apply \cref{etarounding} to $(x^*, \Pi)$.
    By \cref{etafracload}, obtain integral head profiles $(y, \Pi)$ such that 
    \[\begin{aligned}
        & && \Hh(y, \Pi, \ell) \leq \Hh(x^*, \Pi, \ell) \leq \frac{n}{d}\min\{\ell, h\},  & \forall 0 \leq \ell \leq m,\\
        & && \tau(y, \Pi) \leq\tau(x^*, \Pi) \leq \frac{n}{d}\\
         & && \Ii(y, \Pi) \leq \Ii(x^*, \Pi) = n,\\
        &&&\Ii(y, \Pi)\geq \Ii(x^*, \Pi) - \rho(h, X^*) \geq n - \rho(h, nd).
    \end{aligned}\]
    Let $I\subseteq N$ be a set of size $\Ii(y, \Pi)$, where $i \in I$ if and only if $y_i(\bot)  = 0$.
    Let $\pi_i \in \Pi - \bot$ be the profile such that $y_i(\pi_i) = 1$, and $\Pi_I := \{\pi_i\}_{i \in I}$.
    Taking $\eta = (n + 1)/d$, we have for every $0 \leq \ell \leq h$ that $\Pp(\Pi_I, \ell)= \Hh(y, \Pi, \ell) \leq \Hh(x^*, \Pi, \ell) \leq \eta \ell$, and for every $h + 1 \leq \ell \leq m$ that
    \begin{multline*}
        \Pp(\Pi_I, \ell) + (\ell - h)\sum_{i \in I}\tau_i(\pi_i) + |I| = \Hh(y, \Pi, \ell) + (\ell - h)\tau(y, \Pi) + \Ii(y, \Pi) \\
        \leq \Hh(x^*, \Pi, \ell) + (\ell - h)\cdot \tau(x^*, \Pi) + \Ii(x^*, \Pi) \leq\frac{n}{d}\cdot h + (\ell - h)\cdot \frac{n}{d} + n \\
        < \frac{n}{d}\cdot \ell + \frac{\ell}{d} = \frac{n + 1}{d}\cdot \ell = \eta \ell.
    \end{multline*}
    By construction, for every $i \in I$ we have that $T_i = 0$ implies $\Delta(\pi_i) = 0$, as $y_i(\pi_i) = 1$ implies $x_i(\pi_i) > 0$, hence $\pi_i$ must be some MMS profile of agent $i$.
    Then, we can apply \cref{etatailcomplete} and obtain a collection $\Psi_I = \{\psi_i\}_{i \in I}$ of tail profiles such that  for every $0 \leq \ell \leq m$  we have  \[
        \Pp(\Pi_I, \ell) + \Pp(\Psi_I, \ell) \leq \eta \ell,
    \]
    and for every $i \in I$ it holds $v(\pi_i) + v(\psi_i) \geq 1$.
    Hence, \cref{etaload} applied to $(I, \Pi_I, \Psi_I)$ produces some set $F\subseteq M$, such that all profiles $\{\pi_i, \psi_i\}_{i \in I}$ are simultaneously satisfied using distinct items of $F$, and for every $0 \leq \ell \leq m$ it holds $|F\cap [\ell]|\leq \eta \ell$.

    Every agent $i \in I$ receives a disjoint bundle $F_i\subseteq F$ such that $v_i(F_i) \geq v(\pi_i) + v(\psi_i) \geq 1$.
    It remains to allocate the agents in $U := N - I$.
    Since $|I| = \Ii(y, \Pi) \geq n - \rho(h, nd)$, we have $|U| \leq \rho(h, nd)$.
    At the same time, \cref{prefixbound} guarantees us that for every $u \in U$ it holds $v_u(F) \leq \eta d = n + 1$.
    Hence, as long as $|U|\leq (d - n)/2$, \cref{leftoveralloc} allocates agents $U$ with items $M - F$ so that every $u \in U$ receives a bundle of value at least $1$.
    Then, to guarantee $|U|\leq (d - n)/2$ it suffices to take $d$ such that 
    \[
        \rho(h, nd) \leq (d - n)/2 \quad \iff \quad d \geq n + 2\rho(h, nd).
    \]
    We now estimate $\rho(h, nd)$.
    Since $h , X^* \leq nd$ and $d \leq 2n - 1$ (\cref{leftovers} solves $d > 2n-1$),
    \begin{multline*}
        \rho(h, nd) = 8 + 4\log (nd)\cdot (1 + \ln (nd) + \ln (nd))\\
        \leq 8 + 4\log (nd) + 8\ln2\log^2(nd) \leq 8 + 4\log(2n^2) + 8\ln2 \log^2(2n^2).
    \end{multline*}
    Setting $z = 1 + \log n$, the bound above is
    \begin{multline*}
        \rho(h, nd) \leq 8 + 4(2z - 1) + 8\ln2 (2z - 1)^2\\
        = 32\ln 2z^2  + (8 - 32\ln 2)z + 4 + 8\ln 2 < 32\ln2 z^2 < 23(1 + \log n)^2.
    \end{multline*}
    As a result, taking $d \geq n + 46(1 + \log n)^2$ suffices.
\end{proof}
\section{$(1 - \delta)$-MMS allocation for $(1 - \eps)n$ agents}
\label{sec:large}
In this section, we prove \cref{thm:largen}.
We assume $m \geq n$, otherwise pad items with zeroes.
\begin{theorem}[\cref{thm:largen} restated]\label{dchoiceproof}
    For every fixed $\eps \in (0, 1)$ and integer $q \geq 3$, there exists $n(\eps, q)$ such that for every $n \geq n(\eps, q)$ the following holds.
    
    Consider a fair allocation instance with $m$ items $M$ and $n$ agents $N$, where for every $i \in N$ valuation $v_i$ is additive.
    Suppose that every agent $i \in N$ has $\mms(M, v_i, d) = 1$, where $1 \leq d \leq (1 - \eps)n$.
    Then there exists a choice $I = \{i_1, \ldots, i_d\}\subseteq N$ of $|I| = d$ distinct agents and an allocation $(B_{i_1},\ldots, B_{i_d})$ of $M$ to $I$, such that for every $i \in I$ it holds $v_i(B_i) \geq 1 - 2/q$.
\end{theorem}

The core of our approach will be essentially the same as in the proof of \cref{dmms}.
That is, we would like to find a collection of profiles $\Pi_I = \{\pi_i\}_{i \in I}$ assigned to agent $I\subseteq N$, such that $|I|\geq d$, for every $i \in I$ we have $v(\pi_i) \geq 1 - 2/q$, and for every $0 \leq \ell \leq m$ it holds
\[
    \Pp(\Pi_I, \ell) \leq \ell.
\]
Then, we obtain the desired $(1 - 2/q)$-MMS allocation by applying \cref{loadequiv} to $I, \Pi_I$.

\subsection{Head and tail profiles with different thresholds}

As before, we will first find a collection $\Pi_I$ of head profiles $\pi_i \in \Pi_I$, and then complete it to full profiles with a collection of tail profiles $\psi_i \in \Psi_I$, so that for every $i \in I$ we have $v(\pi_i) + \psi(\pi_i) \geq 1- 2/q$, and agent $i$ gets $\{\pi_i, \psi_i\}$.
However, unlike the proof of \cref{dmms}, each agent $i \in N$ now will have an individual \textit{head threshold} $h_i$.


\begin{definition}
    For an integer $q > 0$, ordered items $M = [m]$ and agent $i \in N$, let 
    \[
        h_i := \left|\left\{e \in M : v_i(e) \geq 1/q\right\}\right|.
    \]
    We refer to items $\{1, \ldots, h_i\}$ as \textbf{head items} of agent $i$, and items $\{h_i + 1, \ldots, m\}$ as \textbf{tail items}.
    For agent $i \in N$, the \textbf{tail value} of $i$ is
    \[
        T_i = \sum_{e > h_i}v_i(e).
    \]
\end{definition}

In the proof of \cref{dmms}, it was of no difference for us how many different head profiles there are for agents $i \in N$, and how large these profiles can be.
Each head demand could take essentially arbitrary value in $(0, 1]$, and there could be up to $nd = O(n^2)$ different demands in a head profile.
However, for the proof of \cref{dchoiceproof}, it will be crucial for us that 1) each head profile has at most constant size, and 2) there is only a constant number of different head demands.
We achieve this by rounding down the values of head items to multiples of $1/q^2$.

\begin{definition}
    For an integer $q > 0$, a \textbf{head demand} is a positive value $\alpha \in (0, 1]$ such that
    \[
        \alpha = \frac{q + k}{q^2} \;\text{ for integer  }\;k\in \{0, 1, 2, \ldots, q^2 - q\}.
    \]
    That is, a head demand $\alpha$ takes values in
    \[
        \frac{1}{q},\;\; \frac{1}{q} + \frac{1}{q^2},\;\; \frac{1}{q} + \frac{2}{q^2},\;\; \ldots,\;\; \frac{1}{q} + \frac{k}{q^2},\;\;\ldots,\;\; 1.
    \]
    A \textbf{head profile} is a finite multiset $\pi = \{\alpha_1, \ldots, \alpha_{|\pi|}\}$ of head demands.
    
    Similarly, a \textbf{tail demand} is a positive value $\beta \in (0, 1]$ such that $\beta \leq 1/q$ (we allow $\beta = 1/q$).
    A \textbf{tail profile} is a finite multiset $\psi =\{\beta_1, \ldots, \beta_{|\psi|}\}$ of tail demands.
\end{definition}
Note that for every agent $i \in N$, a bundle $B\subseteq M$ such that $v_i(B) = 1$ can contain at most $q$ head items.
Hence, if we round down all head items of $B$ to the multiple of $1/q^2$, its value remains at least $1 - q\cdot (1/q^2) = 1 - 1/q$.
So, these head profiles cost us only $1/q$ loss so far.

Due to rounding down, we will measure our deficit with respect to $1 - 1/q$ instead of $1$.
\begin{definition}
    A \textbf{deficit} of a head profile $\pi$ is
    \[
        \Delta(\pi) := (1 - 1/q - v(\pi))_+.
    \]
\end{definition}

As in the proof of \cref{dmms}, if an agent $i \in N$ gets assigned a head profile $\pi_i \neq \bot$ with deficit $\Delta(\pi_i) > 0$, we will need to compensate for the missing value using the tail.
We prove an analogue of \cref{etatailcomplete}, giving sufficient conditions on the prefix load of the head profiles $\pi_i$ that allows adding tail profiles $\psi_i$ without violating the total prefix load constraints.

There are two key differences from the original \cref{etatailcomplete}.
First, for every agent $i \in I$ there is a different head threshold $h_i$ and tail, so instead of a common tail-load multiple $(\ell - h)$ we will have individual tail multiples $(\ell - h_i)$.
Second, we relax the slackness requirements by removing the term $|I|$.
Since now the prefix load of head profiles can be larger, we have less room and will have to ``sacrifice'' up to one tail item, similar to faithful rounding \cite{BEF22BoBW,EG26-2}.
We achieve this by taking the \textit{last} item in each segment of length $1/\tau_i(\pi_i)$ instead of the first, which causes one less unit of load per agent of $I$, but costs us another additive term of $1/q$.
\begin{lemma}[Tail completion]\label{tailcomplete}
    Let $I\subseteq N$, let $\pi_i$ be a head profile assigned to agent $i \in I$, and let $\Pi_I = \{\pi_i\}_{i \in I}$.
    Assume that $T_i = 0$ implies $\Delta(\pi_i) = 0$ for every $i \in I$.
    If for every $0 \leq \ell \leq m$:
    \[
        \Pp(\Pi_I, \ell) \leq \ell - \sum_{i \in I}\tau_i(\pi_i)(\ell - h_i)_+ ,
    \]
    there exists a collection $\Psi_I := \{\psi_i\}_{i \in I}$ of tail profiles, where agent $i$ is assigned $\psi_i$, such that
    \[
        \forall 0 \leq\ell \leq m:\; \Pp(\Pi_I, \ell) + \Pp(\Psi_I, \ell) \leq \ell
    \]
    and $v(\pi_i) + v(\psi_i) \geq 1 - 2/q$ for every $i \in I$.
\end{lemma}
\begin{proof}
    Consider $i\in I$, and denote $\tau_i := \tau_i(\pi_i)$.
    If $\tau_i = 0$, put $\psi_i = \varnothing$.
    Otherwise, we construct the tail profile $\psi_i$ similarly to \cref{etatailcomplete}: break the agent $i$'s tail $\{h_i + 1, \ldots, m\}$ into segments of length roughly $ 1/\tau_i$, and in each segment add the value of the last item to $\psi_i$.
    That is, for every $t = 1,\ldots, \lfloor \tau_i\cdot (m - h_i)\rfloor$, add $v_i(e_{i, t})$ to $\psi_i$, where
    \[
        e_{i,t} = h_i + \left\lceil \frac{t}{\tau_i}\right\rceil.
    \]
    Note that in this construction, for every item we take, its value is at least as large as if we took a $\tau_i$-fraction of every tail item in the segment of length $1/\tau_i$ \textit{immediately after} the taken item.
    So, the total value of selected items is at least as large as the $\tau_i$-fraction of the total tail value $T_i$, minus the very first segment, from which we lose at most $1/q$ (the first tail item).

    We will now prove this formally.
    Similar to \cref{etatailcomplete}, the construction of $\psi_i$ can equivalently be described as follows.
    For $1 \leq k \leq m - h_i$, let $\beta_{i,k} := v_i(h_i + k)$.
    Then, for every $k \in [m - h_i]$, we add exactly
    \[
        \lfloor \tau_i\cdot k\rfloor - \lfloor \tau_i\cdot (k - 1)\rfloor
    \]
    copies of $\beta_{i,k}$ to the profile $\psi_i$.
    To see this, observe that
    \[
        e_{i,t} \leq h_i + k \iff \left\lceil \frac{t}{\tau_i}\right\rceil \leq k \iff t \leq \tau_i\cdot k \iff t\leq \lfloor \tau_i\cdot k\rfloor.
    \]
    So, the first $k$ tail items contain exactly $\lfloor \tau_i\cdot k\rfloor$ selected items from $\psi_i$, and the number of copies of tail item $k$ is exactly $\lfloor \tau_i\cdot k\rfloor - \lfloor \tau_i\cdot (k - 1)\rfloor$.
    We note that here $\tau_i > 1$ is allowed, in which case agent $i$ adds to $\psi_i$ multiple copies of the same demand.
    Furthermore, later in the proof of \cref{dchoiceproof} we will see that it will indeed happen for some $i \in I$ that $\tau_i > 1$ substantially.
    For example, if agent $i \in I$ gets assigned $\pi_i = \varnothing$, while $T_i > 0$ is extremely small.
    This is still allowed, because we are working with value profiles and not actual items.
    As long as the conditions on the total prefix load are preserved, \cref{loadequiv} eventually produces an allocation satisfying all the profiles.
    So, in case $\pi_i = \varnothing$, the tail profile $\psi_i$ will have to compensate almost all the value, and may as well get satisfied by some head items (while having only tail demands).
    
    A straightforward computation shows that our choice of $\psi_i$ covers the deficit of agent $i$:
    \begin{multline*}
        v(\psi_i) = \sum_{k = 1}^{m - h_i}\beta_{i,k}\cdot \big(\left\lfloor \tau_i\cdot k\right\rfloor - \left\lfloor \tau_i\cdot (k -1)\right\rfloor\big)  = \sum_{k = 1}^{m - h_i}\lfloor \tau_i\cdot k\rfloor \cdot (\beta_{i,k} - \beta_{i,k + 1}) \\\geq \sum_{k = 1}^{m - h_i}(\tau_i\cdot k - 1)\cdot (\beta_{i,k} - \beta_{i,k+1})
        = \tau_i \sum_{k = 1}^{m - h_i}\beta_{i,k} - \beta_{i,1} = \Delta(\pi_i) - \beta_{i,1} \geq \Delta(\pi_i) - 1/q.
    \end{multline*}
    Since $\Delta(\pi_i) = (1 - 1/q - v(\pi_i))_+$, it immediately follows that $v(\pi_i) + v(\psi_i) \geq 1 - 2/q$.

    We will now bound the additional prefix load generated by $\Psi_I = \{\psi_i\}_{i \in I}$.
    As in \cref{etatailcomplete}, for every $0 \leq \ell \leq h_i$ we have $\Pp_i(\psi_i, \ell) = 0$, for every agent $i \in I$.
    When $\ell > h_i$, as shown earlier the number of selected items from $\psi_i$ within the first $k = \ell - h_i$ tail items is exactly $\lfloor\tau_i\cdot k\rfloor \leq \tau_i\cdot k$.
    Hence, the additional load generated by tail profiles $\Psi_I$ at prefix $\ell$ is
    \[
        \Pp(\Psi_I, \ell) = \sum_{i \in I}\Pp_i(\psi_i, \ell)\leq \sum_{i \in I}\tau_i\cdot (\ell - h_i)_+.
    \]
    The lemma follows.
\end{proof}
Note that the main difference of \cref{tailcomplete} from the usual faithful rounding setting is that every agent $i \in I$ has a different tail, which can overlap heads of other agents in $I$.
Nevertheless, our methods of \cref{etatailcomplete} extend to the different tails in \cref{tailcomplete}.

\subsection{Fractional head profiles and the large tail case}

We now focus on finding a collection of head profiles $\Pi_I$ of agents $I\subseteq N$ satisfying the requirements of \cref{tailcomplete}.
As in the proof of \cref{dmms}, MMS partitions of agents $i \in N$ induce a collection of head profiles from which the agents will choose from.
The profiles will differ slightly, as now 1) head thresholds are different across different across the agents, and 2) we will have to round the items down to the multiples of $1/q^2$.
\begin{definition}
    For every agent $i \in N$, let $\Bb_i = \{B_{i,1},\ldots, B_{i,d}\}$ denote the $d$-partition of $M$ witnessing $\mms(M, v_i, d)$, for every $B_{i,j} \in \Bb_i$ it holds $v_i(B_{i,j}) = 1$.
    For every $j \in [d]$, define 
    \[
        \pi_{i,j}:= \left\{\frac{\lfloor q^2\cdot v_i(e)\rfloor}{q^2} :\; e \in B_{i,j},\; v_i(e) \geq 1/q\right\}.
    \]
    Let $\Pi := \{\varnothing, \bot\} + \bigcup_{i \in N}\{\pi_{i,j} : j \in [d]\}$.
\end{definition}
As in the proof of \cref{dmms}, we will move to fractional head profiles and use rounding to get an integral collection.
All the definitions of fractional head profiles $(x, \Pi)$, fractional $\ell$-prefix head load $\Hh(x, \Pi, \ell)$ and fractional number of assigned profiles $\Ii(x, \Pi)$ are exactly the same as in \cref{sec:fracpro}.
However, due to the fact that instead of a common head threshold $h$ we have separate head thresholds $h_i$, we will need to define fractional tail load separately for every agent $i \in N$ and every prefix $0\leq \ell\leq m$.
\begin{definition}
    For agent $i \in N$ with fractional head profile $(x_i, \Pi)$, his \textbf{fractional $\ell$-prefix tail load} of $\Pi$ is 
    \[ 
        \Tt_i(x_i, \Pi, \ell):= \sum_{\pi \in \Pi}x_i(\pi)\cdot \tau_i(\pi)(\ell - h_i)_+ = \tau_i(x_i, \Pi)\cdot (\ell - h_i)_+.
    \]
    Assume that every agent $i \in N$ has fractional head profile $(x_i, \Pi)$ for the shared set of profiles $\Pi$, and let $x = \{x_i\}_{i \in N}$.
    For $0 \leq \ell \leq m$, the \textbf{total fractional $\ell$-prefix tail load} of $\Pi$ for $x$ is
    \[
        \Tt(x, \Pi, \ell) := \sum_{i \in N}\Tt_i(x_i, \Pi, \ell)  = \sum_{i \in N}\tau_i(x_i, \Pi)\cdot (\ell - h_i)_+.
    \]
\end{definition}

We define the fractional MMS head profiles a bit differently from \cref{sec:fracpro}: instead of taking a uniform distribution over $\pi_{i,j}$ for every agent $i \in N$, a profile $\pi_{i,j}$ is picked with probability $1/n$, and the remaining weight $1 - d/n$ goes into the no-profile $\bot$.
\begin{definition}
    Let $\Pi := \{\varnothing, \bot\} + \bigcup_{i \in N}\{\pi_{i,j} : j \in [d]\}$.
    For every agent $i \in N$, consider the following fractional head profile $(x_i^*, \Pi)$: set $x_i(\bot) = 1 - d/n$, and for every $\pi \in \Pi-\bot$ let
    \[
        x_i^*(\pi) = \frac{|\{j \in [d]: \pi_{i,j} = \pi\}|}{n}.
    \]
    We will refer to $(x_i^*, \Pi)$ as a \textbf{fractional MMS head profile} of agent $i$.
\end{definition}

We prove an analogue of \cref{etafracload}, showing that $(x^*, \Pi)$ satisfy \cref{tailcomplete} fractionally.
\begin{lemma}\label{fracload}
    Consider the fractional MMS head profiles $(x^*, \Pi)$ of agents $i \in N$.
    It holds that
    \begin{itemize}
        \item $\Ii(x^*, \Pi) = d$;
        \item for every $0 \leq \ell \leq m$,
        \[
            \Hh(x^*, \Pi, \ell) \leq \ell - \frac{1}{n}\sum_{i \in N}(\ell - h_i)_+;
        \]
        \item for every $i \in N$,
        \[
            \sum_{j = 1}^d\Delta(\pi_{i,j}) \leq (1 - 1/q)T_i;
        \]
        \item for every $0 \leq \ell \leq m$,
        \[
            \Tt(x^*, \Pi, \ell) \leq \frac{1 - 1/q}{n}\sum_{i \in N}(\ell - h_i)_+.
        \]
    \end{itemize}
    The $(1 - 1/q)$-fraction in tail load appears due to rounding head values down to multiples of $1/q^2$, and deficit counting from $1 - 1/q$.
    This extra slackness will be useful later in the rounding.
\end{lemma}
\begin{proof}
    The identity $\Ii(x^*, \Pi)=d$ is trivial: the total sum of non-$\bot$ weights of every agent is $d/n$.
    Next, by the same reasoning as in \cref{etafracload}, for every $i \in N$ and every $0\leq \ell \leq m$ we have
    \[
        \sum_{j = 1}^d\Pp_i(\pi_{i,j}, \ell) \leq \min\{\ell, h_i\},
    \]
    which immediately gives by definition of $\Hh(x^*, \Pi, \ell) = \frac{1}{n}\sum_{i \in N}\sum_{j = 1}^d\Pp_i(\pi_{i,j}, \ell)$ that
    \[
        \frac{1}{n}\sum_{i \in N}\sum_{j = 1}^d\Pp_i(\pi_{i,j}, \ell) + \frac{1}{n}\sum_{i \in N}(\ell - h_i)_+
        \leq \frac{1}{n}\sum_{i \in N}\big(\min\{\ell, h_i\} +  (\ell - h_i)_+\big)= \ell.
    \]

    Now, fix agent $i \in N$.
    Recall that for every $j \in [d]$ we have $v(B_{i,j}) = 1$, so the non-rounded head items of $B_{i,j}$ have total value $1 - T_{i,j}$, where 
    \[
        T_{i,j}:= \sum_{\substack{e \in B_{i,j}\\e>h_i}}v_i(e).
    \]
   Since each demand in $\pi_{i,j}$ is at least $1/q$, we have 
   \[
        \frac{|\pi_{i,j}|}{q} \leq v(\pi_{i,j}) \leq 1 - T_{i,j}.
   \]
   For every head item $e \in [h_i]$ consider its \textit{rounded value}
    \[
        \hv_i(e) :=\frac{\lfloor q^2\cdot v_i(e)\rfloor}{q^2}.
    \]
    We round each $v_i(e)$ for $e \leq h_i$ down to the multiple of $1/q^2$, hence $0 \leq v_i(e) - \hv_i(e) \leq 1/q^2$, and the total loss from rounding down head items to their demands in $\pi_{i,j}$ is
    \[
        \rho_{i,j} := \sum_{\substack{e \in B_{i,j}\\e \leq h_i}}\big(v_i(e) - \hv_i(e)\big) \leq \frac{|\pi_{i,j}|}{q^2} \leq \frac{1 - T_{i,j}}{q}.
    \]
    At the same time, 
    \[
        v(\pi_{i,j}) =\hv_i(B_{i,j}) = \sum_{\substack{e \in B_{i,j}\\e \leq h_i}}\hv_i(e) = \sum_{\substack{e \in B_{i,j}\\e \leq h_i}}v_i(e) - \rho_{i,j} = 1 - T_{i,j} - \rho_{i,j}.
    \]
    As a result, the deficit of $\pi_{i,j}$ is
    \[
        \Delta(\pi_{i,j}) = (1 - 1/q - v(\pi_{i,j}))_+ = (T_{i,j} +\rho_{i,j} - 1/q)_+ \leq \left(T_{i,j} +\frac{1 - T_{i,j}}{q} - \frac{1}{q}\right)_+ = (1 - 1/q)T_{i,j},
    \]
    and the total deficit of MMS profiles of agent $i \in N$ is
    \[
            \sum_{j = 1}^d\Delta(\pi_{i,j}) \leq (1 - 1/q)\sum_{j = 1}^dT_{i,j} = (1 - 1/q)T_i.
        \]
    It immediately follows that
    \[
        \Tt(x^*, \Pi, \ell) = \frac{1}{n}\sum_{i \in N}\sum_{j = 1}^d\tau_i(\pi_{i,j})(\ell - h_i)_+ = \frac{1}{n}\sum_{\substack{i \in N\\T_i > 0}}\frac{(\ell - h_i)_+}{T_i}\sum_{j = 1}^d\Delta(\pi_{i,j}) \leq\frac{1 - 1/q}{n}\sum_{i \in N}(\ell - h_i)_+. 
    \]
\end{proof}

As in \cref{sec:fracpro}, we will round the fractional MMS head profiles $(x^*, \Pi)$ to some integral collection of profiles $\Pi_I = \{\pi_i\}_{i \in I}$.
The rounding algorithm will be slightly different.
On the one hand, the size of the profiles $\pi \in \Pi$ is bounded by constant $q$, which allows stronger bounds.
On the other hand, instead of a single tail load constraint, we have additional $m + 1$.
Because of this, unlike \cref{etarounding}, in this case we will not preserve all tail load constraints exactly. 
Our rounding algorithm will 1) keep all head load constraints satisfied, 2) violate all tail load constraints by at most a small factor of $\lambda > 1$, and 3) select at least $\Ii(x^*, \Pi) - O_{q}(\log n)$ agents with assigned profiles.
The additional slackness of $(1 - 1/q)$ in the fractional tail load of $(x^*, \Pi)$ will absorb the factor $\lambda$, and the rounded solution will satisfy the requirements of \cref{loadequiv}.

\begin{theorem}[Head and tail rounding]\label{rounding}
    Let $\Pi$ be a shared set of head profiles, such that $|\pi|\leq q$ for some integer $q \geq 3$.
    For every $i \in N$, let $(x_i, \Pi)$ be a fractional profile, denote $x = \{x_i\}_{i \in N}$.

    Suppose that for every $\pi \in \Pi - \bot$, either $\Delta(\pi) = 0$ or $\Delta(\pi) \geq \delta > 0$ for some $\delta \in (0, 1]$.
    For any $\lambda > 1$, there exists a collection $y = \{y_i\}_{i \in N}$ of integer head profiles $(y_i, \Pi)$, such that for every $i, \pi$ we have $y_i(\pi) \in \{0, 1\}$ and $y_i(\pi) = 1$ for $\pi \neq \bot$ only if $x_i(\pi) > 0$, and 
    \[\begin{aligned}
        & && \Hh(y, \Pi, \ell) \leq \Hh(x, \Pi, \ell),  & \forall 0 \leq \ell \leq m,\\
        & && \Tt(y, \Pi, \ell) \leq \lambda\Tt(x, \Pi, \ell),&\forall 0 \leq \ell \leq m,\\
        &  && \Ii(y, \Pi)\geq \Ii(x, \Pi) - \rho(q, \delta, X), 
    \end{aligned}\]
    where, for for $\tau(x, \Pi)=\sum_{i \in N}\sum_{\pi \in \Pi}x_i(\pi)\cdot \tau_i(\pi)$ and $X= |\{(i, \pi):0 < x_i(\pi) <1 \}|$,
    \[
        \rho(q, \delta, X) = 8 + 4\ \log_\lambda\max\left\{1, \frac{1}{\delta}h_{\max}T_{\max}X\tau(x, \Pi)\right\}+ 8q\log \max\{1, X\}.
    \]
    Here $h_{\max} := \max_{i \in N}h_i$ and $T_{\max} := \max_{i \in N}T_i$.
\end{theorem}
We prove \cref{rounding} in \cref{appendix:rounding} (\cref{roundingproof}), similarly to \cref{etarounding}.

We will show that combining \cref{fracload} with \cref{rounding} guarantees us a set of at least $d - O_{\eps,q}(\log n)$ agents $I$ satisfying the requirements of \cref{loadequiv}.
The reader may have noticed an arising issue: we have not explained at all how we would allocate the missing $O_{q}(\log n)$ agents, for the promised total of $d$.
Unlike in the proof of \cref{etarounding} for $\mms(d)$, we do not have additional $\eta$-slackness in the prefix loads, so we cannot use the earlier method.

We resolve this by splitting \cref{dchoiceproof} into two cases, based on the average tail value
\[
    \oT := \frac{1}{n}\sum_{i \in N}T_i.
\]
\begin{enumerate}
    \item when $\oT$ is large, we will modify the fractional MMS profiles $(x^*_i, \Pi)$ by shifting some probability from $\bot$ to empty head profile $\varnothing$, as long as agent $i$ has $T_i > 0$.
    This creates a new solution $\hx$.
    The weight we give to $\varnothing$ is chosen so that in total, the fractional number of assigned profiles w.r.t $\hx$ is $\Ii(\hx, \Pi) \geq d + O_{\eps,q}(\log n)$, enough to compensate for the rounding loss of \cref{rounding} and get at least $d$ agents in the integral solution.
    The tradeoff is that in the rounding, some of the $d$ chosen agents $i \in I$ will be assigned an empty profile $\pi_i = \varnothing$.
    However, since average tail $\oT$ is large, and we assign $\varnothing$ only to agents with nonempty tail, we should be able to accommodate selected agents with $\varnothing$ by giving them enough value with tail items only in  tail completion \cref{tailcomplete}.

    \item the case then $\oT$ is small is significantly more complicated. 
    First, we will split item set into $M = E + S$, where $|S| = O_{\eps,q}(\log n)$, and put items $S$ aside.
    We will use \cref{rounding} on the remaining items $E$ in order to obtain $d - |S| - O_{\eps,q}(\log n)$ agents with profiles that have no deficit and are satisfied exclusively by items $E$.
    After that, we will choose a small set of unselected agents and allow them to use saved items $S$ in their profiles, to get a collection of $d$ agents satisfying the analogue of tail completion \cref{tailcomplete}.
    The main complication with this approach is to guarantee that newly selected agents 1) fit together with the agents selected by the rounding in terms of $E$-prefix loads, and 2) their $S$-demands can be simultaneously satisfied by distinct items of $S$.
\end{enumerate}

To prove \cref{dchoiceproof} in the case of large $\oT$, we use \cref{largetail} below  with \cref{tailcomplete}.
\begin{theorem}[Large tail]\label{largetail}
    Let $\gamma := \min(\eps, 1/(2q))$ and $\lambda = \frac{2q}{2q-1}$.
    Suppose that agents $N$ satisfy
    \[
        \oT := \frac{1}{n}\sum_{i \in N}T_i \geq \frac{\rho_{q,n}}{\gamma},
    \]
    where 
    \[
        \rho_{q,n} = 8+ 4 \log_\lambda\big(q^3d^2(d + 2)n\big)+ 8q\log \big((d + 2)n\big).
    \]
    There exists $I\subseteq N$ of size $|I| \geq d$ and head profiles $\Pi_I:= \{\pi_i\}_{i \in I}$, where agent $i \in I$ is assigned $\pi_i$, such that for every $i$ with $T_i = 0$ we have $\Delta(\pi_i) = 0$, and for every $0 \leq \ell \leq m$ it holds
    \[
        \Pp(\Pi_I, \ell) \leq \ell - \sum_{i \in I}\tau_i(\pi_i)\cdot (\ell - h_i)_+.
    \]
\end{theorem}
\begin{proof}
    Costruct MMS head profiles $\pi_{i,j}$ for every $i \in N, j \in [d]$, and let $\Pi := \{\varnothing, \bot\} + \bigcup_{i \in N}\{\pi_{i,j} : j \in [d]\}$.
    Consider the fractional MMS head profiles $(x^*, \Pi)$ as defined earlier.
    We modify these fractional profiles by transferring some weight from $\bot$ to the empty profile $\varnothing$.

    More specifically, for every agent $i \in N$ we create a fractional profile $(\hx_i, \Pi)$, where
    \[
        \hx_i(\varnothing) = x_i^*(\varnothing) + \frac{\gamma T_i}{n},\qquad \hx_i(\bot) = x_i^*(\bot) - \frac{\gamma T_i}{n},\qquad \forall \pi \notin \{\varnothing, \bot\}:\; \hx_i(\pi) = x_i^*(\pi).
    \]
    These are valid fractional profiles: $\sum_{\pi \in \Pi}\hx_i = 1$, and, since $d \leq (1 - \eps)n$ and $T_i \leq v_i(M) = d$,
    \[
        \hx_i(\bot) = 1- \frac{d}{n } - \frac{\gamma T_i}{n} \geq \eps - \frac{\gamma T_i}{n} \geq \gamma\left(1 - \frac{T_i}{n}\right) \geq 0.
    \]
    Note that only the agents $i \in N$ with $T_i > 0$ are allowed to choose $\pi = \varnothing$.
    If $T_i = 0$, then all positive-valued items in $M$ are head items for $i$, and (by \cref{fracload}) every MMS profile of agent $i$ has $\Delta(\pi_{i,j}) = 0$.
    As a result, if $T_i = 0$ then for every profile $\pi \in \Pi - \bot$ with $\hx_i(\pi) > 0$ it holds $\Delta(\pi)= 0$ --- exactly the property we need in \cref{tailcomplete}.

    It is easy to see that
    \[
        \Ii(\hx, \Pi) = \sum_{i \in N}\sum_{\pi\neq \bot}\hx_i(\pi) = \sum_{i \in N}\left(\frac{d}{n} + \frac{\gamma T_i}{n}\right) = d + \gamma \oT \geq d + \rho_{q,n}.
    \]
    In addition, since $\Pp_i(\varnothing, \ell) = 0 = \Pp_i(\bot, \ell)$ for every $i, \ell$, it holds $\Hh(\hx, \Pi, \ell) = \Hh(x^*, \Pi, \ell)$ for every $0 \leq \ell \leq m$.
    When it comes to tail load, by \cref{fracload} for every $i \in N$ with $T_i > 0$:
    \begin{multline*}
        \tau_i(\hx_i, \Pi) = \sum_{\pi \in \Pi}\hx_i(\pi)\cdot \tau_i(\pi) = \frac{1}{nT_i}\sum_{j = 1}^d\Delta(\pi_{i,j}) + \frac{\gamma T_i}{n}\cdot \frac{\Delta(\varnothing)}{T_i} \leq \frac{1 - 1/q}{n} + \frac{\gamma(1 - 1/q)}{n}\\
        \leq \frac{1}{n}\left(1 - \frac{1}{2q} - \frac{1}{2q^2}\right) \leq \frac{1}{\lambda n},
    \end{multline*}
    using $\gamma \leq 1/(2q)$ and $\lambda = \frac{2q}{2q - 1}$.
    So, $\tau(\hx, \Pi) = \sum_{i \in N}\tau_i(\hx_i, \Pi) \leq 1/\lambda$ and for every $0 \leq \ell \leq m$,
    \[
        \Tt(\hx, \Pi, \ell) = \sum_{i \in N}\tau_i(\hx_i, \Pi)\cdot (\ell - h_i)_+ \leq \frac{1}{\lambda n}\sum_{i \in N}(\ell - h_i)_+.
    \]

    By construction, for every $\pi \in \Pi$ we have $|\pi|\leq q$ (as each demand in $\pi_{i,j}$-s is at least $1/q$).
    In addition, for every $\pi \in \Pi - \bot$, both $v(\pi)$ and $1 - 1/q$ are integer multiples of $1/q^2$.
    So, if $\Delta(\pi) = (1 - 1/q - v(\pi)) > 0$, then it must hold $\Delta(\pi) \geq 1/q^2$.
    We apply \cref{rounding} to $(\hx, \Pi)$ with $\lambda = \frac{2q}{2q -1}$ and $\delta = 1/q^2$, and obtain integral profiles $(y, \Pi)$ such that
    \[\begin{aligned}
        & && \Hh(y, \Pi, \ell) \leq \Hh(\hx, \Pi, \ell),  & \forall 0 \leq \ell \leq m,\\
        & && \Tt(y, \Pi, \ell) \leq \lambda\Tt(\hx, \Pi, \ell),&\forall 0 \leq \ell \leq m,\\
        &  && \Ii(y, \Pi)\geq \Ii(\hx, \Pi) - \rho(q, 1/q^2, \hX),
    \end{aligned}\]
    where $\hX = |\{(i, \pi) : 0 < \hx_i(\pi) < 1\}|$.
    
    Let $I\subseteq N$ be a set of size $\Ii(y, \Pi)$, where $i \in I$ if and only if $y_i(\bot) = 0$.
    Let $\pi_i \in \Pi - \bot$ be the profile such that $y_i(\pi_i) = 1$, and $\Pi_I := \{\pi_i\}_{i \in I}$.
    \cref{rounding} guarantees us that if $T_i = 0$ then $\Delta(\pi_i) = 0$, as $T_i = 0$ implies $\hx_i(\pi) > 0$ only if $\Delta(\pi) = 0$, and $y_i(\pi) > 0$ only when $x_i(\pi) > 0$.
    In addition, for every prefix $0 \leq \ell \leq m$, using $\Hh(\hx, \Pi, \ell) = \Hh(x^*, \Pi, \ell)$ and \cref{fracload},
    \begin{multline*}
        \Pp(\Pi_I, \ell) + \sum_{i \in I}\tau_i(\pi_i)\cdot (\ell - h_i)_+ = 
        \Hh(y, \Pi, \ell) + \sum_{i \in I}\sum_{\pi \in \Pi}y_i(\pi)\cdot \tau_i(\pi)(\ell - h_i)_+\\
        =\Hh(y, \Pi, \ell) + \Tt(y, \Pi, \ell) \leq \Hh(\hx, \Pi, \ell) + \lambda \Tt(\hx, \Pi, \ell) \\
        \leq \ell - \frac{1}{n}\sum_{i \in N}(\ell - h_i)_+ + \frac{1}{n}\sum_{i \in N}(\ell - h_i)_+
        = \ell.
    \end{multline*}
    It remains to estimate the number of selected agents $|I|$.
    Clearly, 
    \[
        |I| = \Ii(y, \Pi) \geq \Ii(\hx, \Pi) - \rho(q, 1/q^2, \hX) \geq d + \rho_{q,n} - \rho(q, 1/q^2, \hX).
    \]
    The total number $\hX$ of strictly fractional entries in $\hx$ is at most $(d + 2)\cdot n$, as every agent contributes at most $d$ from MMS profiles $\pi_{i,j}$, as well as $\varnothing$ and $\bot$.
    In addition, for every agent $i \in N$ it holds $h_i \leq qd$, as $v_i(M) = d$ and every head item has value at least $1/q$.
    Combining this with $T_i \leq v_i(M) \leq d$, we get
    \begin{multline*}
        \rho(q, 1/q^2, \hX) = 8 + 4\ \log_\lambda\max\big\{q^2h_{\max}T_{\max}\hX\tau(\hx, \Pi)\big\}+ 8q\log \max\{1, \hX\}\\
        \leq 8+ 4 \log_\lambda\big(q^3d^2(d + 2)n\big)+ 8q\log \big((d + 2)n\big) = \rho_{q,n}.
    \end{multline*}
    Thus, $|I|\geq d + \rho_{q,n} - \rho(q, 1/q^2, \hX) \geq d$.
    This finishes the proof.
\end{proof}
\section{Small tail case}

\label{sec:small}
When average tail $\oT$ is small, i.e $\oT \leq O_{\eps,q}(\log n)$, the available tail value may not be enough to compensate a selected the agent $i \in N$ if we assign him an empty profile $\varnothing$.
For example, if $\oT = 0$, i.e there are no tail items at all, using the rounding of \cref{rounding} on the fractional MMS profiles $(x^*, \Pi)$ will leave us with only $d - O_{\eps, q}(\log n)$ agents.
We will have to somehow get another $O_{\eps,q}(\log n)$ agents for the desired $d$, without violating the prefix load constraints.

Nevertheless, \cref{rounding} provides us with a good starting point, as we can get integral profiles for \textit{most} agents by rounding the fractional MMS profiles.
As we will show below, it is actually possible to guarantee something even stronger: \cref{rounding} can give most agents profiles with deficit $0$, while using only a subset of items.

As mentioned earlier, on a high level our approach in the small case is as follows.
First, we will split the item set $M$ into $M = E + S$, where $S$ is relatively small, and put items $S$ aside.
Then, we will use \cref{rounding} to choose $d - O_{\eps, q}(\log n)$ agents with profiles that are satisfied exclusively by items $E$, and have no deficit.
After that, we will choose a small set of unselected agents to get a total of $d$.
These agents will be selected carefully, so that 1) there additional prefix load w.r.t items $E$ fits together with agents assigned earlier, and 2) all their $S$ demands can be satisfied using distinct items of $S$.

\subsection{Partial assignment}

First, consider an arbitrary partition $M = E + S$.
We extend the definition of MMS profiles $\pi_{i,j}$ to sub-profiles that use only items from $E$ or $S$ respectively.
\begin{definition}
    For every agent $i \in N$, let $\Bb_i = \{B_{i,1},\ldots, B_{i,d}\}$ denote the $d$-partition of $M$ witnessing $\mms(M, v_i, d)$, for every $B_{i,j} \in \Bb_i$ it holds $v_i(B_{i,j}) = 1$.
    For a given partition $M = E + S$, for every $j \in [d]$, define 
    \begin{gather*}
        \pi_{i,j}^E:= \left\{\frac{\lfloor q^2\cdot v_i(e)\rfloor}{q^2} :\; e \in B_{i,j}\cap E,\; v_i(e) \geq 1/q\right\},\\ \pi_{i,j}^S:= \left\{\frac{\lfloor q^2\cdot v_i(e)\rfloor}{q^2} :\; e \in B_{i,j}\cap S,\; v_i(e) \geq 1/q\right\}.
    \end{gather*}
    It is clear that for every $\pi_{i,j}$ it holds $\pi_{i,j} = \pi_{i,j}^E \uplus \pi_{i,j}^S$, where $\uplus$ is union with multiplicities.
\end{definition}
Let $V\subseteq N$ be an arbitrary set of at least $d$ agents.
The following theorem establishes that, as long as the average tail is small, for any partition $M = E + S$ it is always possible to select at least $d - |S| - O_{\eps,q}(\log n)$ agents $A\subseteq V$ and assign to them head profiles so that
\begin{enumerate}
    \item all profiles have zero deficit (so agents are satisfied already from the head items only);
    \item all profiles are simultaneously satisfiable using distinct items of set $E$ only.
\end{enumerate} 
We prove it by observing that, in the small tail case, the number of MMS-profiles with positive deficit is small, and then apply of the rounding \cref{rounding} to the profiles $\pi_{i,j}$ that 1) have zero deficit and 2) consist of only $E$-demands, i.e $\pi_{i,j} = \pi_{i,j}^E$.
\begin{theorem}[Partial selection]\label{Erounding}
    Let $M = E + S$ be any partition of the items, and let $V\subseteq N$ satisfy $|V|\geq d$.
    There exists set $A\subseteq V$, and for every $i \in A$ a profile choice $j_i \in [d]$, such that
    \begin{itemize}
        \item profile $\pi_i := \pi_{i,j_i}$ assigned to $i$ satisfies $\Delta(\pi_{i,j_i}) = 0$ and $\pi_{i,j_i}^S = \varnothing$ (i.e, $\pi_{i,j_i}^E = \pi_{i,j_i}$);
        \item all profiles $\{\pi_i\}_{i \in A}$ can be simultaneously satisfied by distinct items of $E$; and for every $0 \leq \ell \leq m$ the total $\ell$-prefix load of $\{\pi_i\}_{i \in A}$  satisfies
        \[
            \sum_{i \in A}\Pp_i(\pi_i, \ell) \leq \frac{1}{|V|}\sum_{i \in V}\sum_{j = 1}^d\Pp_i(\pi_{i,j}^E, \ell);
        \]
        \item taking $\rho_A := 8 + 16q \log n$, it holds
        \[
            |A| \geq d - |S| - \frac{q(q - 1)}{|V|}\sum_{i \in V}T_i - \rho_A.
        \]
        I.e, when $|S| = O_{\eps, q}(\log n)$, $\oT = O_{\eps,q}(\log n)$ and $|V|\geq 3n/4$ have $|A| \geq d- O_{\eps,q}(\log n)$.
    \end{itemize}
\end{theorem}
\begin{proof}
    For every agent $i \in V$, take the set $J_i$ of zero deficit profiles $\pi_{i,j}$ that only use items $E$:
    \[
        J_i := \{j \in [d]: \Delta(\pi_{i,j}) = 0\text{ and }\pi_{i,j}^S = 0\}.
    \]
    First we bound the size of $J_i$.
    If $\pi_{i,j}^S \neq \varnothing$, then $B_{i,j}$ contains an item from $S$.
    Since all bundles $B_{i,j}$ are disjoint, removing all profiles $j$ with $\pi_{i,j}^S \neq \varnothing$ leaves us with at least $d - |S|$ profiles.
    Next, since both $v(\pi_{i,j})$ and $1 - 1/q$ are integer multiples of $1/q^2$, if $\Delta(\pi_{i,j}) = (1 - 1/q - v(\pi_{i,k})) > 0$, then necessarily $\Delta(\pi_{i,j}) \geq 1/q^2$.
    By \cref{fracload} we have $\sum_{j = 1}^d\Delta(\pi_{i,j}) \leq (1 - 1/q)T_i$, implying that the number of $j \in [d]$ such that $\Delta(\pi_{i,j}) > 0$ is at most $q^2(1 - 1/q)T_i = q(q - 1)T_i$.
    Thus,
    \[
        |J_i| \geq d - |S| - q(q -1)T_i.
    \]
    
    We now construct the fractional profiles $(x^*, \Pi)$, where $\Pi =  \{\varnothing, \bot\}+\bigcup_{i\in N}\{\pi_{i,j} : j \in [d]\}$ is the common set of profiles of all agents $N$.
    We define $x^*$ so that the rounding algorithm may choose only agents from $V$, and each of these agents $i$ can be assigned only a profile from $J_i$.
    First, for $i \notin V$, set $x^*_i(\bot) = 1$ and $x^*_i(\pi) = 0$ for every other $\pi  \neq \bot$ --- this ensures that we only choose agents from $V$.
    Then, for $i \in V$, set
    \[
        x_i^*(\bot) = 1 - \frac{|J_i|}{|V|},\qquad\qquad \forall \pi \neq\bot:\; x_i^*(\pi) = \frac{|\{j \in J_i : \pi_{i,j} = \pi\}|}{|V|}.
    \]
    As $|J_i|\leq d \leq |V|$, these are valid fractional profiles.
    The expected number of selected agents is
    \[
        \Ii(x^*, \Pi) = \sum_{i \in N}\sum_{\pi \neq \bot}x^*_i(\pi) = \sum_{i \in V}\frac{|J_i|}{|V|} \geq d - |S| - \frac{q(q - 1)}{V}\sum_{i \in V}T_i.
    \]
    Next, let $i \in V$ and $0 \leq \ell \leq m$.
    Consider some demand $\alpha \in \pi_{i,j}$ for $j \in J_i$, the item $e \in B_{i,j}$ inducing this $\alpha$ belongs $E$.
    Every demand $\alpha$ counted in $\Pp_i(\pi_{i,j}, \ell) = |\{\alpha \in \pi_{i,j} : m_i(\alpha) \leq \ell\}|$ has breakpoint at most $\ell$, so the item $e$ inducing this $\alpha$ also lies in $\{1, \ldots, \ell\}$.
    Since all such items $e$ are different across the MMS bundles $B_{i,j}$, we get that
    \[
        \sum_{j \in J_i}\Pp_i(\pi_{i,j}, \ell) \leq |E\cap [\ell]|.
    \]
    Averaging over $V$, we get that for every $0 \leq \ell \leq m$ the fractional $\ell$-prefix head load of $x^*$ satisfies
    \[
        \Hh(x^*, \Pi, \ell) = \frac{1}{|V|}\sum_{i \in V}\sum_{j \in J_i}\Pp_i(\pi_{i,j}, \ell)\leq |E\cap [\ell]|.
    \]
    
    As all deficits are zero, $(x^*, \Pi)$ generates no tail load at all, and $\tau(x, \Pi) = 0$ as well as $\Tt(x^*, \Pi, \ell) = 0$ for all $0 \leq \ell \leq m$.
    We apply the rounding \cref{rounding} to $(x^*, \Pi)$.
    The number $X^*$ of strictly fractional entries of $x^*$ is at most $(d + 1)\cdot|V|\leq (d + 1)n \leq n^2$.
    So, \cref{rounding} produces an integral profile $(y, \Pi)$ such that, taking $\rho(q, 1/q^2, X^*) \leq 8 + 16q \log n =: \rho_A$,
    \[
        \forall 0 \leq \ell \leq m: \Hh(y, \Pi, \ell) \leq \Hh(x^*, \Pi, \ell)\qquad\text{and}\qquad \Ii(y, \Pi) \geq \Ii(x, \Pi) - \rho_A.
    \]
    Let $A = \{i \in N: y_i(\bot) = 0\}$ be the set of $\Ii(y, \Pi)$ selected agents.
    Since $y$ preserves the support of $x^*$, we must have $A\subseteq V$, and the profile $\pi_i \in \Pi$ such that $y_i(\pi_i) = 1$ must be of the form $\pi_i = \pi_{i,j_i}$ for some $j_i \in J_i$.
    Denoting $\Pi_A := \{\pi_{i,j_i} : i \in A\}$, we have for every $0 \leq \ell \leq m$:
    \[
        \Pp(\Pi_A, \ell) = \Hh(y, \Pi, \ell) \leq |E\cap [\ell]|,
    \]
    so by \cref{loadequiv} we can satisfy $\Pi_A$ by distinct items of $E$. In addition, for every $0\leq \ell \leq m$,
    \[
            \sum_{i \in A}\Pp_i(\pi_i, \ell) \leq \frac{1}{|V|}\sum_{i \in V}\sum_{j = 1}^d\Pp_i(\pi_{i,j}^E, \ell).
        \]
\end{proof}
Consider some partition $M = E + S$ and some large set $V\subseteq N$, and apply \cref{Erounding}.
If the resulting set $A\subseteq V$ satisfies $|A|\geq d$, we are done.
Unfortunately, we do not know how to pick such $E$ and $V$, and after the rounding we will most likely have $|A| < d$.
However, assuming $|S|$ and $\oT$ are at most $O_{\eps,q}(\log n)$, we have $d - |A| \leq O_{\eps,q}(\log n)$ too, so we need to add at most $O_{\eps,q}(\log n)$ additional agents to reach $d$.
We will do it as follows.

Before applying \cref{Erounding}, we will fix some set $U\subseteq N - V$ of unused agents.
Then, having obtained agents $A\subseteq V$, we will select some set $A_0 \subseteq A$ of them with assigned profiles $\{\pi_i\}_{i \in A_0}$ from rounding, and add a small set of agents $U_0\subseteq U$ with assigned profiles $\{\pi_u\}_{u \in U_0}$.
These $A_0, U_0$ will satisfy $|A_0| + |U_0| = d$, and for every prefix $0 \leq \ell \leq m$ we will have 
\[
    \sum_{i \in A_0}\Pp_i(\pi_i, \ell) + \sum_{u \in U_0}\Pp_u(\pi_u, \ell) \leq \ell.
\]
For now we will not worry about additional slackness for tail completion, this will be handled later.
The main question is, even without tails, how to select agents $U_0$ and assign to them profiles to guarantee the prefix load bound above, so that to ensure that both $A_0, U_0$ can be simultaneously satisfied using distinct items (i.e the condition of \cref{loadequiv}).

The key point is that agents $A_0$ will use only items from $E$, so it suffices to ensure that the $E$-demands of added agents $U_0$ fit together with $A_0$.
At the same time, agents $U_0$ will be able to use items $S = M - E$ freely, and we need to pick $U_0$ so that their $S$-demands can be simultaneously satisfied by $S$.
In the following subsection, we explain how exactly we define the partitions $M = E + S$ and $N = U + V + W$ to guarantee us such set $U_0\subseteq U$ of sufficient size.

\subsection{Saved items}

\label{sec:saved}

Suppose that we selected agents $A_0 \subseteq A$, keeping their profiles from \cref{Erounding}, and additional agents $U_0\subseteq U$ with newly assigned profiles $\{\pi_{u}\}_{u \in U_0}$, so that $|A_0| + |U_0| = d$.
As mentioned earlier, for the profiles of $A_0, U_0$ together to be simultaneously satisfiable by distinct head items, these agents must preserve the prefix load condition.
That is, for every $0\leq \ell \leq m$,
\[
    \sum_{i \in A_0}\Pp_i(\pi_i, \ell) + \sum_{u \in U_0}\Pp_u(\pi_u, \ell) \leq \ell.
\]

Jumping ahead, for every agent $u \in U_0$ the assigned profile will be of the form $\pi_{u_*,j}$ for some special agent $u_* \in U$ (not necessarily $u_* = u$) and some $j \in [d]$.
Because of this, the decomposition $\pi_u := \pi_u^E \uplus \pi_u^S$ is well-defined also for profiles of agents $u \in U_0$.

Observe that, for the demands of $A_0$ and $U_0$ to fit together into a prefix $\ell$, it suffices that the $E$-parts of their profiles fit in $E\cap [\ell]$, and the $S$-parts of their profiles fit in $S\cap [\ell]$.
To see this, recall that agents $i \in A_0$ keep their assigned profiles $\pi_i = \pi_{i,j_i}$ such that $\pi_{i,j_i}^S = \varnothing$ (i.e $\pi_{i} = \pi_i^E$), so they will only use items from $E\cap [\ell]$.
Then, writing separate prefix load conditions on $E$-parts of $A_0, U_0$ and $S$-parts of $U_0$ only, we have for every $0\leq \ell \leq m$,
\begin{gather}
    \sum_{i \in A_0}\Pp_i(\pi_i^E, \ell) + \sum_{u \in U_0}\Pp_u(\pi_u^E, \ell) \leq |E\cap [\ell]|;\tag{E}\label{E}\\
    \sum_{u \in U_0}\Pp_u(\pi_u^S, \ell) \leq  |S\cap [\ell]|.\tag{S}\label{S}
\end{gather}
Since $M = E + S$ and $\Pp_u(\pi_u, \ell) = \Pp_u(\pi_u^E, \ell) + \Pp_u(\pi_u^S, \ell)$, together (\ref{E}) and (\ref{S}) imply
\[
    \sum_{i \in A_0}\Pp_i(\pi_i, \ell) + \sum_{u \in U_0}\Pp_u(\pi_u, \ell) \leq |E\cap [\ell]| + |S\cap [\ell] = \ell.
\]

We will first focus on condition (\ref{E}).
\cref{Erounding} guarantees us that for every $0 \leq \ell \leq m$,
\[
    \sum_{i \in A_0}\Pp_i(\pi_i^E, \ell)\leq  \frac{1}{|V|}\sum_{i \in V}\sum_{j = 1}^d\Pp_i(\pi_{i,j}^E, \ell).
\]
We would like to pick sets $E\subseteq M$ and $V\subseteq N$ in such a way, that the expression above has enough slackness with respect to $|E\cap [\ell]|$ to absorb additional load $\sum_{u \in U_0}\Pp_u(\pi_u^E, \ell)$.
We can give the following crude upper bound on the additional load of $U_0$.
For every $0 \leq \ell \leq m$,
\[
   \sum_{u \in U_0}\Pp_u(\pi_u^E, \ell) \leq \sum_{u \in U_0}|\pi_u^E| \leq q|U_0|,
\]
since MMS profiles have at most $q$ head demands.
Then, to ensure condition (\ref{E}), it suffices to pick $E\subseteq M$ and $V\subseteq N$, and then select set $U_0\subseteq U$, so that for all $0 \leq \ell \leq m$, there is sufficient slackness at prefix $\ell$:
\[
    \frac{1}{|V|}\sum_{i \in V}\sum_{j = 1}^d\Pp_i(\pi_{i,j}^E, \ell) \leq |E\cap [\ell]| - q|U_0|.
\]
This condition turns out to be too strong.
While the demands of $\pi_{i,j}^E$ originate from items of $E$, they \textit{need not} have their breakpoints in $E$.
So, as $\ell$ increases, the lefthand side of the expression above may increase substantially, while the righthand side stays unchanged (i.e, when $\ell \in S$).

Here is when the choice of $S$ comes into play.
Using $S$, we will define a special set $\Ll_S \subset S$ of \textbf{protected prefixes}, of size at most $2q^2$.
For each prefix $\ell_s \in \Ll_S$, we will enforce
\begin{equation}
    \sum_{i \in A_0}\Pp_i(\pi_i^E, \ell_s) + \sum_{u \in U_0}\Pp_u(\pi_u^E, \ell_s) \leq |E\cap [\ell_s]|.\tag{$*$}\label{*}
\end{equation}
The sets $S$ and $\Ll_S$ will be chosen in such a way that, as long as the prefix load is preserved at every protected prefix $\ell_s \in \Ll_S$, for some integer $\eta \simeq q^{2q^2}|S|$ we have the following.
For every $\ell \in E$ that is a \textit{breakpoint} of some agent $i \in V$, we get a lot of slackness in the fractional solution:
\[
    \frac{1}{|V|}\sum_{i \in V}\sum_{j = 1}^d\Pp_i(\pi_{i,j}^E, \ell) \leq |E\cap [\ell]| - \frac{\eta}{3}.
\]
We will show that, with this choice of $S$ and $\Ll_S$, in order to guarantee that condition (\ref{E}) holds for every $0 \leq \ell \leq m$ it suffices to choose $U_0\subseteq U$ of size $|U_0|\leq \eta/(3q)$ that satisfies (\ref{*}) for all protected prefixes $\ell_s \in \Ll_S$.
At the same time, we will need $|U_0| = d - |A_0| \geq d - |A|$.
When $|S|$ and $\oT$ are $O_{\eps,q}(\log n)$, we get $d - |A|\leq c_{\eps,q}\log n$ for some constant $c_{\eps,q}$, so we need $U_0$ such that $c_{\eps,q}\log n \leq |U_0|\leq \eta/3q = O_{\eps,q}(|S|)$.
Taking $n$ large enough will guarantee that such $U_0$ exists.

\subsubsection{Construction of item/agent partitions}
Our set $S\subseteq M$ will be of the form
\[
    S = \{a_1, a_1 + 1,\ldots, b_1\} \cup \ldots \cup \{a_t, a_t + 1, \ldots, b_t\},
\]
where each $[a_s, b_s]$ for $s \in [t]$ is a maximal consecutive interval of items from $S$.
We will have that $t \leq 2q^2$, and the length of each consecutive interval and the total size of $S$ are both at most $O_{\eps,q}(\log n)$.
In addition, it will hold that $a_1 = 1$, i.e $S$ includes some prefix $\{1, \ldots, b_1\}$.
The protected prefixes $\Ll_s$ will be the endpoints $b_s$ for $s \in [t]$.

Put $E = M - S$.
What is arguably more important for slackness bound is not necessarily the $S$ itself, but a \textbf{breakpoint-free region} $F\subseteq E$ defined by $S$.
Specifically, after every interval $[a_s, b_s]\subset S$, $s \in [t]$, we will have another interval $\{b_s + 1, \ldots, b_s + \eta\}$ of length $\eta \geq q^{2q^2} |S|$, and
\[
    F = E\cap (\{b_1 + 1, \ldots, b_1 + \eta\} \cup \ldots \cup \{b_t + 1, \ldots, b_t + \eta\}).
\]
The set $S$ and parameter $\eta$ are picked in such a way that
\begin{itemize}
    \item the set of agents $W\subset N$ that have any breakpoint in $F$ satisfies $|W|\leq \eps n / 8$;
    \item if $\ell \in E - F$ is a breakpoint of some agent from $N - W$, then the whole interval $(\ell - \eta, \ell] = \{\ell - \eta + 1, \ldots, \ell\}$ lies in $E$, and there are at most $n/2$ agents with breakpoints in $(\ell - \eta, \ell]$.
\end{itemize}
In a sense, set of saved items $S$ defines a breakpoint-free region $F\subseteq E$ that serves as a ``buffer'' for prefix load with respect to items $E$.
The last property essentially says that the $E$-demands of agents from $N - W$ are ``well-spread'' across the whole set of items $E$.
It is exactly this property that will ensure that the total fractional load of agents $V\subseteq N - W$ has slackness at least $\eta/3$.

We will also have partition agents $N$ into sets $N = U + V + W$, where
\begin{enumerate}
    \item $W$ contains all agents with breakpoints in $F$, $|W|\leq  \eps n/8$.
    We throw $W$ away.
    \item $U$ contains $\eps n/8$ agents of $N - W$ with \textit{largest heads} $h_i$.
    The heads property will be needed later for tail completion, if there are no tails, $U$ can be arbitrary subset of $N - W$.
    \item $V = N - U - W$ contains most of the agents, i.e $|V|\geq (1 - \eps/4)n$, and will participate in \cref{Erounding} together with $E = M - S$, which gives us the set $A\subseteq V$ described earlier.
\end{enumerate}

We will now establishes our choices of set $W\subset N$ and $S\subset M$ with desired properties.
Consider the set $Q$ of possible values of head demands:
\[
    Q :=\left\{\frac{q + k}{q^2} : k \in \{0, 1, 2, \ldots, q^2 - q\right\}.
\]
\begin{lemma}[Saved items]\label{saved}
    Let $\eps \in (0, 1)$, and let $r_1, C_1 \geq 2$, $r_1 \leq n$ be integers.
    There exist a set $W\subset N$ of size $|W|\leq  \eps n/8$ and a set $S\subset M$ of size $r_1 \leq |S|\leq 2q^2C_1^{8q^2/\eps}r_1$ and of the form \[
        S = \{a_1, a_1 + 1, \ldots, b_1\} \cup \ldots\cup  \{a_t, a_t +1, \ldots, b_t\}
    \]
    for $t \leq 2q^2$, where $1 \in S$ and each $[a_s, b_s]\subset S$, $s\in[t]$, is a maximal consecutive interval, with the following property.
    Let $E = M - S$.
    There exists integer $\eta$ satisfying
    \[
        \eta\geq \frac{C_1 - 1}{2q^2} |S|,
    \]
    such that for every $i \in N - W$ and every $\alpha \in Q$, if the breakpoint $\ell := m_i(\alpha)$ belongs to $E$, then
    \begin{itemize}
        \item $\ell \geq \eta$, and the $\eta$ preceding items all belong to $E$, i.e $\{\ell -\eta + 1, \ldots, \ell\} \subseteq E$;
        \item the interval $(\ell - \eta, \ell]$ contains at most $ n/2 $ breakpoint occurrences, with repetitions.
        I.e,
        \[
            |\{(i, \alpha) :\; i \in N, \alpha \in Q \text{ and }\ell - \eta < m_i(\alpha) \leq \ell\}| \leq  n/2.
        \]
    \end{itemize}
\end{lemma}
\begin{proof}
For every agent $i \in N$, we list all possible positive breakpoints of $i$, i.e all $m_i(\alpha) > 0$ for $\alpha \in Q$, and combine them into a multiset $\Mm(Q)$:
\[
    \Mm(Q) := \biguplus_{i \in N}\biguplus_{\substack{\alpha \in Q\\m_i(\alpha) > 0}}\{(i, m_i(\alpha))\}.
\]
Since $|Q|\leq q^2$, $|\Mm(Q)|\leq q^2n$.
We sort $\Mm(Q)$ in non-decreasing order w.r.t $m_i(\alpha)$, and mark every $(\lfloor n/2\rfloor + 1)$-th breakpoint.
That is, we create a set of \textit{marked breakpoints}
\[
        \Mm_0 = \{(i_{k(\lfloor n/2\rfloor + 1)}, m_{k(\lfloor n/2\rfloor + 1)}) : k = 1, 2\ldots\}.
\]
Clearly, $|\Mm_0|\leq 2q^2 - 1$.
Note that any sub-interval of $\Mm(Q)$ with no marked breakpoints has length at most $n/2$. 
Let set $S_0$ contain $\{1\}$ and every unique position in $M$ marked by $\Mm_0$, i.e
\[
    S_0 = \{1\}\cup \big\{s \in \{1, \ldots, m\} : (i, s) \in \Mm_0 \text{ for some }i \in N\big\}.
\]
As $|\Mm_0|\leq 2q^2 - 1$, we have $|S_0|\leq 2q^2$.

We will now construct a sequence of \textbf{saved regions}.
For integer radius $r \geq 1$, saved region $S(r)$ will contain all items of $M$ within distance $r$ from $S_0$ (to the right):
\[
        S(r) := \bigcup_{s \in S_0}\{s, s + 1, \ldots, s + r -1\}\cap M.
\]
We will also have a sequence of \textbf{protected regions} $F(r) := S(C_1 \cdot r) - S(r)$.
In addition, let 
\[
        \Mm_F(r) := \{(i_k, m_{i_k}) \in \Mm(Q): m_{i_k} \in F(r)\}
\]
be the multiset of all breakpoints in $\Mm(Q)$ whose value falls into the protected region $F(r)$.

We start with $S(r_1)$ for then given radius $r_1$, and for $k \geq 1$ keep increasing radius multiplicatively by a given factor of $C_1$, i.e $r_{k + 1} = C_1\cdot r_k$.
So, by definition, $S(r_{k + 1}) = S(r_k) + F(r_k)$.
We keep increasing $r_k \to r_{k + 1}$ until we reach some $k^*$ such that at most $\lfloor\eps n/8\rfloor$ breakpoints fall into the $k^*$-th protected region $F(r_{k^*})$, i.e until
\[
    |\Mm_F(r_{k^*})| \leq \lfloor\eps n/8\rfloor.
\]
Since $|\Mm(Q)|\leq q^2n$, we will increase radius at most $\lfloor 8q^2/\eps\rfloor$ times, so $r_*:= r_{k^*}$ satisfies 
\[
    r_* \leq C_1^{8q^2/\eps}r_1.
\]
For this $r_*$, we define set $S$ as $S = S(r_*)$, set $F$ as $F = F(r_*)$, and set $W$ as
\[
    W = \{i \in N :\; \exists \alpha \in Q \text{ such that } m_i(\alpha) \in F\}.
\]
Clearly, $|W|\leq |\Mm_F(r_{_*})| \leq \eps n /8$.
In addition,
\[
    S = \bigcup_{s \in S_0}\{s, s + 1, \ldots, s + r_* -1\}\cap M,
\]
so $S$ is a union of at most $|S_0| \leq 2q^2$ consecutive intervals, each of length at most $r_*$, and $S$ contains at least $S(r_1).$
Hence, $r_1 \leq |S|\leq 2q^2r_*\leq 2q^2C_1^{8q^2/\eps}r_1$, and taking $\eta := (C_1 - 1)\cdot r_*$ gives
\[
    \eta \geq \frac{C_1 - 1}{2q^2}|S|.
\]

We will now prove that integer $\eta = (C_1 - 1)\cdot r_*$ has the desired property.
Take representation
\[
        S = \{a_1, a_1 + 1, \ldots, b_1\} \cup \ldots\cup  \{a_t, a_t +1, \ldots, b_t\}
\]
for $t \leq 2q^2$, where each $[a_s, b_s]\subset S$, $s \in [t]$, is a maximal consecutive interval.
Observe that
\[
    S(C_1\cdot r_*) = S(r_* + \eta) = M\cap \bigcup_{s = 1}^t\{a_s, a_s + 1,\ldots, b_s, b_s + 1, \ldots, b_s + \eta\},
\]
therefore
\[
    F= F(r_*) = S(C_1\cdot r_*) - S(r_*) = E\cap \bigcup_{s = 1}^t\{b_s + 1, \ldots, b_s + \eta\}.
\]

Consider some agent $i \in N - W$ and a demand $\alpha \in Q$ whose breakpoint $\ell = m_i(\alpha)$ belongs to $E$.
Let $[a_s, b_s]\subset S$ be the last maximal interval of $S$ preceding $\ell$, i.e such that $b_s < \ell$ and $b_s$ is the largest possible.
Such interval exists, as $1 \in S_0\subseteq S$.

By definition of $W$, we have $\ell \notin F$.
Then, it must also hold $\ell > b_s + \eta$, as otherwise $\ell \in E\cap \{b_s + 1, \ldots, b_s + \eta\}\subseteq F$.
Choice of $[a_s, b_s]$ implies $\{b_s + 1 ,\ldots, b_s + \eta, \ldots, \ell\}\subseteq E$, and 
\[
    \{\ell - \eta + 1, \ldots, \ell\} \subseteq \{b_s + 1, \ldots, b_s + \eta, \ldots, \ell\}\subseteq E.
\]
In particular, $\ell - \eta + 1 \geq 1 \implies \ell \geq \eta$.

It remains to bound the number of different breakpoint occurrences in $(\ell - \eta, \ell]$.
The breakpoints of $\Mm(Q)$ that fall into $(\ell - \eta, \ell]$ form a consecutive block in $\Mm(Q)$.
If this block contained at least $\lfloor n/2 \rfloor + 1$ breakpoints, then by construction it would contain at least one marked breakpoint from $\Mm_0$.
But then, the position of this marked breakpoint would be in $S_0 \cap (\ell - \eta, \ell]$, which is impossible as $(\ell - \eta, \ell]$ lies entirely in $E$.
It follows that
\[
            |\{(i, \alpha) :\; i \in N, \alpha \in Q \text{ and }\ell - \eta < m_i(\alpha) \leq \ell\}| \leq \left\lfloor n/2\right\rfloor \leq n/2.
        \]
Note that all repetitions are counted separately.
Breakpoints that fall on the same position cause no issue, as $(\ell - \eta, \ell]$ would include either all breakpoint occurences or none of them.
\end{proof}

\subsubsection{Protected prefixes}

Given \cref{saved}, we define the partitions $M = E + S$ and $N = U + V + W$ explicitly.
\begin{definition}
    Given the sets $W$ and $S$ from \cref{saved}, we define $E := M - S$, $U$ as the set of $\lfloor \eps n/8\rfloor$ agents $i \in N - W$ with the largest head thresholds $h_i$, and $V := N - U - W$.
\end{definition}

In addition, we will explicitly define the set of protected prefixes $\Ll_S\subset S$ and rigorously prove that, for $A_0\subseteq A$ and $U_0\subseteq U$ to satisfy prefix load constraints (\ref{E}) for every prefix $0 \leq \ell \leq m$, it suffices to satisfy them only on protected prefixes $\ell_s \in \Ll_s$, provided that $|U_0|$ is not too large.
\begin{definition}
    Decompose set $S$ from \cref{saved} as
    \[
        S = \{a_1, a_1 + 1, \ldots, b_1\} \cup \ldots\cup  \{a_t, a_t +1, \ldots, b_t\},
    \]
    where each $[a_s, b_s]$ for $s \in [t]$ is a maximal list of consecutive integers from $\{1, \ldots, m\}$.
    We define a set $\Ll_S\subset S$ of \textbf{protected prefixes} as
    \[
        \Ll_S := \{b_s : s \in [t]\}.
    \]
    Note that as $t \leq 2q^2$, we have $|\Ll_S|\leq 2q^2$ --- a constant number of prefixes.
\end{definition}
\begin{corollary}\label{Ebound}
    Let $A\subseteq V$ be agents with assigned profiles $\{\pi_i\}_{i \in A}$, obtained from \cref{Erounding} applied to the defined $E, S$ and $V$, and let $A_0\subseteq A$.
    Let $U_0\subseteq U$ be some additional set of agents with assigned integral profiles $\{\pi_u\}_{u \in U_0}$.
    Suppose that for every $\ell_s \in \Ll_S$,
    \begin{equation}
        \sum_{i \in A_0}\Pp_i(\pi_i^E, \ell_s) + \sum_{u \in U_0}\Pp_u(\pi_u^E, \ell_s) \leq |E\cap [\ell_s]|\tag{$\mathrm{E_{\Ll_S}}$},\label{ES}
    \end{equation}
    as well as $|U_0|\leq \eta/(3q)$ for $\eta$ defined in \cref{saved}.
    Then, for every $0 \leq \ell \leq m$ it holds
    \begin{equation}
        \sum_{i \in A_0}\Pp_i(\pi_i^E, \ell) + \sum_{u \in U_0}\Pp_u(\pi_u^E, \ell) \leq |E\cap [\ell]|. \tag{E}\label{E2}
    \end{equation}
\end{corollary}
\begin{proof}
We will prove the statement separately for prefixes from $E$ and from $S$.
First, take $\ell \in S$.
If $\ell \in \Ll_S$, condition (\ref{E2}) is already satisfied by assumption (\ref{ES}), so let $\ell \notin \Ll_S$.
In this case, $\ell \in [a_s, b_s]$ for some maximal consecutive interval $[a_s, b_s]\subset S$.
Then, 
\[
    \forall i \in A_0:\; \Pp_i(\pi_i^E, \ell)\leq \Pp_i(\pi_i^E, b_s) \qquad\text{and}\qquad \forall u \in U_0:\; \Pp_u(\pi_u^E, \ell) \leq \Pp_u(\pi_u^E, b_s).
\]
Since $b_s \in \Ll_s$ by definition, and every item in $(\ell, b_s]$ belongs to $S$, it holds
\[
    \sum_{i \in A_0}\Pp_i(\pi_i^E, \ell) + \sum_{u \in U_0}\Pp_u(\pi_u^E, \ell) \leq |E\cap [b_s]| = |E\cap [\ell]|.
\]

Next, consider $\ell \in E$.
Assume first that $\ell$ is a breakpoint of some agent $i \in N - W$.
In this case, by \cref{saved} it holds $\{\ell - \eta + 1, \ell\}\subseteq E$ and interval $(\ell - \eta, \ell]$ contains at most $n/2$ breakpoint occurrences.
In addition, the rounding of \cref{Erounding} guarantees that
\[
    \sum_{i \in A_0}\Pp_i(\pi_i^E, \ell) \leq \frac{1}{|V|}\sum_{i \in V}\sum_{j = 1}^d\Pp_i(\pi_{i,j}^E, \ell).
\]
Let $V_{\eta,\ell} \subset V$ be the set of agents $i \in V$ that have a breakpoint in $(\ell - \eta, \ell]$, it holds $|V_{\eta,\ell}|\leq  n/2 $.
For every $i\in V_{\eta,\ell}$ we have (say, by \cref{fracload} or \cref{Erounding}) that
\[
    \sum_{j = 1}^d\Pp_i(\pi_{i,j}^E, \ell) \leq |E\cap [\ell]|.
\]
On the other hand, for every agent $i \in V- V_{\eta,\ell}$, as there are no breakpoints in $(\ell - \eta, \ell]$, it holds
\[
    \sum_{j = 1}^d\Pp_i(\pi_{i,j}^E, \ell)  = \sum_{j = 1}^d\Pp_i(\pi_{i,j}^E, \ell - \eta) \leq |E\cap [\ell - \eta]| = |E\cap [\ell]| - \eta.
\]
The last equality is due to $(\ell - \eta, \ell] \subset E$.
Combining the bounds for $V_{\eta,\ell}$ and $V - V_{\eta,\ell}$, we get
\begin{multline*}
    \frac{1}{|V|}\sum_{i \in V}\sum_{j = 1}^d\Pp_i(\pi_{i,j}^E, \ell) \leq \frac{|V| - |V_{\eta,\ell}|}{|V|}\cdot (|E\cap [\ell] - \eta) + \frac{|V_{\eta,\ell}|}{|V|}\cdot |E\cap [\ell]| \\
    = |E\cap [\ell]| - \eta \cdot \left(1 - \frac{|V_{\eta,\ell}|}{|V|}\right) \leq |E\cap [\ell]| - \frac{\eta}{3}.
\end{multline*}
We used the fact that $|V_{\eta,\ell}|\leq n/2$ and $|V|\geq (1 - \eps/4)n \geq 3n/4$.
As mentioned earlier,
\[
    \sum_{u \in U_0}\Pp_u(\pi_u^E, \ell) \leq \sum_{u \in U_0}|\pi_u^E|\leq qU_0,
\]
hence under the assumption $|U_0|\leq \eta/(3q)$ we get condition (\ref{E2}) at $\ell$, combining the bounds:
\[
    \sum_{i \in A_0}\Pp_i(\pi_i^E, \ell) + \sum_{u \in U_0}\Pp_u(\pi_u^E, \ell) \leq |E\cap [\ell]|.
\]

Now, consider $\ell \in E$ that is not a breakpoint of any agent $i \in N - W$.
Let $\ell_s \in \Ll_s$ be the largest protected prefix preceding $\ell$ (such exists as $1 \in S$).
If there is some $\ell_* \in (\ell_s, \ell]$ that is a breakpoint of some $i \in N - W$, then
\[
    \sum_{i \in A_0}\Pp_i(\pi_i^E, \ell) + \sum_{u \in U_0}\Pp_u(\pi_u^E, \ell)  =  \sum_{i \in A_0}\Pp_i(\pi_i^E, \ell_*) + \sum_{u \in U_0}\Pp_u(\pi_u^E, \ell_*) \leq |E\cap [\ell_*]| \leq |E\cap[\ell]|,
\]
since prefix load of an agent is constant between consecutive breakpoints.
If there are no breakpoints in $(\ell_s, \ell]$, then the load at $\ell$ is the same as the load at $\ell_s$, hence
\[
    \sum_{i \in A_0}\Pp_i(\pi_i^E, \ell) + \sum_{u \in U_0}\Pp_u(\pi_u^E, \ell)  =  \sum_{i \in A_0}\Pp_i(\pi_i^E, \ell_s) + \sum_{u \in U_0}\Pp_u(\pi_u^E, \ell_s) \leq |E\cap [\ell_s]| \leq |E\cap[\ell]|,
\]
by assumption (\ref{ES}).
Finally, for $\ell = 0$, since the first interval of $S$ is $[1, b_1]$ for $b_1 \in \Ll_S$, it holds $|E\cap [b_1]| = 0$.
As a result, by (\ref{ES}) we have
\[
    \sum_{i \in A_0}\Pp_i(\pi_i^E, 0) + \sum_{u \in U_0}\Pp_u(\pi_u^E, 0) \leq \sum_{i \in A_0}\Pp_i(\pi_i^E, b_1) + \sum_{u \in U_0}\Pp_u(\pi_u^E, b_1) \leq |E\cap[b_1]| = 0.
\]
\end{proof}


We show how to find sets $A_0\subseteq A$ and $U_0\subseteq U$ satisfying \cref{Ebound} in the next section.

\section{Supplier and recipients}

\label{sec:supp}
Following \cref{Ebound}, we now have a concrete goal to satisfy conditions (\ref{E}) and (\ref{S}): find  sets $A_0\subseteq A$ and $U_0\subseteq U$, such that $|A_0| + |U_0|= d$ as well as
\begin{align}
    &&&\sum_{i \in A_0}\Pp_i(\pi_i^E, \ell_s) + \sum_{u \in U_0}\Pp_u(\pi_u^E, \ell_s) \leq |E\cap [\ell_s]|,&\forall \ell_s \in \Ll_S\tag{$\mathrm{E_{\Ll_S}}$}\label{ES2}\\
    &&&|U_0|\leq \frac{\eta}{3q} \tag{U}\label{U2},\\
    &&&\sum_{u \in U_0}\Pp_u(\pi_u^S, \ell) \leq |S\cap[\ell]| ,&\forall 0 \leq \ell \leq m.\tag{S}\label{S2}
\end{align}

Our key idea is the following.
Instead of directly choosing agents $U_0\subseteq U$ satisfying conditions (\ref{ES2}), (\ref{U2}) and (\ref{S2}),
we will first pick a single \textit{supplier} agent $u_* \in U$, and fill existing agents $A$ up to $d$ total agents by using different copies of $u_*$.

More specifically, let $\{\pi_{u_*,j} : j \in [d]\}$ be the MMS profiles of the agent $u_* \in U$.
We want to select a set $J_0\subseteq [d]$ and create $|J_0|$ copies of agent $u_*$, each copy $j \in J_0$ gets assigned a distinct supplier profile $\pi_{u_*,j}$.
Then, we select $A_0\subseteq A$ such that $|A_0| + |J_0| = d$ and analogous conditions (\ref{ES2}), (\ref{U2}) and (\ref{S2}) hold for $A_0$ and the copies $J_0$ of supplier $u_* \in U$.
That is,
\begin{align*}
    &&&\sum_{i \in A_0}\Pp_i(\pi_i^E, \ell_s) + \sum_{j \in J_0}\Pp_{u_*}(\pi_{u_*,j}^E, \ell_s) \leq |E\cap [\ell_s]|,&\forall \ell_s \in \Ll_S;\tag{$\mathrm{E}_{\Ll_S}*$}\label{ES3}\\
    &&&|J_0|\leq \frac{\eta}{3q},\tag{$\mathrm{U}*$}\label{U3}\\
    &&&\sum_{j\in J_0}\Pp_{u_*}(\pi_{u_*,j}^S, \ell) \leq  |S\cap [\ell]|,&\forall 0 \leq \ell \leq m.\tag{S$*$}\label{S3}
\end{align*}
Note that condition (\ref{S3}) holds automatically.
Indeed, for every $j \in J_0$, demands of $\pi_{u_*,j}^S$ are satisfied by their originating items from $B_{u_*,j}\cap S$.
Since all MMS bundles $B_{u_*,j}$ for $j \in J_0$ are disjoint, for every $\ell$-prefix profiles $\{\pi_{u_*,j}:j\in J_0\}$ will use at most $\ell$ distinct items from $S\cap [\ell]$.
So, the total $S$-load at $\ell$ is at most $|S\cap [\ell]|$, and we only need to control (\ref{ES3}) and (\ref{U3}).

Given such set $A_0\subseteq A$ and a supplier $u_* \in U$ with copies $J_0\subseteq [d]$, we then pick a set of recipients $U_0\subseteq U$ of size $|J_0|$.
Each of recipients $u \in U_0$ will receive a designated copy of the \textit{supplier's profile} $\pi_{u_*,j}$, for different $j \in J_0$.
The supplier agent $u_* \in U$ will be chosen in such a way that there exists a sufficiently large set of recipients $U_0\subseteq U$ that can accept distinct copies $J_0$ of $u_*$, while satisfying all the conditions (\ref{ES2}), (\ref{U2}) and (\ref{S2}) with the assigned supplier profiles.

\subsection{Selecting supplier and recipients}

Assume $|A|<d$.
We will now show that, given an arbitrary supplier agent $u_* \in U$, we can pick copies $J_0\subseteq [d]$ of $u_*$ and agents $A_0\subseteq A$ so that $|A_0| + |J_0| = d$, condition (\ref{ES3}) holds and a variation of condition (\ref{U3}) holds.
We refer to it as a ``single supplier'' theorem.

\begin{theorem}[Single Supplier]\label{supplier}
    There exists a global constant $C_{t,q}$ such that the following holds.
    
    For $u_* \in U$, let $\{\pi_{u_*,j} : j \in [d]\}$ be his MMS profiles.
    In addition, let $\Ll_S \subseteq \{0, 1, \ldots, m\}$ be a set of $t$ \textbf{protected prefixes}.
    There exist $A_0\subseteq A$ and $J_0\subseteq [d]$ such that $|A_0| + |J_0| = d$ and
    \begin{enumerate}
        \item $|J_0| \leq C_{t,q}(d- |A|)$, and 
        \item for every protected prefix $\ell_s \in \Ll_S$, the $E$-load is
            \[
                \sum_{i \in A_0}\Pp_i(\pi_i^E, \ell_s) + \sum_{j \in J_0}\Pp_{u_*}(\pi_{u_*,j}^E, \ell_s) \leq |E\cap [\ell_s]|.
            \]
    \end{enumerate}
\end{theorem}
We prove \cref{supplier} in \cref{appendix:supplier} (\cref{supplierproof}).

Note that the bound given by \cref{supplier} on $|J_0|$ is $C_{t,q}(d - |A|)$ for some constant $C_{t,q}$, while the desired condition (\ref{U3}) needs $|J_0|\leq \eta/(3q)$.
This will not be an issue.
We will show that in the small tail case, for an appropriate choice of values $r_1, C_1$ in \cref{saved}, we can guarantee that $d - |A|\leq 2|S|$.
Since $\eta \geq \frac{C_1 - 1}{2q^2}|S|$, as long as constants match, we will get the desired upper bound on $|J_0|$ in terms of $\eta$.

We will now describe how to replace the $J_0$ copies of a single agent $u_*$ with distinct recipient agents $U_0\subseteq U$, and assign them certain profiles.
Our plan is to choose for every $j \in J_0$ some recipient $u(j) \in U$ and assign him the \textit{supplier}'s profile:
\[
    \forall j \in J_0 :\; \text{take $u(j) \in U_0$ and assign }\pi_{u(j)} :=   \pi_{u_*,j}. 
\]
This automatically gives
\[
    |U_0| = |J_0| \leq C_{t,q}(d- |A|),
\]
and we only need to worry about the two following conditions.
First, we need to ensure that the prefix load of $\pi_{u_*,j}$ for agent $u(j)$ still satisfies the bound (\ref{ES4}) for protected prefixes:
\begin{align*}
    &&&\sum_{i \in A_0}\Pp_i(\pi_i^E, \ell_s) + \sum_{j \in J_0}\Pp_{u(j)}(\pi_{u_*,j}^E, \ell_s) \leq |E\cap [\ell_s]|,&\forall \ell_s \in \Ll_S.\tag{$\mathrm{E}_{\Ll_S}$}\label{ES4}
\end{align*}
By \cref{supplier}, this bound is guaranteed to hold if we consider the prefix load of a single agent $u_*$ (condition (\ref{ES3})).
However, swapping $u_*$ with $u(j)$ affects the demand breakpoints, which can change the prefix load drastically if $u_*$ and $u(j)$ are not chosen carefully.
So, condition (\ref{ES4}) is not guaranteed to hold if we just give profiles $\pi_{u_*,j}$ to arbitrary $u(j)$ for every $j \in J_0$.

Second, when we considered copies $J_0$ of a single agent $u_*$, condition (\ref{S3.5}) held automatically, as all $\pi_{u_*,j}$ originate from different bundles of $u_*$'s MMS partition:
\begin{align*}
    &&&\sum_{j\in J_0}\Pp_{u_*}(\pi_{u_*,j}^S, \ell) \leq  |S\cap [\ell]|,&\forall 0 \leq \ell \leq m.\tag{S$*$}\label{S3.5}
\end{align*}
This does not necessarily hold when we give profiles $\pi_{u_*,j}$ to $u(j)$, as agents $u(j)$ for $j \in J_0$ most likely will have different MMS partitions.
So, we still need to ensure that we can simultaneously satisfy the $S$-parts of profiles $\pi_{u_*,j}$ for $j \in J_0$ when treated as profiles of different recipient agents $u(j)$, which is exactly the original condition (\ref{S2}):
\begin{align*}
    &&&\sum_{j\in J_0}\Pp_{u(j)}(\pi_{u_*,j}^S, \ell) \leq  |S\cap [\ell]|,&\forall 0 \leq \ell \leq m.\tag{S}\label{S4}
\end{align*}

Finally, $U_0$ must contain exactly $|J_0|$ different eligible agents that would get a supplier profile, so that $|A_0| + |U_0| = |A_0| + |J_0| = d$.
The following lemma establishes how to choose a supplier $u_* \in U$, a set of $|J_0|$ recipient agents $U_0\subseteq U$, and how to assign supplier profiles to different recipients, so that we can guarantee that conditions (\ref{ES4}) and (\ref{S4}) hold.

\begin{lemma}[Supplier and recipients choice]\label{recipients}
    Let $\Ll_S\subseteq \{0, 1, \ldots, m\}$ be the set of $t$ protected prefixes.
    There exists a set $U_0$ of size at least
    \[
        |U_0|\geq \left\lceil\frac{|U|}{(t + |S| + 1)^{q^2}}\right\rceil
    \]
    with the following property.
    Let $u_* \in U_0$ be arbitrary, and let $J_0\subseteq [d]$ be any set of size $|J_0|\leq |U_0|$.
    Consider MMS profiles $\{\pi_{u_*,j} : j \in J_0\}$ of $u_*$, and assign them arbitrarily to $|J_0|$ distinct agents of $U_0$, where agent let $u(j)\in U_0$ receives $\pi_{u_*,j}$.
    Then
    \begin{align*}
    &&& \sum_{j \in J_0}\Pp_{u(j)}(\pi_{u_*,j}^E, \ell_s) = \sum_{j \in J_0}\Pp_{u_*}(\pi_{u_*,j}^E, \ell_s),&\forall \ell_s\in \Ll_S,\\
    &&&\sum_{j \in J_0}\Pp_{u(j)}(\pi_{u_*,j}^S, \ell) \leq |S\cap [\ell]|,&\forall 0 \leq \ell \leq m.
    \end{align*}
\end{lemma}
\begin{proof}    
    For each agent $u \in U$ and each possible head value $\alpha \in Q$, we record
    \begin{enumerate}
        \item whether $m_u(\alpha) \leq \ell_s$, for each protected prefix $\ell_s \in \Ll_S$, maintaining a binary vector
        \[
        \One\{m_u(\alpha)\leq \ell_1\}, \ldots, \One\{m_u(\alpha) \leq \ell_t\};
        \]
        \item which items of $S$ can satisfy demand $\alpha$ for $u$, i.e set
        \[
            S\cap [m_u(\alpha)].
        \]
    \end{enumerate}
    It is easy to see that for every fixed $\alpha \in Q$, there are $t + |S| + 1$ possible combined records (w.r.t $\Ll_S$ and $S$).
    Indeed, take prefixes $\{0, 1, \ldots, m\}$ and keep increasing $m_u(\alpha)$, starting from $0$.
    A record will change only when $m_u(\alpha)$ crosses one of the protected prefixes $\ell_s \in \Ll_S$, or when it crosses one of the items in $S$.
    Hence, there can be at most $t + |S|$ changes in the combined record, and $t + |S| + 1$ total possible records, for a fixed $\alpha \in Q$.
    There are $|Q|\leq q^2$ possible values of $\alpha$, so in total there are at most $(t + |S| + 1)^{q^2}$ possible records overall.
    
    We partition all agents $U$ into groups, two agents are in the same group if their records above are exactly the same.
    Now, when it comes to satisfying just constraints on protected prefixes $\Ll_S$ and on prefix loads for $S$, all agents within the same group look identical from our perspective.
    There are at most $(t + |S| + 1)^{q^2}$ different groups, hence taking $U_0$ as the largest such group gives
    \[
        |U_0|\geq \left\lceil\frac{|U|}{(t + |S| + 1)^{q^2}}\right\rceil.
    \]
    For completeness, we show rigorously that all agents within group $U_0$ can indeed be replaced with copies of just one arbitrary group representative, i.e supplier, and vice-versa.
    
    Pick and arbitrary $u_* \in U_0$, any $J_0\subseteq [d]$ satisfying $|J_0|\leq |U_0|$, and assign each of the profiles $\pi_{u_*,j}$ to arbitrary distinct agents $u(j)$ for $j \in J_0$.
    Since $u_*$ and $u(j)$ belong to the same group, for every $\alpha \in Q$ we have
    \[
        m_{u(j)}(\alpha) \leq \ell_s \iff m_{u_*}(\alpha) \leq \ell_s,\qquad\qquad \forall\ell_s \in \Ll_S.
    \]
    It follows that, for every protected prefix $\ell_s \in \Ll_S$,
    \begin{multline*}
        \sum_{j \in J_0}\Pp_{u(j)}(\pi_{u_*,j}^E, \ell_s) = \sum_{j \in J_0}\sum_{\alpha\in \pi_{u_*,j}^E}\One\{m_{u(j)}(\alpha)\leq \ell_s\} \\= \sum_{j \in J_0}\sum_{\alpha\in \pi_{u_*,j}^E}\One\{m_{u_*}(\alpha)\leq \ell_s\} = \sum_{j \in J_0}\Pp_{u_*}(\pi_{u_*,j}^E, \ell_s).
    \end{multline*}

    Next, consider some $\alpha \in Q$ and items $S\cap [m_{u(j)}(\alpha)]$.
    Every demand in $\pi_{u_*,j}^S$ originates from a distinct item $e \in B_{u_*,j}\cap S$.
    If $v_{u_*}(e) \geq \alpha$, then $e \in S\cap [m_{u_*}(\alpha)]$ and hence $e \in S\cap [m_{u(j)}(\alpha)]$.

    Assign these distinct items $e \in B_{u_*,j}\cap S$ to corresponding recipients $u(j)$, then all profiles $\pi_{u(j)}^S = \pi_{u_*,j}^S$ are simultaneously satisfied for all agents $u(j)$, $j \in J_0$, by distinct items of $S$.
    Then, every demand counted in $\Pp_{u(j)}(\pi_{u_*,j}^S, \ell)$ must receive a distinct item from $S\cap [\ell]$.
    It follows that
    \[
        \sum_{j \in J_0}\Pp_{u(j)}(\pi_{u_*,j}^S, \ell) \leq |S\cap [\ell]|,\qquad\qquad \forall 0 \leq \ell \leq m.
    \]
\end{proof}
\begin{remark}
    The denominator in \cref{recipients} can be improved to $O((t|S|)^q)$.
    In order to achieve this, instead of classifying just $u \in U$, we classify pairs of agents $u \in U$ and their MMS profiles $\pi_{u,j}$ for $j \in [d]$.
    Two pairs are in the same group if and only if both of their $E$ and $S$ profile parts are the same, their profile $E$-prefix loads are the same, and their $S$-demands can be satisfied by the same items of $S$.
    This splits $U$ into a roughly $G = O((t|S|)^q)$ groups.
    Within each group, reserve up to $|U|/G$ agents, and choose some non-reserved supplier $u_*$.
    For every $j \in J_0$, take the group of $(u_*, j)$ and pick an arbitrary agent within this group as $u(j)$.
    With this construction, the bound on $n_{\eps,q}$ in \cref{savedconst} and \cref{appendix:constant} becomes $2^{O(q^5/\eps)\log q}$ instead of $2^{O(q^6/\eps)\log q}$.
    We do not provide a rigorous proof of this improvement.
\end{remark}

It follows that for sufficiently large number of agents $n$, in the small tail case we can always choose a recipient set $U_0\subseteq U$ and a supplier $u_* \in U_0$ so that all $J_0$ copies of the supplier that we get from \cref{supplier} can be assigned to distinct agents of $U_0$, whole preserving constraints (\ref{ES4}), (\ref{U2}) and (\ref{S4}) with corresponding set $A_0\subseteq A$ from the single supplier theorem.
\begin{corollary}\label{savedconst}
    For every fixed $\eps \in (0, 1)$ and integer $q \geq 3$, there exists $n_{\eps,q}$ such that the following holds for every $n \geq n_{\eps,q}$.
    For $\gamma = \min(\eps, 1/(2q))$ and $\lambda = (1 - 1/(2q))^{-1}$ suppose that agents $N$ satisfy
    \[
        \oT := \frac{1}{n}\sum_{i \in N}T_i < \frac{\rho_{q,n}}{\gamma},
    \]
    where 
    \[
        \rho_{q,n} := 8+ 4 \log_\lambda\big(q^3d^2(d + 2)n\big)+ 8q\log \big((d + 2)n\big).
    \]
    Let $A\subseteq V$ be the assigned agents from \cref{Erounding}, each receiving some profile $\pi_i$, $i \in A$.
    If $|A| < d$, there exists a choice of $U_0\subseteq U$ with the following property.
    For any $u_* \in U_0$, consider its MMS profiles $\{\pi_{u_*,j} : j \in [d]\}$.
    We can choose $A_0\subseteq A$ and $J_0\subseteq [d]$ with $|J_0| \leq |U_0|$ and $|A_0| + |J_0| = d$,
    and for every $j \in J_0$ assign $\pi_{u_*,j}$ to distinct $u(j) \in U_0$, so that
    \[
    \sum_{i \in A_0}\Pp_i(\pi_i, \ell) + \sum_{j\in J_0}\Pp_{u(j)}(\pi_{u_*,j}, \ell) \leq \ell,\qquad \forall 0 \leq \ell \leq m.
    \]
\end{corollary}
\begin{proof}
    We will first bound the number $d - |A|$ of missing agents  after rounding.
    By \cref{Erounding}, taking $\rho_A = 8 + 16q\log n$ and using $|V|\geq (1 - \eps/4)n \geq 3n/4$,
    \[
        d - |A| \leq |S| + \frac{q(q -1)}{|V|}\sum_{i \in V}T_i + \rho_A \leq  |S| + 2q(q - 1)\oT  + \rho_A \leq |S| + \frac{2q(q - 1)}{\gamma}\rho_{q,n} + \rho_A. 
    \]
    The number of protected prefixes $|\Ll_S|$ is $t \leq 2q^2$, and at least $1$, so we can pick a global constant 
    \[
        C_q := \max\{1, \max_{1 \leq t \leq 2q^2}C_{t,q},\},
    \]
    where each $C_{t,q}$ is from \cref{supplier}.
    Choose integers
    \[
        C_1 := 1 + \lceil 12q^3 C_q\rceil\qquad \text{and}\qquad r_1 := 1 + \left\lceil \left(\frac{2q(q - 1)}{\gamma} + 2\right)\rho_{q,n}\right\rceil.
    \]
    Using $\rho_A\leq 2\rho_{q,n}$ and the parameters above, we get
    \[
        d - |A| \leq |S| + \frac{2q(q - 1)}{\gamma}\rho_{q,n} + \rho_A \leq |S| + r_1 - 1.
    \]
    By \cref{saved}, $|S|\geq r_1$, hence $d - |A|\leq 2|S|$ and
    \[
        \eta \geq \frac{C_1 - 1}{2q^2}|S|,\qquad\implies\qquad C_q(d - |A|) \leq 2C_q|S| \leq \frac{C_1 - 1}{6q^3}|S|\leq \frac{\eta}{3q}.
    \]
    
    We will now find large enough set of recipients $U_0$ to accommodate any $|J_0|\leq C_q(d - |A|)$ copies supplied by \cref{supplier}.
    Since $C_1 = O_q(1)$ and $r_1 = O_{\eps,q}(\log n)$,we have $|S| = O_{\eps,q}(\log n)$, thus $d - |A| = O_{\eps,q}(\log n)$.
    Hence, there exists constant $C_* = C_*(\eps, q)\geq 1$ such that 
    \[
        t + |S| + 1 \leq C_*\log n,\qquad\text{and}\qquad C_q(d - |A|)\leq C_*\log n.
    \]
    Choose $n_{\eps,q}\geq 16/\eps$ sufficiently large so that for all $n \geq n_{\eps,q}$ it holds
    \[
        \frac{n}{(\log n)^{q^2 + 1}}\geq \frac{16C_*^{q^2 + 1}}{\eps}.
    \]

    Next, by definition, $|U| = \lfloor \eps n/8\rfloor \geq \eps n/16$.
    Applying \cref{recipients}, we get $U_0\subseteq U$ of size
    \[
        |U_0| \geq \frac{|U|}{(t + |S| + 1)^{q^2}} \geq \frac{\eps n}{16(C_*\log n)^{q^2}} \geq C_*\log n \geq C_q(d - |A|).
    \]
    Fix an arbitrary supplier $u_* \in U_0$.
    Apply \cref{supplier} to $A$ and $u_*$, obtaining $A_0\subseteq A$ and $J_0\subseteq [d]$ such that
    \[
        |A_0| + |J_0| = d,\qquad\text{and}\qquad |J_0|\leq C_q(d - |A|)\leq \min\{|U_0|, \eta/(3q)\},
    \]
    where for every protected prefix $\ell_s \in \Ll_S$ it holds
    \[
        \sum_{i \in A_0}\Pp_i(\pi_i^E, \ell_s) + \sum_{j \in J_0}\Pp_{u_*}(\pi_{u_*,j}^E, \ell_s) \leq |E\cap [\ell_s]|.
    \]
    Since $|J_0|\leq |U_0|$, we can assign each supplier copy $\pi_{u_*,j}$ for $j \in J_0$ to a distinct recipient $u(j) \in U_0$, setting $\pi_{u(j)} = \pi_{u_*,j}$.
    As shown in \cref{recipients}, replacing the supplier copies with $u(j)$-s preserves all $E$-loads, as well as guarantees feasibility at $S$-prefixes.
    So,
    \begin{align*}
    &&& \sum_{i \in A_0}\Pp_i(\pi_i^E, \ell_s) + \sum_{j \in J_0}\Pp_{u(j)}(\pi_{u_*,j}^E, \ell_s) \leq |E\cap [\ell_s]|,&\forall \ell_s\in \Ll_S,\\
    &&&\sum_{j \in J_0}\Pp_{u(j)}(\pi_{u_*,j}^S, \ell) \leq |S\cap [\ell]|,&\forall 0 \leq \ell \leq m.
    \end{align*}
    Let $U_0(J_0) := \{u(j) : j \in J_0\}$.
    Since $|U_0(J_0)|= |J_0|\leq \eta/(3q)$, applying \cref{Ebound} with $A_0$ and $U_0(J_0)$ we get that for every prefix $0\leq\ell \leq m$, 
    \[
        \sum_{i \in A_0}\Pp_i(\pi_i^E, \ell) + \sum_{j \in J_0}\Pp_{u(j)}(\pi_{u_*,j}^E, \ell_s) \leq |E\cap [\ell]|. 
    \]
    Since for every $i \in A_0$ we have $\pi_i^E = \pi_i$, and for every $j \in J_0$ we have $\pi_{u_*,j}^E\uplus \pi_{u_*,j}^S = \pi_{u_*,j}$, combining the individual bounds on $E$-loads and $S$-loads we get
    \[
        \sum_{i \in A_0}\Pp_i(\pi_i, \ell) + \sum_{j \in J_0}\Pp_{u(j)}(\pi_{u_*,j}, \ell) \leq |E\cap [\ell]| + |S\cap [\ell]| = \ell
    \]
    for every $0 \leq \ell \leq m$, as desired.
    Note that $|A_0| + |U(J_0)| = |A_0| + |J_0| = d$ by construction.
\end{proof}

\subsection{Common tail completion}

For sets $U_0 \subseteq U$, $J_0\subseteq [d]$ and supplier $u_* \in U_0$ from \cref{savedconst}, denote by $U_0(J_0)$ the chosen collection of agents $u(j) \in U_0$, such that for $j \in J_0$ agent $u(j)$ is assigned a supplier profile $\pi_{u_*,j}$.
\cref{savedconst} gives us a set of $d$ agents $I = A_0 + U(J_0)$ and a collection of head profiles $\Pi_I = \{\pi_i : i \in A_0\} + \{\pi_{u_*,j} : j \in J_0\}$, such that
\[
    \sum_{i \in A_0}\Pp_i(\pi_i, \ell) + \sum_{j\in J_0}\Pp_{u(j)}(\pi_{u_*,j}, \ell) \leq \ell,\qquad \forall 0 \leq \ell \leq m.
\]
So, by \cref{loadequiv}, we can simultaneously satisfy all demands in $\Pi_I$ via distinct items of $M$.
If for every agent $i \in I$ we have $\Delta(\pi_i) = 0$, this would automatically produce an allocation where each $i$ receives a bundle of value at least $1 - 1/q$.

This may not be the case: our construction guarantees only for agents $i \in A_0$ that $\Delta(\pi_i) = 0$, while the copied supplier profiles $\pi_{u_*,j}$ may still have positive deficit.
Recall that in the large tail case, the rounding of \cref{rounding} produced a collection of integral profiles that had enough slackness, for every prefix $\ell$, to allow additional tail load of 
\[
    \sum_{i \in I}\frac{\Delta(\pi_i)}{T_i}(\ell - h_i)_+.
\]
Even if for some $\Delta(\pi_i) > 0$ the tail value $T_i$ is very small, in the large tail case the number of such agents is limited (since the average tail is large), hence the total fractional tail load is bounded.
However, when all tails are small with positive deficits, terms $\frac{\Delta(\pi_i)}{T_i}$ and the fractional tail load of profiles $\Pi_I$ may be unbounded.
To make things worse, at first glance it seems as if in our construction of $I = A_0 + U_0(J_0)$ and $\Pi_I$, we did not control the fractional tail load at all.

In reality, the choice of the original set $U$, as well as the construction of $U_0(J_0)$, gives us some tail guarantees.
In particular, it ensures that all agents in  $I = A_0 + U_0(J_0)$ have a common tail.
\begin{corollary}\label{commonhead}
    All demands in $\Pi_I = \{\pi_i : i \in A_0\} + \{\pi_{u_*,j} : j \in J_0\}$ can be simultaneously satisfied by distinct items in $\{1, \ldots, h_*\}$, where $h_*$ is the head threshold of the supplier $u_*$.
    Furthermore, if $T_*$ is the total tail value of $u_*$, it holds
    \[
        \sum_{i \in A_0}\Delta(\pi_i) + \sum_{j \in J_0}\Delta(\pi_{u_*,j}) \leq T_*.
    \]
\end{corollary}
\begin{proof}
    The bound on the total deficit follows by recalling that $\Delta(\pi_i) = 0 $ for all $i \in A_0$, and then applying \cref{fracload} applied to a subset of MMS profiles of the supplier $u_*$.   

    Now, consider the proof of \cref{savedconst}.
    After applying \cref{supplier} to agent $u_* \in U_0$ and $A$, the resulting copies $J_0\subseteq [d]$ of $u_*$ satisfied the property (\ref{ES5}):
    \begin{align*}
    &&&\sum_{i \in A_0}\Pp_i(\pi_i^E, \ell_s) + \sum_{j \in J_0}\Pp_{u_*}(\pi_{u_*,j}^E, \ell_s) \leq |E\cap [\ell_s]|,&\forall \ell_s \in \Ll_S;\tag{$\mathrm{E}_{\Ll_S}*$}\label{ES5}
    \end{align*}
    Furthermore, \cref{supplier} guarantees $|J_0|\leq C_q(d - |A|)$, while \cref{savedconst} chose $C_q, C_1, r_1$ so that $C_q(d - |A|)\leq \eta/(3q)$.
    Hence, we can apply \cref{Ebound} to the set of agents $A_0$ as well as a \textit{supplier set} $U_*$ which consists of $|J_0|$ copies of $u_*$ with profiles $\pi_{u_*,j}$ for $j \in J_0$, with the exact same construction of all partitions and choice of parameters.
    Indeed, the proof of \cref{Ebound} applies equally to copies of the exactly same agent with different profiles, as long as the corresponding bounds apply.
    Therefore, for every prefix $0 \leq \ell \leq m$ we have
    \begin{align*}
    &&&\sum_{i \in A_0}\Pp_i(\pi_i^E, \ell) + \sum_{j \in J_0}\Pp_{u_*}(\pi_{u_*,j}^E, \ell) \leq |E\cap [\ell]|;&\tag{E$*$}\label{E1c}\\
    &&&\sum_{j\in J_0}\Pp_{u_*}(\pi_{u_*,j}^S, \ell) \leq  |S\cap [\ell]|.&\tag{S$*$}\label{S1c}
    \end{align*}
    Constraints (\ref{S1c}) hold automatically as all $\pi_{u_*,j}$ are MMS profiles of agent $u_*$.

    Let $h_{*}:=h_{u_*}$.
    Recall $U$ was selected as the $\lfloor \eps n/8\rfloor$ agents of $N - W = U + V$ with largest head thresholds $h_u$.
    Then, as $u_* \in U$, for every $i \in A_0$ we have $h_i \leq h_*$, which implies that any demand $\alpha \in Q$ satisfies $m_i(\alpha) \leq h_*$.
    As every demand of $u_*$ has breakpoint at most $h_*$, 
    \[
        \sum_{i \in A_0}|\pi_i^E| + \sum_{j \in J_0}|\pi_{u_*,j}^E|= \sum_{i \in A_0}\Pp_i(\pi_i^E, h_*) + \sum_{j \in J_0}\Pp_{u_*}(\pi_{u_*,j}^E, h_*)   \leq |E\cap[h_*]|.
    \]
    as well as
    \[
        \sum_{j \in J_0}|\pi_{u_*,j}^S| =  \sum_{j \in J_0}\Pp_{u_*}(\pi_{u_*,j}^S, h_*) \leq |S\cap [h_*]|.
    \]
    Hence, the total number of demands in $\Pi_I$ is
    \[
       K:= \sum_{i \in A_0}|\pi_i| + \sum_{j \in J_0}|\pi_{u_*,j}| \leq |E\cap[h_*]| + |S\cap[h_*]| = h_*.
    \]
    When assigning profiles $\pi_{u_*,j}$ to recipients $u(j)$ for $j \in J_0$, the total number of demands does not change, and by \cref{savedconst} the collection $\Pi_I$ still remains feasible with agents $u(j)$, $j \in J_0$:
    \[
    \sum_{i \in A_0}\Pp_i(\pi_i, \ell) + \sum_{j\in J_0}\Pp_{u(j)}(\pi_{u_*,j}, \ell) \leq \ell,\qquad \forall 0 \leq \ell \leq m.
    \]
    As a result, applying \cref{loadequiv} with $K\leq h_*$, we get that all profiles $\Pi_I$ can be simultaneously satisfied using distinct items in $\{1, \ldots, h_*\}$.
\end{proof}

Given \cref{commonhead}, our next step is to complete head profiles $\Pi_I = \{\pi_i\}_{i \in I}$ of agents $I = A_0 + U_0(J_0)$ with appropriate tail profiles $\Psi_I = \{\psi_i\}_{i \in I}$, so that for every $i \in I$ we have $v(\pi_i) + v(\psi_i) \geq 1 - 2/q$.
If $T_* = 0$, \cref{commonhead} guarantees that all deficits are $0$, and we are done.
Suppose that $T_* > 0$.
We will first try to do something similar to \cref{tailcomplete}.

In the large tail case, agent $i$ with profile $\pi_i$ tried to compensate for his deficit $\Delta(\pi_i)$ by taking a $\tau_i := \tau_i(\pi_i)$-fraction of every tail item, where
\[
        \tau_i(\pi_i) := \begin{cases}
            \Delta(\pi_i)/T_i,& \pi\neq \bot\text{ and }T_i > 0;\\
            0,& \text{otherwise}.
        \end{cases}
\]
So, we construct a tail profile by taking the last item in each tail segment of length roughly $1/\tau_i$.
This costs at most the value of one tail item, which is at most $1/q$.

We would like to repeat this construction on the common tail $\{h_* + 1, \ldots, m\}$.
However, an item after $h_*$ may still be a head item for some agent $i \in I$, hence have value greater than $1/q$.
So, using the old construction directly risks losing more than $1/q$ value total.

Note that from the perspective of tail profiles, all items of value more than $1/q$ are essentially the same, as any tail demand is at most $1/q$.
This gives us the following idea.
In order to keep loss at $1/q$, we will cap the values of all head items to $1/q$, just for the sake of the tail completion step.
Concretely, for every agent $i \in I$ and item $e\in M$, define valuation
\[
    \ov_i(e) := \min\{v_i(e), 1/q\},
\]
and extend it additively.
We will construct tail demands based off the item values according to capped valuations $\ov_i$.
Note that all the definitions of head profiles, breakpoints and prefix loads still refer to original uncapped valuations $v_i$.

Observe that for every $i \in I$ it holds
\[
    \ov_i(M) = \sum_{e = 1}^m\min\{v_i(e), 1/q\} = \frac{h_i}{q} + T_i.
\]
This is the total capped value of items $M$ for agent $i$ that can in theory be used for his tail profile, however, in our case we can only use items $\{h_* + 1, \ldots, m\}$ of the common tail.
Depending on $h_i$ and $h_*$, common tail may contain some head items of agent $i$, or exclude some tail items of agent $i$.
The value $h_i/q + T_i$ in a sense captures the available ``tail completion budget'', and allows us to compare between the two cases.

At the same time, for every agent $i \in I$ the first $h_*$ items have capped value at most $h_*/q$.
So, the the tail completion budget of $i$, i.e the capped value of the common tail, is at least
\[
    \sum_{e > h_*}\ov_i(e) = \ov_i(M) - \ov_i(\{1, \ldots, h_*\}) \geq \frac{h_i}{q} + T_i - \frac{h_*}{q}.
\]
By \cref{commonhead}, the total deficit of agents $i \in I$ satisfies $\sum_{i \in I}\Delta(\pi_i) \leq T_*$.
Hence, in order to guarantee that every agent $i \in I$ has large enough tail completion budget to cover the whole deficit, it suffices to ensure that he values the common tail $\{h_* + 1, \ldots, m\}$ at least $T_*$.
In particular, this is guaranteed when
\[
    \frac{h_i}{q} + T_i \geq \frac{h_*}{q} + T_*.
\]

This is exactly the condition that defines our choice of supplier $u_* \in U_0$ in \cref{savedconst}.
Since both \cref{savedconst} and \cref{commonhead} apply to any choice of supplier $u_* \in U_0$, we take
\[
    u_* = \argmin_{u \in U_0}\left(\frac{h_u}{q} + T_u\right).
\]
Since all $i \in A_0$ have deficit $\Delta(\pi_i) = 0$, the only $i \in I$ that have positive deficit come from $U_0(J_0)\subseteq U_0$.
So, choosing $u_*$ as above, we guarantee that for any agent $i \in I$ with $\Delta(\pi_i) > 0$,
\[
    \ov_i(\{h_* + 1, \ldots, m\}) \geq  \frac{h_i}{q} + T_i - \frac{h_*}{q} \geq T_*.
\]
It follows that, in order to cover the deficit $\Delta(\pi_i) > 0$ of agent $i \in I$, instead of picking $\tau_i$-fraction of every common tail item, he can pick $\tau_{i,*}$-fraction, where
\[
    \tau_{i,*} := \frac{\Delta(\pi_i)}{T_*}.
\]
Indeed, the derivations above imply
\[
    \sum_{e > h_*}\tau_{i,*}\cdot \ov_i(e)\geq \tau_{i,*}\cdot T_* = \Delta(\pi_i).
\]
Furthermore, \cref{commonhead} guarantees us that
\[
    \sum_{i \in I}\tau_{i,*} = \sum_{i \in I}\frac{\Delta(\pi_i)}{T_*} \leq 1,
\]
so all fractional choices of all agents $i \in I$ fit within the common tail.
This is essentially the classical setting of faithful rounding (\cite{BEF22BoBW, EG26-2}), where one would assign every tail item $e > h_*$ to agent $i \in I$ with probability $\tau_{i,*}$, and get an integral allocation while losing up to the value of one item, in our case at most $1/q$.

To keep the proof self-contained, instead of relying on faithful rounding we will prove the common tail version of earlier \cref{tailcomplete}, which gives us exactly the same guarantees.
\begin{lemma}[\cref{tailcomplete} for a common tail]\label{commontailcomplete}
    Let $I\subseteq N$, let $\pi_i$ be some head profile of agent $i \in I$, and let $\Pi_I := \{\pi_i\}_{i \in I}$.
    If there exist integer $0 \leq h_* \leq m$ and value $T_* \geq 0$ such that
    \begin{enumerate}
        \item all demands in $\Pi_I$ can be simultaneously satisfied by distinct items in $\{1, \ldots, h_*\}$;
        \item for every $i \in I$ with $\Delta(\pi_i) > 0$ it holds $h_i/q + T_i \geq h_*/q + T_*$;
        \item $\sum_{i \in I}\Delta(\pi_i) \leq T_*$;
    \end{enumerate}
    there exists a collection $\Psi_I := \{\psi_i\}_{i \in I}$ of tail profiles, where agent $i$ is assigned $\psi_i$, such that
    \[
        \forall 0 \leq\ell \leq m:\; \Pp(\Pi_I, \ell) + \Pp(\Psi_I, \ell) \leq \ell
    \]
    and $v(\pi_i) + v(\psi_i) \geq 1 - 2/q$ for every $i \in I$.
\end{lemma}
\begin{proof}
    If $T_* = 0$, we can take $\psi_i = \varnothing$ for every $i \in I$ and we are done, so suppose $T_* > 0$.
    Consider $i \in I$, and denote $\tau_{i,*} := \Delta(\pi_i)/T_*$.
    If $\tau_{i,*} = 0$, also take $\psi_i = \varnothing$, so assume $\tau_{i,*} > 0$.

    Note that $T_* > 0$ implies $h_* < m$.
    We construct the tail profile $\psi_i$ exactly the same way as in \cref{tailcomplete}, but with respect to common tail and $\tau_{i,*}$.
    That is, we break the common tail $\{h_* + 1, \ldots, m\}$ into segments of length roughly $1/\tau_{i,*}$, and in each segment add the \textit{capped} value $\ov_i$ of the last item to $\psi_i$.
    That is, for every $t = 1, \ldots, \lfloor \tau_{i,*}\cdot (m - h_*)\rfloor$, if add $\ov_i(e_{i,t})$ to $\psi_i$, where
    \[
        e_{i,t} = h_* + \left\lceil \frac{t}{\tau_{i,*}}\right\rceil.
    \]
    As in \cref{tailcomplete}, we can represent this construction equivalently as follows.
    For $1 \leq k \leq m - h_*$, let $\beta_{i,k} := \ov_i(h_* + k)$.
    Then, for every $k \in [m - h_*]$, we add exactly
    \[
        \lfloor\tau_{i,*}\cdot k \rfloor - \lfloor \tau_{i,*}\cdot (k - 1)\rfloor
    \]
    copies of $\beta_{i,k}$ to the profile $\psi_i$.
    This way, the first $k$ common tail items contain exactly $\lfloor \tau_{i,*}\cdot k\rfloor$ selected items from $\psi_i$.
    Therefore,
    \begin{multline*}
        v(\psi_i) = \sum_{k = 1}^{m - h_*}\beta_{i,k}\cdot \big(\lfloor \tau_{i,*}\cdot k\rfloor - \lfloor\tau_{i,*}\cdot (k - 1)\rfloor\big) = \sum_{k = 1}^{m - h_*}\lfloor \tau_{i,*}\cdot k\rfloor\cdot \big( \beta_{i,k}- \beta_{i,k + 1}\big)\\
        \geq \sum_{k = 1}^{m - h_*}(\tau_{i,*}k - 1)\cdot \big( \beta_{i,k}- \beta_{i,k + 1}\big) = \tau_{i,*}\sum_{k = 1}^{m - h_*}\beta_{i,k} - \beta_{i,1}.
    \end{multline*}
    As we derived earlier, under the condition $h_i /q + T_i \geq h_*/q + T_*$, we have
    \[
        \tau_{i,*}\sum_{k = 1}^{m - h_*}\beta_{i,k} = \tau_{i,*}\cdot \ov_i(\{h_* + 1,\ldots, m\}) \geq \tau_{i,*}\cdot T_* = \Delta(\pi_i),
    \]
    therefore, using $\beta_{i,1} = \ov_i(h_* + 1) \leq 1/q$,
    \[
        v(\psi_i) \geq \Delta(\pi_i) - \beta_{i,1} \geq \Delta(\pi_i) - 1/q.
    \]
    Since $\Delta(\pi_i) = (1 - 1/q - v(\pi_i))_+$, we get $v(\pi_i) + v(\psi_i) \geq 1 - 2/q$.

    It remains to bound the additional prefix load generated by $\Psi_I = \{\psi_i\}_{i \in I}$.
    As in \cref{tailcomplete}, for every $0 \leq \ell \leq h_*$ we have $\Pp_i(\psi_i, \ell) = 0$, for every agent $i \in I$.
    When $\ell > h_*$, the number of selected items from $\psi_i$ within the first $k = \ell - h_*$ common tail items is exactly $\lfloor\tau_{i,*}\cdot k\rfloor \leq \tau_{i,*}\cdot k$.
    In addition, since $\beta_{i,k} = \ov_i(h_* + k) \leq v_i(h_* + k)$, every selected tail demand that is nonzero satisfies $m_i(\beta_{i,k}) \geq h_* + k$.
    Hence, only demands originating from items at most $\ell$ can contribute to the prefix load at $\ell$, and the additional load generated by tail profiles $\Psi_I$ at prefix $\ell$ is
    \[
        \Pp(\Psi_I, \ell)  = \sum_{i \in I}\Pp_i(\psi_i, \ell) \leq \sum_{i \in I}\tau_{i,*}\cdot (\ell - h_*)_+ =(\ell - h_*)_+\sum_{i \in I}\frac{\Delta(\pi_i)}{T_*}\leq (\ell - h_*)_+.
    \]
    Since all demands in $\Pi_I$ can be simultaneously satisfied by distinct items in $\{1, \ldots, h_*\}$, it holds 
    \[
        \Pp(\Pi_I, \ell) \leq \min\{\ell, h_*\}.
    \]
    for every $0 \leq \ell \leq m$. Consequently,
    \[
        \Pp(\Pi_I, \ell) + \Pp(\Psi_I, \ell) \leq \min\{\ell, h_*\} + (\ell - h_*)_+ = \ell
    \]
    for every $0 \leq \ell \leq m$, as desired.
\end{proof}

\cref{commontailcomplete} gives us the particular choice of supplier $u_* \in U_0$ needed to complete the tail profiles: take $u_* \in U_0$ so that for every agent $i \in I = A_0 \cup U_0(J_0)$ with $\Delta(\pi_i) > 0$ it holds $h_i/q + T_i \geq h_*/q + T_*$.
Since all agents $i \in A_0$ have $\Delta(\pi_i) = 0$, it suffices to choose
\[
    u_* = \argmin_{u \in U_0}\left(\frac{h_u}{q} + T_u\right).
\]
All previous conclusions of \cref{supplier}, \cref{savedconst} and \cref{commonhead} still apply.
We can now give a full proof of the small-tail case.

\begin{theorem}[Small tail]\label{smalltail}
    For every fixed $\eps\in (0, 1)$ and integer $q \geq 3$, there exists $n_{\eps,q}$ such that for every $n \geq n(\eps, q)$ the following holds.
    Let $\gamma := \min(\eps, 1/(2q))$, $\lambda = \frac{2q}{2q-1}$ and 
    \[
        \rho_{q,n} := 8+ 4 \log_\lambda\big(q^3d^2(d + 2)n\big)+ 8q\log \big((d + 2)n\big).
    \]
    If 
    \[
        \oT := \frac{1}{n}\sum_{i \in N}T_i < \frac{\rho_{q,n}}{\gamma},
    \]
    there exists $I\subseteq N$ of size $|I|\geq d$ and a collections $\Pi_I = \{\pi_i\}_{i \in I}$ and $\Psi_I=\{\psi_i\}_{i \in I}$ of head and tail profiles respectively, where agent $i \in I$ is assigned $(\pi_i, \psi_i)$, such that for all $i \in I$:
    \[
        v(\pi_i) + v(\psi_i) \geq 1 - 2/q
    \]
    and for every prefix $0 \leq \ell \leq m$ it holds
    \[
        \Pp(\Pi_I, \ell) + \Pp(\Psi_I, \ell) \leq \ell.
    \]
\end{theorem}
\begin{proof}
    Choose $n_{\eps,q}$ as in \cref{savedconst}, and construct partitions $N = U + V + W$ and $M = E + S$ according to \cref{savedconst}, using corresponding parameters.
    Apply \cref{Erounding} to agents $V$ and items $E$, obtaining a set $A\subseteq V$ with assigned profiles $\{\pi_i\}_{i \in A}$.
    These profiles can be simultaneously satisfied by distinct items of $E$, and $\Delta(\pi_i) = 0$ for all $i \in A$.
    If $|A|\geq d$, any subset of $I\subseteq A$ of size $d$ suffices.
    Assign $\psi_i = \varnothing$ to every $i \in I$, and we are done.

    Suppose that $|A|< d$.
    \cref{savedconst} gives $U_0\subseteq U$.
    Choose supplier $u_*\in U_0$ such that
    \[
    u_* = \argmin_{u \in U_0}\left(\frac{h_u}{q} + T_u\right).
    \]
    For this supplier, \cref{savedconst} provides sets $A_0\subseteq A$ and $J_0\subseteq [d]$, together with assignment $u(j)$ of copies of $u_*$ to agents $u(j) \in U_0$, such that $|A_0| + |J_0| = d$ and
    \[
    \sum_{i \in A_0}\Pp_i(\pi_i, \ell) + \sum_{j\in J_0}\Pp_{u(j)}(\pi_{u_*,j}, \ell) \leq \ell,\qquad \forall 0 \leq \ell \leq m.
    \]

    Set $U_0(J_0) := \{u(j) : j \in J_0\}$, and $I = A_0\cup U_0(J_0)$.
    Assign $\pi_{u(j)}:= \pi_{u_*,j}$ for all $j \in J_0$, and write $\Pi_I := \{\pi_i\}_{i \in I}$.
    By \cref{commonhead}, all demands in $\Pi_I$ can be simultaneously satisfied by distinct items in $\{1, \ldots, h_*\}$, where $h_* = h_{u_*}$.
    Moreover, writing $T_* = T_{u_*}$, we have
    \[
        \sum_{i \in I}\Delta(\pi_i) = \sum_{i \in A_0}\Delta(\pi_i) + \sum_{j \in J_0}\Delta(\pi_{u_*,j})\leq T_*.
    \]
    Since $\Delta(\pi_i) = 0$ for all $i \in A_0$, every agent with positive deficit belongs to $U_0(J_0)\subseteq U_0$.
    By the choice of $u_*$, every such agent satisfies 
    \[
        \frac{h_i}{q} + T_i \geq \frac{h_{u_*}}{q} + T_{u_*} = \frac{h_*}{q} + T_*.
    \]
    So, we can apply \cref{commontailcomplete}.
    We obtain a collection $\Psi_I = \{\psi_i\}_{i \in I}$ of tail profiles, such that 
    \[
        \Pp(\Pi_I, \ell) + \Pp(\Psi_I, \ell) \leq \ell,\qquad \forall 0\leq \ell \leq m,
    \]
    and $v(\pi_i) + v(\psi_i) \geq 1 - 2/q$ for every $i \in I$.
    Noticing $|I| =  |A_0| + |J_0| = d$ finishes the proof.
\end{proof}
Combining \cref{smalltail} with \cref{loadequiv} proves \cref{dchoiceproof} in the small tail case.
\section{Concluding thoughts}
\label{sec:corol}

In the previous sections, we proved our main result \cref{thm:largen}, which we restate below.
\begin{theorem}[\cref{thm:largen} restated]\label{dchoiceend}
    For every fixed $\eps \in (0, 1)$ and integer $q \geq 3$, there exists $n(\eps, q)$ such that for every $n \geq n(\eps, q)$ the following holds.
    
    Consider a fair allocation instance with $m$ items $M$ and $n$ agents $N$, where for every $i \in N$ valuation $v_i$ is additive.
    Suppose that every agent $i \in N$ has $\mms(M, v_i, d) = 1$, where $1 \leq d \leq (1 - \eps)n$.
    Then there exists a choice $I = \{i_1, \ldots, i_d\}\subseteq N$ of $|I| = d$ distinct agents and an allocation $(B_{i_1},\ldots, B_{i_d})$ of $M$ to $I$, such that for every $i \in I$ it holds $v_i(B_i) \geq 1 - 2/q$.
\end{theorem}
An immediate corollary of \cref{dchoiceend} is that there always exist  $(1 - \delta)$-MMS allocations when the number of items $m$ is at most $(2 - \eps)n$, for sufficiently large number of agents $n \geq n_{\eps,\delta}$.
\begin{corollary}[\cref{cor:largen} restated]
    For every fixed $\eps, \delta \in (0, 1)$, there exists $n_{\eps,\delta}$ such that for every $n\geq n_{\eps,\delta}$ the following holds.
    Consider a fair allocation instance with $m$ items $M$ and $n$ agents $N$, where fir every $i \in N$ valuation $v_i$ is additive. 
    If $m \leq (2 - \eps)n$, then there exists an allocation $(A_1, \ldots, A_n)$ of $M$ to $N$ such that for every agent $i \in N$,
    \[
        v_i(A_i) \geq (1 - \delta)\cdot \mms(M, v_i, n).
    \]
\end{corollary}
\begin{proof}
    Using standard reductions, without loss of generality we can assume that
    \begin{itemize}
        \item for every agent $i \in N$ and item $e \in M$ it holds $v_i(e) \in [0, 1]$;
        \item for every $i \in N$, $n$-partition $\Bb_i:= (B_{i,1}, \ldots, B_{i,n})$ of $M$ witnessing $\mms(M, v_i, n)$ satisfies
        \[
            \forall j \in [n]:\; v_i(B_{i,j}) = 1;
        \]
        \item items $M = \{e_1, \ldots, e_m\}$ are ordered so that
        for every agent $i \in N$ it simultaneously holds
        \[
            v_i(e_1) \geq v_i(e_2) \geq \ldots \geq v_i(e_m).
        \]
    \end{itemize}
    
For every $i \in N$, consider his MMS partition $\Bb_i = (B_{i,1}, \ldots, B_{i,n})$.
    Observe that if $s_i$ bundles of $\Bb_i$ are singletons, i.e have exactly one positive item, then every other bundle has at least two positive items, so for every $i \in N$,
    \[
        s_i + 2(n - s_i)  \leq m \qquad \implies \qquad s_i\geq 2n - m \geq \eps n.
    \]
    Each item of value $1$ produces a singleton bundle in agent $i$'s MMS partition, thus the first $s_i$ items of $M$ have value $1$ for $i$.
    Let $s_* := \min_{i \in [n]}s_i$, then items $\{1, \ldots, s_*\}$ have value $1$ for every agent $i \in N$.
    If $s_* = n$, we can give a separate item to each agent.
    Otherwise, remove first $s_*$ items from $M$, and set $d := n - s_*$.
    It is clear that for every $i \in N$, after removing the singleton bundles from $\Bb_i$ corresponding to $[s_*]$, the remaining bundles form a $d$-partition of $\{s_* + 1, \ldots, m\}$ of value $1$. As $d \leq (1 - \eps)n$, we can apply \cref{dchoiceend} with agents $N$ and items $\{s_* + 1, \ldots, m\}$, values $\eps$ and $q:= \left\lceil 2/\delta\right\rceil$, and give some $d$ agents bundles of value at least $1 - 2/q \geq 1 - \delta$.
    The remaining $n - d = s_*$ agents each get a single item from $[s_*]$ of value $1$.
\end{proof}

On a separate topic, in \cite{akrami25aug} the authors related the guarantees for the $1$-out-of-$d$ MMS allocations to the number of duplicated items needed for the exact original MMS.
Applying their construction (\cite{akrami25aug}, Lemma 3.10) to our $\mms(d)$ result (\cref{dmms}), we get the following corollary.
\begin{corollary}
    Consider a fair allocation instance with items $M$ and $n$ additive agents $n$ with valuations $\{v_i\}_{i \in N}$.
    It suffices to add one copy of each of at most
    \[
        O\left(\frac{m\log^2n}{n}\right)
    \]
    items to guarantee that, for the new instance with extra items, there exists an allocation that gives every agent $i \in N$ a bundle of value at least $\mms(M, v_i, n)$.
\end{corollary}
In particular, when $m = O(n)$ it suffices to add only $O(\log^2 n)$ copies.
This connection has been previously observed in the work of \cite{EG26-2}, who obtained a similar implication from their $1$-out-of-$d$ MMS result with a bound of $O(m\sqrt{\frac{\log n}{n}})$ on the number of copies.

\subsubsection*{AI Usage Disclosure}

The authors used OpenAI's ChatGPT-6.1 Sol as a research assistant in some parts of the proof.
In particular, it suggested using Dickson's lemma argument to prove the existence of a replacement step of a constant size in the proof of \cref{supplier} (\cref{supplierproof}), as well as Onn's lemma (\cref{onn}) to simplify the computation of the corresponding constant $C_{t,q}$.
In addition, it was used to help the authors refine the construction of the set $S$ of saved items in \cref{sec:saved} (\cref{saved}).
All mathematical arguments, statements, proofs and exposition in the paper were written and verified by the authors, who take full responsibility for the content of this work.



\bibliographystyle{alpha}
\bibliography{main}

\appendix
\crefalias{section}{appendix}

\section{Proof of the unbounded head rounding theorem}
\label{appendix:etarounding}

\begin{theorem}[\cref{etarounding} restated]\label{etaroundingproof}
    Let $\Pi$ be a shared set of head profiles, such that $|\pi|\leq h$ for some integer $h \geq 1$.
    For every $i \in N$, let $(x_i, \Pi)$ be a fractional profile, denote $x = \{x_i\}_{i \in N}$.

    There exists a collection $y = \{y_i\}_{i \in N}$ of integer head profiles $(y_i, \Pi)$, such that for every $i, \pi$ we have $y_i(\pi) \in \{0, 1\}$ and $y_i(\pi) = 1$ for $\pi \neq \bot$ only if $x_i(\pi) > 0$, and
    \[\begin{aligned}
        & && \Hh(y, \Pi, \ell) \leq \Hh(x, \Pi, \ell),  & \forall 0 \leq \ell \leq m,\\
        & && \tau(y, \Pi) \leq\tau(x, \Pi),\\
        & && \Ii(y, \Pi) \leq \Ii(x, \Pi),\\
        &  && \Ii(y, \Pi)\geq \Ii(x, \Pi) - \rho(h, X), 
    \end{aligned}\]
    where $X = |\{(i, \pi) :\; 0 < x_i(\pi) < 1\}|$ and
    \[
        \rho(h, X) := 8 + 4\log X\cdot (1 + \ln h + \ln X).
    \]
\end{theorem}
\begin{proof}
If $X = 0$, $x$ is already integral, so we can take $y = x$.
From now on, assume $X\geq 2$.
We will round the starting fractional profiles $x$ (which we will refer to as a ``solution'') to an integral $y$.
The rounding algorithm will operate in phases, each phase $k$ will start with the current solution $x^k$ and obtain $x^{k + 1}$.
Every entry in $x^{k + 1}$ corresponding to some profile $\pi \neq \bot$ that reaches $0$ or $1$ will be fixed for subsequent phases.
Throughout the phase, we maintain for every $i \in N$ that $x_i^k(\pi) \geq 0$ for all $\pi \in \Pi$, and $\sum_{\pi \in\Pi}x_i^k(\pi) = 1$.
Let $\Xx_k := \{(i, \pi) : 0 < x_i^k(\pi) < 1\}$ be the set of strictly fractional entries in $x^k$ at the beginning of phase $k$, and let $X_k := |\Xx_k|$.
Initially, we start with $x^1 := x$ and $X_k := X$.

Suppose that $X_k > 16$.
Before rounding some of the entries of $x^k$ to $0$ or $1$, we will create additional slackness in every prefix $0 \leq \ell \leq m$, that can absorb potential increase in the prefix load caused by the rounding.
Denote 
\[
    p_k := \frac{4(1 + \ln (hX_k))}{X_k}.
\]
We create an \textbf{intermediate solution} $\wx^k$ as follows.
For every $(i, \pi) \in \Xx_k$, where $\pi \neq \bot$, we transfer $\min\{x_i^k(\pi), p_k\}$ weight from $x_i^k(\pi)$ into $x_i^k(\bot)$.
That is,
\[
    \forall (i, \pi) \in \Xx_k:\;\wx^k_i(\pi) = x^k_i(\pi) - \min\{x_i^k(\pi), p_k\},\quad\quad \wx_i^k(\bot) = x_i^k(\bot) + \sum_{\substack{(i, \pi) \in \Xx_k\\\pi\neq \bot}}\min\{x_i^k(\pi), p_k\}.
\]

Let $\wXx_k = \{(i, \pi) :0 < \wx^k_i(\pi) < 1\}$ be the strictly fractional entries in $\wx^k$.
If $\wXx_k = \varnothing$, then all entries of $\wx^k$ are integral.
Then we set $x^{k + 1} = \wx^k$ and move to the next phase.

If $\wXx_k \neq \varnothing$, every pair $(i, \pi) \in \wXx_k$ with $\pi\neq \bot$ had $x_i^k(\pi) > p_k$, and lost exactly $p_k$ when moving to $\wx^k$.
For $(i, \pi) \in \wXx_k$ with $\pi \neq \bot$, we list all the breakpoints of $i$ for all demands $\alpha \in \pi$, and combine them into a multiset $\Mm_k$:
\[
    \Mm_k := \biguplus_{\substack{(i, \pi) \in \wXx_k\\\pi \neq \bot}}\{m_i(\alpha) : \alpha \in \pi\}.
\]
\begin{lemma}\label{etatransfer}
    It holds $\Ii(\wx^k, \Pi) \geq \Ii(x^k, \Pi) - p_kX_k$ and $\tau(\wx^k, \Pi) \leq \tau(x^k, \Pi)$.
    Furthermore, 
    \[
        \forall 0 \leq \ell \leq m:\qquad \Hh(\wx^k, \Pi, \ell) + p_k\Pp_k(\ell) \leq \Hh(x^k, \Pi, \ell),
    \]
    where for every $-1 \leq \ell \leq m$,
    \[
        \Pp_k(\ell) = \sum_{(i, \pi) \in \wXx_k}\Pp_i(\pi, \ell)
    \]
    is the total number of breakpoints in $\Mm_k$ that are at most $\ell$, with a convention $\Pp_k(-1) = 0$.
\end{lemma}
\begin{proof}
    Since each non-$\bot$ entry in $x^k$ loses at most $p_k$, the bound $\Ii(\wx^k, \Pi) \geq \Ii(x^k, \Pi) - p_kX_k$ is immediate.
    Similarly, moving weight to $\bot$ can only decrease prefix loads, so $\tau(\wx^k, \Pi) \leq \tau(x^k, \Pi)$.

    Finally, by construction of $\wx^k$, for every $0 \leq \ell \leq m$ we have
\[
    \Hh(x^k, \Pi, \ell) - \Hh(\wx^k, \Pi, \ell) = \sum_{i \in N}\sum_{\pi\neq \bot}\big(x_i^k(\pi) - \wx_i^k(\pi)\big)\cdot \Pp_i(\pi, \ell) \geq \sum_{(i, \pi) \in \wXx_k}p_k\cdot \Pp_i(\pi, \ell),
\]
which is exactly $p_k\cdot \Pp_k(\ell)$ by definition.
\end{proof}

Next, we will create a small set of \textbf{head checkpoints}, i.e a subset of prefixes $\ell$ for which we would like to preserve the inequality $\Hh(x^{k + 1}, \Pi, \ell) \leq \Hh(x^k, \Pi, \ell)$ no matter what happens in the rounding.
We do it as follows.
First, sort all the breakpoints in $\Mm_k$ in non-decreasing order, and let $m_{i_j}$ denote the $j$-th breakpoint in $\Mm_k$.
For every $1 \leq j \leq |\Mm_k|$, we associate a weight $1/j$ with a breakpoint $m_{i_j}$.
We start with a virtual checkpoint $t = -1$.
Then, we find a smallest prefix $\ell \in \{0, \ldots, m\}$ such that $\ell > t$ and the total weight of breakpoints between $t + 1$ and $\ell$ in $\Mm_k$ is at least $p_k$.
That is, $\ell > t$ is the smallest satisfying
\[
    \sum_{j = \Pp_k(t) + 1}^{\Pp_k(\ell)}\frac{1}{j} > p_k.
\]
We put a checkpoint at $\ell$, increase $t$ to $\ell$, and repeat the process until no such $\ell > t$ exists.
We denote the set of obtained checkpoints as $\Ll_k$ (we do not include $-1$).

\begin{lemma}\label{etacheckpoints}
    The total number of checkpoints at phase $k$ is $|\Ll_k|\leq X_k/4$.
    In addition, for any solution $z$ satisfying $z_i(\pi) = \wx_i^k(\pi)$ for all $(i, \pi)$ such that $\wx_i^k(\pi) \in \{0, 1\}$, the following holds.

    Suppose that for every $\ell_k \in \Ll_k$ we have $\Hh(z, \Pi, \ell_k) \leq \Hh(\wx^k, \Pi, \ell_k)$.
    Then, for every $0 \leq \ell \leq m$,
    \[
        \Hh(z, \Pi, \ell) \leq \Hh(\wx^k, \Pi, \ell) + p_k\Pp_k(\ell).
    \]
\end{lemma}
\begin{proof}
If $\Mm_k = \varnothing$, take $\Ll_k = \varnothing$, then all $(i, \pi) \in \wXx_k$ have no breakpoints at any prefix $\ell$, in which case lemma holds trivially.
From now on, assume $\Mm_k \neq \varnothing$, so $\Pp_k(m) \geq 1$.
We will start with the number of checkpoints.
By construction of $\Ll_k$, we have
\[
    |\Ll_k|\cdot p_k \leq \sum_{j = 1}^{\Pp_k(m)}\frac{1}{j} \leq 1 + \ln\Pp_k(m)\quad\implies\quad |\Ll_k|\leq \frac{1 + \ln\Pp_k(m)}{p_k}.
\]
By assumption, every profile $\pi \in \Pi - \bot$ has $|\pi|\leq h$.
In addition, every coordinate $(i, \pi) \in \wXx_k$ with $\pi \neq \bot$ was strictly fractional in solution $x^k$ as well.
Therefore
\[
    \Pp_k(m) = |\Mm_k| = \sum_{\substack{(i, \pi) \in \wXx_k\\\pi\neq\bot}}|\pi| \leq \sum_{\substack{(i, \pi) \in \Xx_k\\\pi\neq\bot}}|\pi| \leq hX_k,
\]
and substituting the definition of $p_k = (1/X_k)\cdot 4(1 + \ln (hX_k))$, we get
\[
    |\Ll_k| \leq \frac{1 + \ln\Pp_k(m)}{p_k} \leq \frac{X_k(1 + \ln(hX_k))}{4(1 + \ln (hX_k))} = \frac{X_k}{4}.
\] 

Next, for $0 \leq \ell \leq m$, let $\ell_k \in \Ll_k$ be the largest checkpoint preceding $\ell$, i.e satisfying $\ell_k \leq \ell$ (set $\ell_k = -1$ if none exists).
Since $z$ coincides with $\wx^k$ on all coordinates outside of $\wXx_k$, it holds
\begin{multline*}
    \Hh(z, \Pi, \ell) - \Hh(\wx^k, \Pi, \ell) = \Hh(z, \Pi, \ell_k) - \Hh(\wx^k, \Pi, \ell_k) \\
    + \sum_{(i, \pi) \in \wXx_k}\big(z_i(\pi) - \wx_i^k(\pi)\big)\cdot \big(\Pp_i(\pi, \ell) - \Pp_i(\pi, \ell_k)\big).
\end{multline*}
By assumption, $\Hh(z, \Pi, \ell_k) \leq \Hh(\wx^k, \Pi, \ell_k)$.
As both $z_i(\pi)$ and $\wx^k_i(\pi)$ lie in $[0, 1]$, we have
\begin{multline*}
    \Hh(z, \Pi, \ell) - \Hh(\wx^k, \Pi, \ell)\leq \sum_{(i, \pi) \in \wXx_k}|z_i(\pi) - \wx_i^k(\pi)|\cdot \big(\Pp_i(\pi, \ell) - \Pp_i(\pi, \ell_k)\big)\\
    \leq \sum_{(i, \pi) \in \wXx_k} \big(\Pp_i(\pi, \ell) - \Pp_i(\pi, \ell_k)\big) = \Pp_k(\ell) - \Pp_k(\ell_k).
\end{multline*}
If $\ell_k = \ell$ then $\Pp_k(\ell) - \Pp_k(\ell_k) = 0$, and we are done.
Otherwise, by construction of $\Ll_k$ we have
\[
    \sum_{j = \Pp_k(\ell_k) + 1}^{\Pp_k(\ell)}
    \frac{1}{j} \leq p_k.
\]
When $\Pp_k(\ell) > 0$, each summand above is at least $1/\Pp_k(\ell)$, hence 
\[
    \frac{\Pp_k(\ell) - \Pp_k(\ell_k)}{\Pp_k(\ell)} \leq \sum_{j = \Pp_k(\ell_k) + 1}^{\Pp_k(\ell)}
    \frac{1}{j} \leq p_k \quad\implies\quad \Pp_k(\ell) - \Pp_k(\ell_k) \leq p_k\Pp_k(\ell).
\]
We conclude that $\Hh(z, \Pi, \ell) - \Hh(\wx^k, \Pi, \ell) \leq p_k\Pp_k(\ell)$, which finishes the proof.
\end{proof}

It follows from \cref{etatransfer} and \cref{etacheckpoints} that, to preserve the prefix load constraints at every prefix $0 \leq \ell \leq m$ at phase $k$, it suffices to preserve only the checkpoint prefixes $\Ll_k$.

We will now construct an auxiliary polytope (P) with a relatively small number of constraints, and partially round the solution $\wx^k$ to one of its extreme points, i.e some $z^*$, that will satisfy \cref{etacheckpoints}.
The polytope (P) is defined by the following constraints.
\begin{equation}
        \begin{aligned}
        & && \sum_{\pi \in \Pi}z_i(\pi) =1,&\forall i \in N,\\
        & && z_i(\pi) = \wx^{k}_i(\pi),&\forall (i, \pi): \wx_i^k(\pi) \in \{0, 1\},\\
        & && \tau(z, \Pi) \leq \tau(\wx^k, \Pi),\\
        & && \Hh(z, \Pi, \ell_k) \leq \Hh(\wx^k, \Pi, \ell_k),&\forall \ell_k \in \Ll_k,\\
        & && 0 \leq z_i(\pi) \leq 1,&\forall i \in N, \pi \in \Pi.
    \end{aligned}\tag{P}\label{P}
    \end{equation}
Clearly, $z = \wx^k$ is feasible for (P).
We choose an extreme point $z^*$ of (P) maximizing 
\[
    \Ii(z^*, \Pi) = \sum_{i \in N}\sum_{\pi\neq \bot}z_i^*(\pi).
\]
Such an extreme point must exist, because (P) is nonempty and compact, i.e bounded and closed.
Then, by maximality $\Ii(z^*, \Pi) \geq \Ii(\wx^k, \Pi)$.
\begin{lemma}\label{etaextreme}
    Let $z^*$ be an extreme point of the polytope $\mathrm{(P)}$.
    Then, the number of strictly fractional entries in $z^*$ is at most $X_k/2 + 2$.
\end{lemma}
\begin{proof}
    Let $Z^* := |\{(i, \pi) : 0 < z_i^*(\pi) < 1\}|$ be the total number of strictly fractional coordinates in $z^*$.
    Note that if for some $i \in N$ we have $0 < z_i^*(\pi') < 1$ for some $\pi' \in \Pi$, there must be another $\pi''\in \Pi$ with $0 < z_i^*(\pi'') < 1$, as $\sum_{\pi \in \Pi}z_i^*(\pi)= 1$.
    Hence, there can be at most $Z^*/2$ different agents $i \in N$ with $0 < z_i^*(\pi) < 1$ for some $\pi \in \Pi$, and for the remaining agents $z_i^*(\pi) \in \{0, 1\}$ for every coordinate $(i, \pi)$.

    Suppose that $Z^* > X_k/2 + 2$.
    Restrict the equation-matrix of the polytope (P) to the entries $(i, \pi)$ with $0 < z_i^*(\pi) < 1$, it is a system of equations with $Z^*$ variables.
    It has at most $Z^*/2$ constraints of the form $\sum_{\pi \in \Pi}z_i^*(\pi) = 1$ for some $i \in N$.
    We have an additional constraint on $\tau(z^*, \Pi)$, and the number of constraints on $\Hh(z^*, \Pi, \ell_k)$ for $\ell_k \in \Ll_k$ is $|\Ll_k|\leq X_k/4$ by \cref{etacheckpoints}.
    So, if $Z^* > X_k/2 + 2$, the total number of constraints in this restriction is at most
    \[
        \frac{Z^*}{2} + 1 + |\Ll_k|\leq \frac{Z^*}{2}  + \frac{4 + X_k}{4} < \frac{Z^*}{2} + \frac{2Z^*}{4} = Z^*.
    \]
    Therefore, the restricted matrix has rank less than $Z^*$, implying that there is a non-zero solution supported on strictly fractional entries of $z^*$ that preserves all constraints.
    All these entries are strictly between $0$ and $1$, hence small perturbations of both signs still keep $z^*$ inside the polytope.
    This contradicts the extremality of $z^*$.
    The lemma follows.
\end{proof}
Take this extreme point $z^*$ of (P) and set $x^{k + 1} = z^*$.
By \cref{etatransfer} and \cref{etacheckpoints},
\[
    \forall 0 \leq \ell \leq m:\; \Hh(x^{k + 1}, \Pi, \ell)\leq \Hh(x^k, \Pi, \ell),\qquad\text{and}\qquad \tau(x^{k+ 1}, \Pi) \leq \tau(x^k, \Pi),
\]
as well as
\[
    \Ii(x^k, \Pi) -\Ii(x^{k + 1}, \Pi)\leq p_kX_k= 4(1 + \ln(hX_k)).
\]

By \cref{etaextreme}, $X_{k + 1} \leq X_k/2 + 2$, hence by induction on $k$ it holds for every $k \geq 1$ that
\[
    X_k \leq 4 + (X - 4)\cdot 2^{-(k -1)}.
\]
Then, if at phase $K + 1$ we have $X_{K + 1} \leq 16$, it must hold $K \leq \log X$.
At this phase, solution $x^{K + 1}$ has at most $16$ strictly fractional coordinates, hence there are at most $8$ agents with a strictly fractional coordinate.
We create a new solution $y$ as follows: for every $i \in N$ with $0 < x_i^{K + 1}(\pi) < 1$ for some $\pi \in \Pi$, set $y_i(\bot) = 1$, i.e throw these agents away.
For every other agent $i \in N$ set $y_i(\pi) = x_i^{K + 1}(\pi)$ for all $\pi \in \Pi$.
Clearly, $y$ is integral,  every head/tail load can only decrease, and
\[
    \Ii(y, \Pi) \geq \Ii(x^{K + 1}, \Pi) - 8.
\]
Summing up the loss in $\Ii(x^k, \Pi)$ over all phases $k \in [K]$, we get (using $X_k \leq X$)
\begin{multline*}
    \Ii(x, \Pi) - \Ii(y, \Pi) \leq 8 + \sum_{k = 1}^{K}\big(\Ii(x^k, \Pi) - \Ii(x^{k + 1}, \Pi)\big) \leq 8 + 4\sum_{k = 1}^K(1 + \ln(hX_k))\\
    \leq 8 + 4\log X\cdot (1 + \ln(hX)) = \rho(h, X).
\end{multline*}
To ensure that $\Ii(y, \Pi) \leq \Ii(x, \Pi)$, in case $\Ii(y, \Pi) > \Ii(x, \Pi)$ we retain any $\lfloor \Ii(x, \Pi)\rfloor$ agents assigned by $y$, and give $\bot$ to the rest.
Removing agents preserves prefix load constraints, and we still have at least $\Ii(x, \Pi) - 1$ selected agents.
Finally, by construction the agents $i \in N$ with $y_i(\pi) = 1$ for some $\pi \neq\bot$ must have had $x_i(\pi) > 0$.
This finishes the proof.
\end{proof}
\section{Proof of the bounded head and tail rounding theorem}

\label{appendix:rounding}

\begin{theorem}[\cref{rounding} restated]\label{roundingproof}
    Let $\Pi$ be a shared set of head profiles, such that $|\pi|\leq q$ for some integer $q \geq 3$.
    For every $i \in N$, let $(x_i, \Pi)$ be a fractional profile, denote $x = \{x_i\}_{i \in N}$.
    
    Suppose that for every $\pi \in \Pi - \bot$, either $\Delta(\pi) = 0$ or $\Delta(\pi) \geq \delta > 0$ for some $\delta \in (0, 1]$.
    For any $\lambda > 1$, there exists a collection $y = \{y_i\}_{i \in N}$ of integer head profiles $(y_i, \Pi)$, such that for every $i, \pi$ we have $y_i(\pi) \in \{0, 1\}$ and $y_i(\pi) = 1$ for $\pi \neq \bot$ only if $x_i(\pi) > 0$, and 
    \[\begin{aligned}
        & && \Hh(y, \Pi, \ell) \leq \Hh(x, \Pi, \ell),  & \forall 0 \leq \ell \leq m,\\
        & && \Tt(y, \Pi, \ell) \leq \lambda\Tt(x, \Pi, \ell),&\forall 0 \leq \ell \leq m,\\
        &  && \Ii(y, \Pi)\geq \Ii(x, \Pi) - \rho(q, X), 
    \end{aligned}\]
    where, for for $\tau(x, \Pi)=\sum_{i \in N}\sum_{\pi \in \Pi}x_i(\pi)\cdot \tau_i(\pi)$ and $X= |\{(i, \pi):0 < x_i(\pi) <1 \}|$,
    \[
        \rho(q, X) = 8 + 4\ \log_\lambda\max\left\{1, \frac{1}{\delta}h_{\max}T_{\max}X\tau(x, \Pi)\right\}+ 8q\log X.
    \]
    Here $h_{\max} := \max_{i \in N}h_i$ and $T_{\max} := \max_{i \in N}T_i$.
\end{theorem}
\begin{proof}
    Both the statement and the proof of this theorem are very similar to \cref{etaroundingproof}.
    There are three key differences.
    First, instead of one tail load constraint, we have $m + 1$ tail constraints, which we will have to take care off.
    Second, we require every positive deficit to be separated from $0$ by some value $\delta \in (0, 1]$.
    We need this in order to guarantee that tail constraints are violated by at most a factor of $\lambda$, while keeping the $\Ii$-value loss logarithmic in $X$.
    Third, the number of demands in every $\pi \in \Pi$ is at most $q$, which is a constant in our setting, while in \cref{etaroundingproof} the bound was $h = O(n^2)$.
    In the proof of \cref{etaroundingproof}, we had to put $h$ under $\log$-factor, and lost $O(\log(hX)) = O(\log n)$ agents per phase of the rounding algorithm because of this.
    In the rounding algorithm of \cref{roundingproof}, however, we will allow a loss of $O(q)$ agents per phase, which allows us to avoid an additional $O(\log X)$ factor.

    Assume $X \geq 2$, otherwise $x$ is already integral.
    Unlike in \cref{etaroundingproof}, we will slightly modify the starting solution $x$ before the actual rounding.    
    Our first step is to remove all entries $x_i(\pi)$ with $\pi\neq\bot$ that are too small.
    Specifically, for every $(i, \pi)$ such that $\pi \neq \bot$ and  $0 < x_i(\pi) < 1/X$, we transfer all the weight $x_i(\pi)$ into $x_i(\bot)$.
    For the new solution, which we denote by $x^1$, it trivially holds for all $0 \leq \ell \leq m$:
    \[
        \Hh(x^1, \Pi, \ell) \leq \Hh(x, \Pi, \ell)\qquad \text{and}\qquad \Tt(x^1, \Pi, \ell) \leq \Tt(x, \Pi, \ell).
    \]
    Since we cannot remove more than $X$ entries in total, and each removed entry is at most $1/X$, 
    \[
        \Ii(x^1, \Pi) = \sum_{i \in N}\sum_{\pi \neq \bot}x_i^1(\pi)= \sum_{i \in N}\sum_{\pi \neq \bot}x_i(\pi) - \sum_{\substack{(i, \pi)\\0<x_i(\pi) < 1/X}}x_i(\pi) \geq \Ii(x, \Pi) - 1.
    \]
    Let $\Xx_1:= \{(i, \pi) : 0<x_i^1(\pi)<1\}$ and $X_1 := |\Xx_1|$.
    Note that $X_1 \leq X$, as a new entry $0 < x_i^1(\bot) < 1$ can appear for an agent $i \in N$ only if we set $x_i^1(\pi) = 0$ for some $\pi \neq \bot$ with $0 < x_i(\pi) < 1$.
    Hence, the number of strictly fractional entries cannot increase.

    Replacing $x$ with $x^1$ allows us to lower bound $ \Tt(x^1, \Pi, \ell)$ for every $0\leq \ell \leq m$.
    Recall that
    \[
        \tau_i(\pi) := \begin{cases}
            \Delta(\pi)/T_i,& \pi\neq \bot\text{ and }T_i > 0;\\
            0,& \text{otherwise}.
        \end{cases}
    \]
    By assumption, for every $\pi \in \Pi - \bot$ either $\Delta(\pi) = 0$ or $\Delta(\pi) \geq \delta > 0$.
    Let $T_{\max} := \max_{i \in N}T_i$.
    If $T_{\max} = 0$, then all $\tau_i(\pi)$ are $0$, and we do not need to consider the tail loads.
    Otherwise
    \[
        \tau_i(\pi) > 0\quad \implies \quad\tau_i(\pi) \geq \frac{\delta}{T_{\max}}
    \]
    If $\Tt(x^1, \Pi, \ell) > 0$, its positive summands must have positive $x_i^1(\pi), \tau_i(\pi)$ and $(\ell - h_i)_+$.
    Thus
    \[
        \Tt(x^1, \Pi, \ell) > 0 \implies \Tt(x^1, \Pi, \ell) = \sum_{i \in N}\sum_{\pi \neq \bot}x_i^1(\pi)\cdot \tau_i(\pi)(\ell - h_i)_+ \geq \frac{\delta}{T_{\max}X} =: \theta.
    \]

    In order to preserve the tail load constraints (or at least, not violate them by boo much), we will do a trick similar to \cref{etacheckpoints}.
    Specifically, before we even start rounding, we will create a set of \textbf{tail checkpoints}.
    That is, a subset of prefixes $r$ for which we will preserve the inequality $\Tt(x^{k + 1}, \Pi, r) \leq \Tt(x^k, \Pi, r)$ at every iteration $k \geq 1$, without multiplicative error $\lambda$.
    We do it as follows.
    Let $h_{\max} := \max_{i \in N}h_i$.
    For every integer $t \geq 0$, consider the largest prefix $0 \leq r \leq h_{\max}$ such that $\Tt(x^1, \Pi, r)/\theta$ is in $[\lambda^t, \lambda^{t + 1})$.
    That is, we consider a set of prefixes
    \[
        \Rr_t := \big\{r \in \{0, 1, \ldots,h_{\max}\} : \theta\lambda^t \leq \Tt(x^1, \Pi, r) < \theta\lambda^{t + 1}\big\},
    \]
    and take checkpoint $r_t = \max_{r \in \Rr_t}r$.
    Denote the set of tail checkpoints by $\Rr$, with $R := |\Rr|$.
    
    We will now show an analogue of \cref{etacheckpoints}: in order to obtain the desired guarantees for $y$ on $\Tt(y, \Pi, \ell)$, it suffices to preserve the fractional prefix tail load only tail checkpoints $r \in \Rr$.
    \begin{lemma}\label{tailcheckpoints}
        Suppose that a solution $z$ satisfies $z_i(\pi) > 0$ for $\pi \neq \bot$ only if $x_i^1(\pi) > 0$, as well as $\tau(z, \Pi) \leq \tau(x^1, \Pi)$ and $\Tt(z, \Pi, r) \leq \Tt(x^1, \Pi, r)$ for every $r \in \Rr$.
        Then, for every $0 \leq \ell \leq m$,
        \[
            \Tt(z, \Pi, \ell)\leq \lambda\Tt(x^1, \Pi, \ell),
        \]
        and in fact $\Tt(z, \Pi, \ell) \leq \Tt(x^1, \Pi, \ell)$ whenever $\ell \geq h_{\max}$.
        In addition, $R = |\Rr|$ satisfies
        \[
            R \leq 1 + \log_\lambda\max\left\{1, \frac{h_{\max}\tau(x, \Pi)}{\theta}\right\}.
        \]
    \end{lemma}
    \begin{proof}
        If $h_{\max} = 0$, then every tail is exactly the same, and a single inequality $\tau(z, \Pi) \leq \tau(x^1, \Pi)$ preserves all tail loads exactly.
        From now on, assume $h_{\max} > 0$.
        
        Suppose that $\Tt(x^1, \Pi, \ell) = 0$.
        Its every summand is non-negative, so every $\pi \in \Pi$ with $x_i^1(\pi) > 0$ contributes $0$.
        Since $z_i(\pi) > 0$ for $\pi \neq \bot$ only if $x_i^1(\pi) > 0$, then $\Tt(z, \Pi, \ell) = 0$ too.
        
        Next, consider $\ell < h_{\max}$ such that $\Tt(x^1, \Pi, \ell)$.
        By construction, prefix $\ell$ belongs to $\Rr_t$ for $t \geq 0$ such that $\theta \lambda^t \leq \Tt(x^1, \Pi, \ell) < \theta\lambda^{t + 1}$.
        Since $r_t = \max_{r \in \Rr_t}r$, we have $\ell \leq r_t$ and
        \[
            \Tt(z, \Pi, \ell) \leq \Tt(z, \Pi, r_t) \leq \Tt(x^1, \Pi, r_t) \leq \theta \lambda^{t + 1} \leq \lambda \Tt(x^1, \Pi, \ell).
        \]
        Note that if $\Tt(x^1, \Pi, h_{\max}) > 0$, then $h_{\max} \in \Rr$, and $\Tt(z, \Pi, h_{\max}) \leq \Tt(x^1, \Pi, h_{\max})$.
        Similarly, if $\Tt(x^1, \Pi, h_{\max}) = 0$, support preservation gives $\Tt(z, \Pi, h_{\max}) = 0$.
        Then, for every $\ell \geq h_{\max}$,
        \begin{multline*}
            \Tt(z, \Pi, \ell) = \sum_{i \in N}\tau_i(z_i, \Pi)\cdot (\ell - h_i) = \sum_{i \in N}\tau_i(z_i, \Pi)\cdot (h_{\max} - h_i) + \sum_{i \in N}\tau_i(z_i, \Pi)\cdot (\ell - h_{\max}) \\
            = \Tt(z, \Pi, h_{\max}) + (\ell - h_{\max})\tau(z, \Pi) \leq \Tt(x^1, \Pi, h_{\max}) + (\ell - h_{\max})\tau(x^1, \Pi) = \Tt(x^1, \Pi, \ell).
        \end{multline*}

        We now bound $R$.
        We have $\tau(x^1, \Pi) \leq \tau(x, \Pi)$ (only $\bot$-entries gained), so for every $\ell \leq h_{\max}$,
        \[
            \Tt(x^1, \Pi, \ell) = \sum_{i \in N}\tau_i(x_i^1, \Pi)\cdot (\ell - h_i) _+ \leq \ell \cdot \tau(x^1, \Pi) \leq h_{\max}\tau(x, \Pi).
        \]
        If $h_{\max}\tau(x, \Pi) < \theta$, all $\Rr_t$ are empty and $R = 0$.
        Otherwise, for every $t \geq 0$ such that $\Rr_t \neq \varnothing$,
        \[
            \theta\lambda^t \leq h_{\max}\tau(x, \Pi) \quad \implies \quad t \leq \log_\lambda\max\left\{1, \frac{h_{\max}\tau(x, \Pi)}{\theta}\right\}.
        \]
        There is at most one tail checkpoint for each such $t \geq 0$.
        The lemma follows.
    \end{proof}
    We now proceed to actual rounding.
    As in \cref{etaroundingproof}, we start with a solution $x^1$ and will round it to integral $y$ over multiple phases, fixing non-$\bot$ entries of current solution $x^k$ that reach $0$ or $1$ and maintaining for every $i \in N$ that $x_i^k(\pi) \geq 0$ for all $\pi \in \Pi$ and $\sum_{\pi \in \Pi}x_i^k(\pi) = 1$.
    Once again, let $\Xx_k := \{(i, \pi) : 0 < x_i^k(\pi) < 1\}$ be strictly fractional entries of $x^k$, and $X_k := |\Xx_k|$.

    Suppose that $X_k > 8(R + 1)$.
    In addition to the tail checkpoints $\Rr$ that are the same at every phase $k$, during the phase $k$ we create a set of \textbf{head checkpoints} $\ell$, for which we would preserve the inequality $\Hh(x^{k + 1}, \Pi, \ell) \leq \Hh(x^k, \Pi, \ell)$ without any error.
    Instead of a per-phase loss $O(\log qX)$ in \cref{etaroundingproof}, we will save a $\log X$-factor by allowing an additive loss of $O(q)$.

    We create additional slackness in every prefix $0 \leq \ell \leq m$ that will absorb potential rounding loss.
    Define \textbf{intermediate solution} $\wx^k$ as follows.
    For every $(i, \pi) \in \Xx_k$, $\pi \notin\{\varnothing, \bot\}$, let
    \[
        b_i(\pi) := \min_{\alpha \in \pi}m_i(\alpha)
    \]
    be the earliest breakpoint of agent $i$ in profile $\pi$.
    We sort all pairs $(i, \pi) \in \Xx_k$ with respect to $b_i(\pi)$.
    Then, starting from the first pair, we keep transferring weight $x_i^k(\pi)$ to $x_i^k(\bot)$ until either a total weight of $8q$ has been transferred, or all non-$\bot$ weights are set to $0$.
    Every transfer, except for possibly the last one, completely exhausts the corresponding coordinate $x_i^k(\pi)$.
    We denote the resulting solution by $\wx^k$, and $\wXx_k := \{(i, \pi) : 0 < \wx_i^k(\pi) < 1\}$ and $\wX_k := |\wXx_k|$.

    As in the proof of \cref{etaroundingproof}, for $(i, \pi) \in \wXx_k$ with $\pi \neq \bot$, we list all the breakpoints of $i$ for all demands $\alpha \in \pi$, and combine them into a multiset $\Mm_k$:
    \[
        \Mm_k := \biguplus_{\substack{(i, \pi) \in \wXx_k\\\pi \neq \bot}}\{m_i(\alpha) : \alpha \in \pi\}.
    \]
    As before, for $-1 \leq \ell \leq m$ we denote by
    \[
        \Pp_k(\ell) = \sum_{(i, \pi)\in \wXx_k}\Pp_i(\pi, \ell)> 0
    \]
    the total number of breakpoints in $\Mm_k$ that are at most $\ell$, with $\Pp_k(-1) = 0$.
    \begin{lemma}\label{ktransfer}
        It holds $\Ii(\wx^k, \Pi) \geq \Ii(x^k, \Pi) - 8q$ and $\tau(\wx^k, \Pi) \leq \tau(x^k, \Pi)$, as well as  $\Tt(\wx^k, \Pi, \ell) \leq \Tt(x^k, \Pi, \ell)$ and $\Hh(\wx^k, \Pi, \ell) \leq \Hh(x^k, \Pi, \ell)$ for every $0 \leq \ell \leq m$.
        In addition, if $\Pp_k(\ell) > 0$, then
        \[
            \Hh(\wx^k, \Pi, \ell) + 8q \leq \Hh(x^k, \Pi, \ell).
        \]
    \end{lemma}
    \begin{proof}
        Every claim except for the last bound on head prefix loads is immediate.

        Let $\ell$ be a prefix such that $\Pp_k(\ell) > 0$.
        Then, there exists $(i_*, \pi_*) \in \wXx_k$ such that $\Pp_{i_*}(\pi_*, \ell) > 0$, and the earliest breakpoint of $\pi$ satisfies $b_{i_*}(\pi_*) \leq \ell$.
        As $\wx^k_{i_*}(\pi_*) > 0$, we must have $x^k_{i_*}(\pi_*) > 0$, and every pair $(i, \pi) \in \Xx_k$ from which positive weight was transferred into $\bot$ must have occurred before $(i_*, \pi_*)$ (or ended at $(i_*, \pi_*))$.
        Then, for every $(i, \pi)$ such that $x_{i}^k(\pi) - \wx^k_i(\pi) > 0$ it holds $b_{i}(\pi) \leq b_{i_*}(\pi_*) \leq \ell$, and $\Pp_i(\pi, \ell) \geq 1$ for every such pair $(i, \pi)$.
        Hence,
        \[
            \Hh(x^k, \Pi, \ell) - \Hh(\wx^k, \Pi, \ell) = \sum_{i \in N}\sum_{\pi\neq \bot}\big(x_{i}^k(\pi) - \wx^k_i(\pi)\big)\cdot \Pp_i(\pi, \ell) \geq 8q.
        \]
        It is at least $8q$ as $\wx_{i_*}^k(\pi_*) > 0$ implies that the transfer stopped before zeroing-out all weights.
    \end{proof}

    We now construct \textbf{head checkpoints}.
    Sort $\Mm_k$ in non-decreasing order, and then choose every $8q$-th breakpoint in $\Mm_k$ as a head checkpoint.
    Let $\Ll_k$ be the set of chosen checkpoints.
\begin{lemma}\label{headcheckpoints}
    The total number of head checkpoints at phase $k$ is $|\Ll_k|\leq X_k/8$.
    In addition, for any solution $z$ satisfying $z_i(\pi) = \wx_i^k(\pi)$ for all $(i, \pi)$ with $\wx_i^k(\pi) \in \{0, 1\}$, the following holds.

    Suppose that for every $\ell_k \in \Ll_k$ we have $\Hh(z, \Pi, \ell_k) \leq \Hh(\wx^k, \Pi, \ell)$.
    Then, for every $0 \leq \ell \leq m$,
    \[
        \Hh(z, \Pi, \ell) \leq \Hh(\wx^k, \Pi, \ell) + 8q\One\{\Pp_k(\ell) > 0\}.
    \]
\end{lemma}
\begin{proof}
    This is an analogue of \cref{etacheckpoints} for heads.
    By construction, every fractional coordinate in $\wx^k$ was fractional in $x^k$.
    Hence, using the assumption $|\pi|\leq q$ for $\pi \in \Pi$, we have
    \[
        |\Mm_k| = \sum_{\substack{(i, \pi) \in \wXx_k\\\pi\neq\bot}}|\pi| \leq qX_k \quad\implies\quad |\Ll_k| \leq \left\lfloor\frac{|\Mm_k|}{8q}\right\rfloor \leq \frac{X_k}{8}.
    \]

    Next, let $0 \leq \ell \leq m$.
    If $\Pp_k(\ell) = 0$, every fractional entry of $\wx^k$ contributes $0$ to $\Hh(\wx^k, \Pi, \ell)$.
    Since $z$ coincides with $\wx^k$ on all coordinates outside of $\wXx_k$, we must have $\Hh(z, \Pi, \ell) = \Hh(\wx^k, \Pi, \ell)$.
    
    If $\Pp_k(\ell) > 0$, let $\ell_k \in \Ll_k$ be the largest checkpoint preceding $\ell$, i.e satisfying $\ell_k \leq \ell$ (set $\ell_k = -1$ if none exists).
    The exact same reasoning as in \cref{etacheckpoints} gives
    \[
        \Hh(z, \Pi, \ell) - \Hh(\wx^k, \Pi, \ell) \leq \Pp_k(\ell) - \Pp_k(\ell_k).
    \]
    If $\ell_k = \ell$, $\Pp_k(\ell) - \Pp_k(\ell_k) = 0$ and we are done.
    Otherwise, by construction of $\Ll_k$ there are at most $8q$ breakpoints of $\Mm_k$ in $(\ell_k, \ell]$.
    Thus, $\Pp_k(\ell) - \Pp_k(\ell_k) \leq 8q$, and the lemma follows.
\end{proof}
    It follows from \cref{tailcheckpoints}, \cref{ktransfer} and \cref{headcheckpoints} that, in order to preserve all head and tail load constraints at every prefix $0 \leq \ell \leq m$ at phase $k$, it suffices to preserve only tail checkpoints $\Rr$, head checkpoints $\Ll_k$ and inequality $\tau(x^{k + 1}, \Pi) \leq \tau(x^k, \Pi)$.

    Similar to \cref{etarounding}, we construct a polytope (P):
    \begin{equation}
        \begin{aligned}
        & && \sum_{\pi \in \Pi}z_i(\pi) =1,&\forall i \in N,\\
        & && z_i(\pi) = \wx^{k}_i(\pi),&\forall (i, \pi): \wx_i^k(\pi) \in \{0, 1\},\\
        &&& \tau(z, \Pi) \leq \tau(\wx^k, \Pi),\\
        & && \Tt(z, \Pi, r) \leq \Tt(\wx^k, \Pi, r),&\forall r \in \Rr,\\
        & && \Hh(z, \Pi, \ell_k) \leq \Hh(\wx^k, \Pi, \ell_k),&\forall \ell_k \in \Ll_k,\\
        & && 0 \leq z_i(\pi) \leq 1,&\forall i \in N, \pi \in \Pi.
    \end{aligned}\tag{P}\label{P}.
    \end{equation}
    Clearly, $z = \wx^k$ is feasible for (P).
    We choose an extreme point $z^*$ of (P) maximizing
    \[
        \Ii(z^*, \Pi) = \sum_{i \in N}\sum_{\pi \neq \bot}z_i^*(\pi) .
    \]
    It exists, since (P) is nonempty and compact.
    By maximality, $\Ii(z^*, \Pi) \geq \Ii(\wx^k, \Pi)$.
    \begin{lemma}\label{zextreme}
        Let $z^*$ be an extreme point of the polytope $(\mathrm{P})$. Then, the number of strictly fractional entries in $z^*$ is at most $X_k/2$.
    \end{lemma}
    \begin{proof}
        The proof is completely analogous to the proof of \cref{etaextreme}.
        If $Z^*$ is the total number of strictly fractional coordinates in $z^*$, then there can be at most $Z^*/2$ different agents $i \in N$ with fractional $z_i^*(\pi)$ for some $\pi \in \Pi$.

        Suppose that $Z^* > X_k/2$ and restrict the equation-matrix of (P) to the fractional entries of $z^*$.
        There are $Z^*$ variables, at most $Z^*/2$ constraints of the form $\sum_{\pi \in \Pi}z_i(\pi) = 1$ for some $i \in N$, $R$ tail load constraints and at most $|\Ll_k|\leq X_k/8$ head load constraints, as well as $\tau(z, \Pi) \leq \tau(\wx^k, \Pi)$.
        Hence, under the assumption $X_k > 8(R + 1)$, the number of constraints in this restriction is
        \[
            \frac{Z^*}{2} + R + \frac{X_k}{8}  + 1\leq \frac{Z^*}{2} + \frac{X_k}{4} < Z^*,
        \]
        where we used $Z^* >X_k/2$.
        As in \cref{etaextreme}, this contradicts extremality of $z^*$.
    \end{proof}

    We take this extreme point $z^*$ of (P) and set $x^{k + 1} = z^*$.
    It follows from the lemmas above that $X_{k + 1}\leq X_k/2$, $\tau(x^{k + 1}, \Pi) \leq \tau(x^k, \Pi)$, $\Ii(x^{k + 1}, \Pi) \geq \Ii(x^k, \Pi) - 8q$, as well as for every $0 \leq \ell \leq m$ that $ \Hh(x^{k + 1}, \Pi, \ell) \leq \Hh(x^k, \Pi, \ell)$ and for every $r \in \Rr$ that $\Tt(x^{k + 1}, \Pi, r)\leq \Tt(x^k, \Pi, r)$.
    
    We repeat the process until at some phase $K + 1$ we have $X_{K + 1}\leq 8(R + 1)$.
    For this $K$,
    \[\begin{aligned}
        &&& \Ii(x^{K + 1}, \Pi) \geq \Ii(x^1, \Pi) - 8qK,\\\
        &&&\tau(x^{K + 1}, \Pi) \leq \tau(x^1, \Pi),\\
        & && \Hh(x^{K + 1}, \Pi, \ell) \leq \Hh(x^1, \Pi, \ell),  & \forall 0 \leq \ell \leq m,\\
        & && \Tt(x^{K + 1}, \Pi, r) \leq \Tt(x^1, \Pi, r),&\forall r\in \Rr.\\
    \end{aligned}\]
    In addition, at most $X_{K + 1}/2\leq 4(R + 1)$ agents have fractional coordinates.
    We assign $\bot$ to every such agent, while keeping all integral weights unchanged.
    We take $y$ as the resulting integral solution.
    All prefix loads can only decrease, and applying \cref{tailcheckpoints} to $y$, we have
    \[
        \forall 0 \leq \ell \leq m:\quad \Tt(y, \Pi, \ell) \leq \lambda \Tt(x^1, \Pi, \ell) \leq \lambda \Tt(x, \Pi, \ell).
    \]
    At the same time, 
    \begin{multline*}
        \Ii(y, \Pi) \geq \Ii(x^{K + 1}, \Pi)  - 4(R + 1) \geq \Ii(x^1, \Pi) - 8qK - 4(R + 1) \\
        \geq \Ii(x, \Pi)  -1 - 8qK - 4(R + 1).
    \end{multline*}
    The last phase $K$ starts with 
    \[
        8(R + 1) < X_K \leq X\cdot 2^{-(K - 1)},
    \]
    implying $K \leq \log X - 1$.
    Using the bound from \cref{tailcheckpoints} and substituting all parameters, 
    \begin{multline*}
        1 + 8qK + 4(R + 1) \leq 9 + 4 \log_\lambda\max\left\{1, \frac{h_{\max}\tau(x, \Pi)}{\theta}\right\} + 8q(\log X - 1)\\
        \leq 8 + 4\ \log_\lambda\max\left\{1, \frac{1}{\delta}h_{\max}T_{\max}X\tau(x, \Pi)\right\}+ 8q\log X.
    \end{multline*}
    By construction, the agents $i \in N$ with $y_i(\pi) = 1$ for some $\pi \neq \bot$ must have had $x_i(\pi) > 0$.
    This finishes the proof.
\end{proof}

\section{Proof of the single supplier theorem}
\label{appendix:supplier}

\begin{theorem}[\cref{supplier} restated]\label{supplierproof}
    For some global constant $C_{t,q}$, the following holds.
    
    For $u_* \in U$, let $\{\pi_{u_*,j} : j \in [d]\}$ be his MMS profiles.
    In addition, let $\Ll_S \subseteq \{0, 1, \ldots, m\}$ be a set of $t$ \textbf{protected prefixes}.
    There exist $A_0\subseteq A$ and $J_0\subseteq [d]$ such that $|A_0| + |J_0| = d$ and
    \begin{enumerate}
        \item $|J_0| \leq C_{t,q}(d- |A|)$, and 
        the total number of added $E$-demands is
        \[
            \sum_{j \in J_0}|\pi_{u_*,j}^E|\leq qC_{t,q}(d - |A|);
        \]
        \item for every protected prefix $\ell_s \in \Ll_S$, the $E$-load is
            \[
                \sum_{i \in A_0}\Pp_i(\pi_i^E, \ell_s) + \sum_{j \in J_0}\Pp_{u_*}(\pi_{u_*,j}^E, \ell_s) \leq |E\cap [\ell_s]|.
            \]
    \end{enumerate}
\end{theorem}
\begin{proof}
Note that the bound on $ \sum_{j \in J_0}|\pi_{u_*,j}^E|$ follows automatically from $|J_0|\leq C_{t,q}(d - |A|)$.
Indeed, each MMS profile profile $\pi_{u_*,j}$ contains at most $q$ demands, and we add $|J_0|$ new profiles.
Thus, we will focus on showing that there exist $A_0 \subseteq A$ and $J_0\subseteq [d]$ such that
\begin{enumerate}
    \item $|A_0| + |J_0| = d$ and $|J_0| \leq C_{t,q}(d - |A|)$; and
    \item the $E$-load at every protected prefix $\ell_s \in \Ll_S$ is at most $|E\cap [\ell_s]|$;
\end{enumerate}
for some global constant $C_{t,q}$.

The key idea to proving the theorem is to observe that, even though the number of agents in $A$ and the number $d$ of copies of $u_*$ are quite large (both order $n$), the actual number of possible \textbf{load configurations} is significantly smaller, i.e only a constant number.
More concretely, for each agent $i \in A$, its load configuration $p_i$ is a length-$t$ vector
\[
    p_i := \{\Pp_i(\pi_i, \ell_1), \ldots, \Pp_i(\pi_i, \ell_{t})\},
\]
where $\{\ell_1, \ldots, \ell_t\} = \Ll_S$ are protected prefixes.
Similarly, a load configuration $p_{*,j}$ of a copy $j \in [d]$ of the supplier $u_*$ is
\[
    p_{*,j} := \{\Pp_{u_*}(\pi_{u_*,j}^E, \ell_1), \ldots, \Pp_{u_*}(\pi_{u_*,j}^E, \ell_t)\}.
\]
Note that both $p_i$ and $p_{*,j}$ are non-decreasing in its coordinates, due to the fact that $\ell_1 < \ldots, < \ell_t$.
In addition, each of the coordinates can be at most $q$, as every profile contains at most $q$ demands.
Hence, every $p_i, p_{*,j}$ belongs to the set
\[
    P = \{p \in \Z^t_{\geq 0}: 0 \leq p_1 \leq \ldots \leq p_t \leq q\}.
\]
We also define the \textbf{load capacity} vector as
\[
    \Ee := \{|E\cap [\ell_1]|, \ldots, |E\cap [\ell_t]|\}.
\]
We would like to find subsets $A_0 \subseteq A$ and $J_0\subseteq [d]$ such that
\[
    |A_0| + |J_0| = d,\qquad\text{and}\qquad \Pp(A_0, J_0) := \sum_{i \in A_0}p_i + \sum_{j \in J_0}p_{*,j} \leq \Ee,
\]
where the last inequality is component-wise.

We do this iteratively.
At the beginning of each iteration $k \geq 1$, we have candidate sets $A^k \subseteq A$ and $J^k \subseteq [d]$, satisfying the condition $\Pp(A^k, J^k)\leq \Ee$.
We select a set of agents $A'\subseteq A^k$ to remove from $A^k$, and a set of copies $J'\subseteq [d] - J^k$ to add to $J^k$, so that $|J'| = |A'| + 1$ and 
\[
    \Pp(A^k - A', J^k + J') \leq \Ee.
\]
After this, we move to iteration $k + 1$ with $A^{k + 1} = A^k - A'$ and $J^{k + 1} = J^k + J'$.
Initially, we start with $A^1 = A$ and $J^1= \varnothing$, note that these sets are feasible, as by \cref{Erounding} we have
\[
    \Pp(A, \varnothing)[\ell_s] = \sum_{i \in A}\Pp_i(\pi_i^E, \ell_s) \leq |E\cap [\ell_s]| = \Ee[\ell_s], \qquad \forall \ell_s\in \Ll_S.
\]

Suppose at iteration $k$ we have $|A^k| + |J^k| < d$, and $\Pp(A^k, J^k) \leq \Ee$.
Then, we can choose a set $J\subseteq [d] - J^k$ of unused copies, such that $|J| = |A^k| + 1$ --- this is possible as $d - |J^k|\geq |A^k| + 1$.
Observe that if we add $J$ to $J^k$ and remove all of $A^k$, we obtain a feasible collection $(\varnothing, J^k + J)$.
\[
    \Pp(\varnothing, J^k + J) = \sum_{j \in J^k + J}p_{*,j} \leq \sum_{j = 1}^dp_{*,j} = \Pp(\varnothing, [d])\leq \Ee.
\]
The last inequality follows from the fact that all profiles $\{\pi_{u_*,j}^E : j \in [d]\}$ correspond to MMS bundles $\{B_{u_*,j} : j \in [d]\}$, and their originating items partition $E\cap [\ell]$ into disjoint subsets satisfying corresponding profiles, for every $0\leq \ell \leq m$.

The problem with replacement $A^k \to \varnothing$ and $J^k \to J^k + J$ is that the set $J$ is of unbounded size, up to $d = \Theta(n)$.
We, on the other hand, want to add at most a constant number of copies of $u_*$ at every iteration.
We manage to do so by applying Dickson's lemma, as explained below.
\begin{lemma}\label{dicksonchoice}
    For a global constant $C(t, q)$, there exist subsets $A'\subseteq A^k$ and $J'\subseteq J$ such that $|J'| = |A'| + 1$ and $|J'| \leq C_{t,q}$, as well as
    \[
        \Pp(A^k - A', J^k + J') \leq \max\big\{\Pp(A^k, J^k), \Pp(\varnothing, J^k + J)\big\} \leq \Ee.
    \]
\end{lemma}
\begin{proof}
    We will first define an abstract set of integral vectors, that depends only on values $t, q$.

    Let $x, y \in \Z_{\geq0}^P$ be non-negative integer vectors, indexed by load configurations $p \in P$.
    For every pair $x, y\in \Z_{\geq0}^P$, let $z_+ = z_+(x, y) \in \Z_{\geq0}^{t}$ be
    \[
        z_+:= \left(\sum_{p \in P}p\cdot y_p - \sum_{p \in P}p\cdot x_p\right)_+.
    \]
    Denote by $\Rr(t, q)$ the set of all such triples $(x, y, z_+)$ satisfying 
    \[
        \sum_{p \in P}y_p = \sum_{p \in P}x_p + 1.
    \]
    Observe that $\Rr(t, q)$ is a subset of $\Z^{2|P| + t}_{\geq 0}$.
    Furthermore, even though elements its are unbounded, $\Rr(t, q)$ is determined entirely by the values of $t, q$, as $P = \{p \in \Z_{\geq 0}^t : 0 \leq p_1 \leq \ldots \leq p_t\leq q\}$.
    
    While vectors $x, y$ themselves can be arbitrary, we are interested in particular realizations that corresponds to our instance.
    More specifically, vector $x$ will represent the changes in the load configuration of $A^k$ induced by \textbf{removing} $A'\subseteq A^k$.
    In turn, vector $y$ will represent the changes in the load configuration of $J^k$ induced by \textbf{adding} $J'\subseteq J$.
    That is, for every $p \in P$:
    \begin{itemize}
        \item $x_p$ will count how many agents $i \in A^k$ with load configuration $p$ we remove from $A^k$;
        \item $y_p$ will count how many copies $j \in J$ of $u_*$ with load configuration $p$ we add to $J^k$.
    \end{itemize}
    In this case, vector $z_+$ corresponds to the load added to $(A^k, J^k)$ by performing this replacement.
    
    Note that $\Rr(t, q)$ contains the vector $(x^*, y^*, z^*_+)$ corresponding to the earlier mentioned replacement $(\varnothing, J^k + J)$: for every $p \in P$, set
    \[
        x_p^* = |\{i \in A^k : p_i = p\}|,\qquad \text{and}\qquad y_p^* = \{|j \in J : p_{*,j} = p\}|,
    \]
    and
    \[
        z^*_+= \left(\sum_{p \in P}p\cdot y_p^* - \sum_{p \in P}p\cdot x_p^*\right)_+ = \left(\sum_{j \in J}p_{*,j} - \sum_{i \in A^k}p_i\right)_+ = \big(\Pp(\varnothing, J\cup J^k) - \Pp(A^k, J^k)\big)_+.
    \]
    Since $|J| = |A^k| + 1$, we have $\sum_{p \in P}y^*_p = \sum_{p \in P}x^*_p + 1$, so $(x^*, y^*, z^*_+) \in \Rr(t, q)$.
    
    We would like an element of $\Rr(t, q)$ that 1) matches our particular instance, i.e sets $A^k$ and $J^k$, 2) the added load in $z$ fits into the $\Ee$-bound, and 3) the total number of copies added to $J^k$, i.e $\sum_{p \in P}y_p$, is only a constant.
    Dickson's lemma says that such element necessarily must exist.
    \begin{lemma}[\cite{Dickson13, Cox15}]\label{dickson}
        For $K \geq 1$, every subset of $\Z^K_{\geq0}$ has finitely many component-wise minimal elements, and every element of the subset dominates some minimal element.
    \end{lemma}
    For completeness, a short proof of \cref{dickson} is given below (\cref{dicksonproof}).

    Since $\Rr(t, q)$ is a subset of $\Z_{\geq0}^{2|P| + t}$, by \cref{dickson} $\Rr(t, q)$ has only finitely many component-wise minimal elements, let them be $\{(x^r, y^r, z^r_+)\}_{r = 1}^R$.
    Since $\Rr(t, q)$ is determined entirely by the values of $t$ and $q$, the number $R$ of minimal elements is some function of $t$ and $q$.
    Furthermore the coordinates of each $(x^r, y^r, z^r_+)$ for $r \in R$ are also bounded by some function of $t, q$ (otherwise we would obtain an infinite sequence of minimal elements).
    Take $C_{t,q} := \max_{r \in [R]}\sum_{p \in P}y^{r}_p$.
    Note that this is a global constant, dependent only on $\Rr(t, q)$.

    Let $(x^*, y^*, z^*_+) \in \Rr(t, q)$ be defined from $A^k$ and $J$ as above.
    By \cref{dickson}, there exists a minimal element $(x^r, y^r, z^r_+) \in \Rr(t, q)$ such that $(x^r, y^r, z^r_+) \leq (x^*, y^*, z^*_+)$ component-wise and
    \[
        \sum_{p \in P}y^{r}_p\leq C_{t,q}.
    \]
    
    Since $x^r \leq x^*$, for every load configuration $p \in P$ we have $x^r_p \leq x^*_p$.
    In other words, if $x^r$ removes some agent with load configuration $p$, he can remove only agents from $A^k$ with configuration $p$ ($0$-s remain $0$-s), and he can remove no more than the number of corresponding agents in $A^k$.
    Then there exists a subset $A'\subseteq A^k$ containing exactly $x^r_p$ agents $i \in A^k$ with load configuration $p$, for every $p \in P$.
    Similarly, $y^r \leq y^*$ implies that there exists a subset $J'\subseteq J$ containing exactly $y^r_p$ copies $j \in J$ with load configuration $p$, for every $p \in P$.
    These subsets satisfy $|J'| = |A'| + 1$, $|J'|= \sum_{p \in P}y^{r}_p\leq C(t, q)$, and (by $z^r_+ \leq z^*_+$)
    \[
        \left(\sum_{j \in J'}p_{*,j} - \sum_{i \in A'}p_i\right)_+ \leq   \big(\Pp(\varnothing, J^k + J) - \Pp(A^k, J^k)\big)_+.
    \]
    It follows that
    \begin{multline*}
        \Pp(A^k - A', J^k + J') \leq  \Pp(A^k, J^k) + \left(\sum_{j \in J'}p_{*,j} - \sum_{i \in A'}p_i\right)_+ \\
        =\Pp(A^k, J^k) +\big(\Pp(\varnothing, J^k + J) - \Pp(A^k, J^k)\big)_+ = \max\big\{\Pp(A^k, J^k), \Pp(\varnothing, J^k + J)\big\} \leq \Ee.
    \end{multline*}
\end{proof}
At iteration $k \geq 1$, given current sets $(A^k, J^k)$, we apply \cref{dicksonchoice} and obtain sets $A'\subseteq A^k$, $J'\subseteq J\subseteq [d]-J^k$ such that $|J'| = |A'| + 1$ and $|J'|\leq C_{t,q}$, as well as $\Pp(A^k - A', J^k + J') \leq \Ee$.
Then, by setting $A^{k + 1} = A^k - A'$ and $J^{k + 1} = J^k + J'$, we preserve load feasibility and increase the total number of selected profiles by $1$, while adding at most $C_{t,q}$ copies of $u_*$.
Thus, after exactly $d - |A|$ iterations, we obtain sets $A_0 := A^{d - |A| + 1}$ and $J_0 := J^{d - |A| + 1}$ such that $\Pp(A_0, J_0) \leq \Ee$ and $|A_0| + |J_0| = d$, with a total number of added supplier copies at most $|J_0|\leq C_{t,q}(d - |A|)$.
\end{proof}

\begin{lemma}[Short proof of Dickson's lemma, \cite{Dickson13, Cox15}]\label{dicksonproof}
        For $K \geq 1$, every subset of $\Z^K_{\geq0}$ has finitely many component-wise minimal elements, and every element of the subset dominates some minimal element.
\end{lemma}
\begin{proof}
    Let $S\subseteq \Z^K_{\geq 0}$.
    Element $y\in S$ is minimal if there is no $x \in S-\{y\}$ satisfying $x \leq y$.

    It is well-known that every infinite sequence of non-negative integers has an infinite non-decreasing subsequence.
    Then, every infinite sequence $x^1, x^2, \ldots \in \Z^K_{\geq 0}$ has an infinite subsequence that is non-decreasing in every coordinate: choose an infinite subsequence non-decreasing in the first coordinate, then its infinite sub-sequence non-decreasing in the second coordinate, and so on, up until the $K$-th coordinate.
    This gives an infinite sub-sequence $x^{i_1}\leq x^{i_2} \leq \ldots \in \Z^K_{\geq 0}$.

    Suppose that $S$ has infinitely many component-wise minimal elements.
    Choose an infinite sequence of distinct minimal elements.
    Then, we can obtain its infinite sub-sequence non-decreasing in every coordinate.
    This subsequence then contains two distinct minimal elements $x, y \in S$ with $x \leq y$, which is not possible.
    Hence, $S$ must have finitely many minimal elements.

    Next, fix $x \in S$ and consider $S_x := \{y \in S : y \leq x\}$.
    This set is nonempty ($x \in S_x$) and finite, as every coordinate of $y \in S_x$ is an integer between $0$ and corresponding coordinate of $x$, and there are $K$ coordinates.
    Choose $y \in S_x$ minimizing $\sum_{k = 1}^Ky_k$.
    If some $z \in S - \{y\}$ satisfied $z \leq y$, then $z \leq x$ and $z \in S_x$.
    Furthermore, every coordinate of $z$ would be at most the corresponding coordinate of $y$, with at least one strict inequality.
    Hence, $\sum_{k = 1}^Kz_k \leq \sum_{k = 1}^Ky_k$, leading to a contradiction.
    Thus, $y$ is minimal in $S$, and $y \leq x$.
    The lemma follows.
\end{proof}

\section{Constant estimate}
\label{appendix:constant}

\begin{theorem}\label{constant}
    For $t \geq 1$ and $q \geq 2$, the constant $C_{t,q}$ in \cref{supplierproof} can be chosen as
    \[
        C_{t,q} = \binom{t + q}{q}(t + 1)q^{t}.
    \]
\end{theorem}
\begin{proof}
    Let $P = \{p \in \Z_{\geq0}^t : 0 \leq p_1\leq \ldots \leq p_t \leq q\}$ be the set of all possible load configurations, and let $\Ll_S = \{ \ell_1, \ldots, \ell_t\}$ be the set of $t$ protected prefixes from \cref{supplierproof}.
    It is clear that
    \[
        |P| = \binom{t + q}{q}.
    \]
    Recall from \cref{dicksonchoice} that set $\Rr(x, y)$ contains all triples $(x, y, z_+)$ such that
    \[
        \sum_{p \in P}y_p = \sum_{p \in P}x_p + 1,\qquad\qquad z_+ = \left(\sum_{p \in P}p\cdot y_p - \sum_{p \in P}p\cdot x_p\right)_+.
    \]
    It suffices to show that every minimal element $(x^r, y^r, z^r_+)$ of $\Rr(t, q)$ satisfies
    \[
        \sum_{p \in P}y_p^r \leq \binom{t + q}{q}(t + 1)q^{t}.
    \]
    To prove this bound, we will need the following result on norm of indecomposable kernel vectors.
\begin{theorem}[\cite{Onn2010}, Lemma 3.20]\label{onn}
    Let $M$ be an integer matrix with $k$ columns and rank $t$.
    Define 
    \[
        \Delta(M) := \max\{|\det M'|:\; M'\text{ is a square submatrix of }M\}.
    \]
    Suppose some $v \in \Z_{\geq 0}^k$, $v\neq 0$ satisfies $Mv = 0$ and is indecomposable, meaning that if for some $u \in \Z_{\geq 0}^k$ such that $0 \leq u \leq v$ it also holds $Mu = 0$, then  $u = 0$ or $u = v$.
    Then
    \[
        \|v\|_1 \leq (k - t)(t + 1)\Delta(M).
    \]
\end{theorem}

We will define a certain matrix $M$ with $k = |P| + t$ columns and rank $t$, such that $\Delta(M) \leq q^t$.
Then, we will show that every minimal element of $\Rr(t, q)$ corresponds to some indecomposable kernel vector of $M$, which will allow us to use the bound of \cref{onn}.

Fix the minimal element $(x^r, y^r, z^r_+) \in \Rr(t, q)$.
First, observe that for every $p \in P$ we have $x_p^ry_p^r = 0$, otherwise we can change both coordinates by $1$ and obtain a smaller vector in $\Rr(t, q)$.
Let $z^r = \sum_{p \in P}p\cdot y_p^r - \sum_{p \in P}p\cdot x_p^r$, so $(z^r)_+ = z_+^r$.
For each $p \in P$, define
\[
    a_p := \begin{cases}
        -p, &x_p^r > 0;\\
        p, & x_p^r = 0.
    \end{cases}
\]
Note that $(x_p^r + y_p^r)\cdot a_p = (y_p^r - x_p^r)\cdot p$.
In addition, for every $s \in [t]$, let $e_s$ be the $s$-th standard basis vector, and define
\[
    b_s := \begin{cases}
        -e_s,& z_s^r > 0;\\
        e_s, & z_s^r \leq 0.
            \end{cases}
\]
This way, $\sum_{s =1}^tb_s\cdot |z_s^r| = -z^r$.
Define the $|P| + t$ columns of integer matrix $M_0$ and vector $v$ as
\[
    M_0 = [\{a_p\}_{p \in P}, b_1, \ldots, b_t],\qquad\qquad v = [\{x_p^r + y_p^r\}_{p \in P}, |z_1^r|,\ldots, |z_t^r|].
\]
The vector $v$ is non-zero, and by construction, $M_0v = 0$ and the rank of $M_0$ is $t$.
\begin{lemma}\label{indecom}
    Vector $v$ is indecomposable w.r.t $M_0$ (as per \cref{onn}).
\end{lemma}
\begin{proof}
    Suppose the contrary, i.e that $v = v' + v'' $, where both $v', v'' \neq 0$ are non-negative integer vectors satisfying $M_0v' = 0 = M_0v''$.
    We will show that this must contradict the minimality of $(x^r, y^r, z^r_+)$, by presenting a triple $(\hx, \hy, \hz_+) \in \Rr(t, q)$ strictly smaller than $(x^r, y^r, z^r_+)$.

    We do it as follows.
    For $p \in P$, define vectors $x', x''$ and $y', y''$ as
    \[
        x'_p = v'_p\One\{x_p^r > 0\},\qquad x''_p = v''_p\One\{x_p^r > 0\},\qquad y'_p = v'_p\One\{y_p^r > 0\},\qquad y''_p = v''_p\One\{y_p^r > 0\}.
    \]
    It is clear that $x'+ x'' = x^r$ and $y'+ y'' = y^r$.
    Next, define vectors $z', z'' \in \Z^t$ as 
    \[
        z'= \sum_{p \in P}p\cdot y_p' - \sum_{p \in P}p\cdot x_p',\qquad \qquad z''= \sum_{p \in P}p\cdot y_p'' - \sum_{p \in P}p\cdot x_p''.
    \]
    Since $M_0v' = 0 = M_0v''$, by definition of $b_s$ for every $s \in [t]$ we have 
    \[
        z'_s = \begin{cases}
            v'_s,& z_s^r > 0;\\
            -v'_s,& z_s^r \leq 0;
        \end{cases}\qquad\qquad z''_s = \begin{cases}
            v''_s,& z_s^r > 0;\\
            -v''_s,& z_s^r \leq 0.
        \end{cases}
    \]
    Since $v'_s \leq v_s = |z_s^r|$ and $v''_s \leq v_s = |z_s^r|$, it holds $(z')_+ \leq (z^r)_+$ and $(z'')_+ \leq (z^r)_+$.

    Next, we claim that $v'$ satisfies $v'_p \geq 1$ for some $p \in P$.
    Indeed, otherwise $v'= [0, v'_1, \ldots, v'_t]$ and $0 =M_0v' = \sum_{s = 1}^tv'_s\cdot b_s$.
    Since each of $b_s$ is $\pm e_s$, the equality above implies $v'_s = 0$ for all $s \in [t]$, making $v'= 0$.
    By the same reasoning, $v''$ satisfies $v''_p \geq 1$ for some $p \in P$.
    Consequently,
    \[
        \sum_{p \in P}(x_p'+ y_p') \geq 1,\qquad\qquad \sum_{p \in P}(x_p''+ y_p'')\geq 1.
    \]
    Define (possibly negative) integers
    \[
        g' := \sum_{p \in P}y_p' - \sum_{p \in P}x_p',\qquad\qquad g'' := \sum_{p \in P}y_p'' - \sum_{p \in P}x_p''.
    \]
    Since $x'+ x'' = x$, $y' + y'' = y$ and $\sum_{p \in P}y_p = \sum_{p \in P}x_p + 1$, it holds $g'+ g''  = 1$.
    
    Without loss of generality, assume $g'\geq 1$, then $g'' = 1 - g'\leq 0$, implying
    \[
         \sum_{p \in P}y_p'' \leq \sum_{p \in P}x_p''.
    \]
    Since $\sum_{p \in P}(x_p''+ y_p'')\geq 1$, it must be the case that $\sum_{p \in P}x_p'' \geq 1$, so for some $p \in P$ we have $x''_p > 0$, and $x'_p = x_p^r - x''_p < x_p^r$.
    Then, the inequality $x'\leq x^r$ is strict for at least one $p \in P$.

    It remains to create an element of $\Rr(t, q)$ with the first part $x'$ that is dominated by $(x^r, y^r, z^r_+)$.
    We do it as follows.
    Remove $g' - 1$ additions from $y'$, it is always possible as $\sum_{p \in P}y_p' = \sum_{p \in P}x_p' + g'$ and $g'\geq 1$.
    The resulting non-negative integer vector $\hy \leq y'$ satisfies
    \[
        \sum_{p \in P}\hy_p = \sum_{p \in P}x_p' + 1.
    \]
    Since removing $y$-additions only decreases $z$-coordinates, we have
    \[
        \hz := \sum_{p \in P}p\cdot \hy_p - \sum_{p \in P}p\cdot x_p' \leq z',
    \]
    It follows that $(\hz)_+ \leq (z')_+ \leq (z^r)_+ = z^r_+$.
    As a result, we have a triple $(x', \hy, \hz_+) \in \Rr(t, q)$ such that $(x', \hy, \hz_+)\leq (x^r, y^r, z^r_+)$, and there is at least one $p \in P$ where $x'_p < x^r_p$.
    So, $(x', \hy, \hz_+)$ is strictly dominated by $(x^r, y^r, z^r_+)$ in $\Rr(t, q)$, contradicting the minimality of $(x^r, y^r, z^r_+)$.
\end{proof}

We will now deal with matrix $M_0$.
First, we change $M_0$ into a matrix $M$ with small column norms.
Define an invertible matrix $L$ so that for any vector $u$,
\[
    Lu = [u_1, u_2 - u_1, \ldots, u_t - u_{t - 1}],
\]
and set $M:= LM_0$.
The rank of $M$ is still $t$, and $\ker M = \ker M_0$.
So, vector $v$ remains indecomposable w.r.t matrix $M$ as well.
For every $p \in P$ vector $Lp = (p_1, p_2 - p_1, \ldots, p_t - p_{t - 1})$ has non-negative coordinates summing up to $p_t \leq q$, hence for every $p \in P$ we have
\[
    \|L\cdot a_p\|_2 = \|L\cdot p\|_2 \leq \|L\cdot p\|_1 = p_t \leq q.
\]
For the remaining columns, for every $s \in [t]$ we have
\[
    Le_s =\begin{cases}
        e_s - e_{s + 1},& s < t\\
        e_t,& s = t.
    \end{cases}
\]
So, $Lb_s = \pm Le_s$ and $\|Lb_s\|_2 \leq \sqrt{2} \leq q$.
Thus, the norm of every column of $M$ is at most $q$.

Applying the bound of \cref{onn}, we have
\[
    \|v\|_1 \leq \big((|P| + t) - t\big)(t + 1)\Delta(M) = |P|(t + 1)\Delta(M).
\]
By definition, $\Delta(M)$ is the largest absolute value of a determinant of a square submatrix of $M$.
By Hadamard's inequality, for any matrix $A$ with $k$ columns $(A^1, \ldots, A^k)$ we have
\[
    |\det A|\leq \prod_{j = 1}^k\|A^j\|_2.
\]
Let $M'$ be some square submatrix of $M$.
If $M'$ has more than $t$ columns, then $\det M'= 0$, since $M$ has rank $t$.
If $M'$ has size at most $t$, then, since the norm of a submatrix column cannot exceed the norm of the original column, Hadamard inequality gives $|\det M'|\leq q^t$.
Therefore
\[
    \|v\|_1 \leq |P|(t + 1)\Delta(M) \leq \binom{t + q}{q}(t + 1)q^{t}.
\]
Finally, by definition of $v$, $\sum_{p \in P}y_p^r \leq \|v\|_1$, which finishes the proof.
\end{proof}

\begin{corollary}\label{bign}
    For every $\eps \in (0, 1)$ and $q \geq 3$, we can choose $n_{\eps,q}$ in \cref{savedconst} as
    \[
        n_{\eps,q} = \left\lceil 2^{64K} \right\rceil\qquad\text{for}\qquad K = \frac{q^6\log q}{\eps}.
    \]
\end{corollary}
\begin{proof}
    The estimates will be crude.
    Recall our parameters $\gamma = \min(\eps, 1/(2q))$, $\lambda = \frac{2q}{2q - 1}$ and
    \[
        \rho_{q,n} := 8+ 4 \log_\lambda\big(q^3d^2(d + 2)n\big)+ 8q\log \big((d + 2)n\big).
    \]
    Assume for now that $n \geq \max\{q, 16/\eps\}$.  
    The number of protected prefixes $|\Ll_S|$ is $t \leq 2q^2$, so by \cref{constant} the constant $C_q$ in \cref{savedconst} can be chosen as
    \[
        C_q = \binom{2q^2 + q }{q}(2q^2 + 1)q^{2q^2}.
    \]
    For $q \geq 3$, $\binom{2q^2 + q }{q}\leq q^{3q}$ and $2q^2 + 1\leq q^3$, consequently by the choice of $C_1$,
    \[
        C_q\leq q^{4q^2}\qquad\text{and}\qquad C_1 = 1 + \lceil 12q^3 C_q\rceil \leq q^{5q^2}.
    \]
    Since $d \leq n$, $q^3d^2(d + 2)n \leq 2q^3n^4$ and $(d + 2)n\leq 2n^2$.
    Using this with $\log \lambda  \geq 1/(2q)$, we have
    \[
        \rho_{q,n} \leq 9 + 8q\log(2q^3n^4) + 8q(2\log n + 2) \leq 80q\log n.
    \]
    Therefore, by the choice of $r_1$,
    \[
        r_1 := 1 + \left\lceil \left(\frac{2q(q - 1)}{\gamma} + 2\right)\rho_{q,n}\right\rceil \leq \frac{256q^2(q - 1)}{\gamma} \log n \leq \frac{512q^4}{\eps}\log n.
    \]
    By \cref{saved} we have the bound on $|S|$:
    \[
        |S|\leq 2q^2C_1^{8q^2/\eps}r_1\leq \frac{1024q^6}{\eps}\cdot q^{40q^4/\eps}\log n\leq q^{42q^4/\eps}\log n, 
    \]
    where we used
    \[
        \frac{1024q^6}{\eps}\leq q^{14/\eps}\leq q^{2q^4/\eps}.
    \]
    Now, as $t\leq |S|$ and $d - |A|\leq 2|S|$, to guarantee in \cref{savedconst} that both
    \[
        t + |S| + 1 \leq C_*\log n\qquad\text{and}\qquad C_q(d - |A|)\leq C_*\log n,
    \]
    due to $C_q \leq q^{5q^2}$ we can choose $C_*$ as
    \[
        C_* = q^{48q^4/\eps}.
    \]
    By \cref{recipients}, the set $U_0\subseteq U$ satisfies
    \[
        |U_0| \geq \frac{\eps n}{16(C_*\log n)^{q^2}},
    \]
    so in \cref{savedconst} it suffices to ensure
    \[
        n\geq \frac{16}{\eps}(C_*\log n)^{q^2 + 1}.
    \]
    We will compute the logarithms of the righthand side expression, to estimate the value of $n$.
    That is, we need to find some value $n_0$ such that
    \[
        \log n_0 \geq \log(16/\eps) + (q^2 + 1)\log C_* + (q^2 + 1)\log\log n_0.
    \]
    Take
    \[
        K := \frac{q^6\log q}{\eps},\qquad\qquad n_{0} = 2^{64K}.
    \]
    Since $q\geq 3$, we have $q^2 + 1\leq \frac{5}{4}q^2$, hence 
    \[
        (q^2 + 1)\log C_* \leq \frac{5}{4}q^2\cdot \frac{48q^4\log q}{\eps} = 60 K.
    \]
    In addition, $\log (16/\eps) \leq K$ and 
    \[
        \log\log n_0 = \log (64K) = 6 + 6\log q + \log\log q + \log(1/\eps) \leq 11\log q + \log(1/\eps)\leq \frac{12\log q}{\eps},
    \]
    hence 
    \[
        (q^2 + 1)\log\log n_0\leq \frac{15q^2\log q}{\eps}\leq K.
    \]
    It follows that
    \[
       \log\left( \frac{16}{\eps}(C_*\log n_0)^{q^2 + 1}\right)\leq K + 60K + K \leq 64K = \log n_0.
    \]
    Clearly, $n_0 \geq \max\{q, 16/\eps\}$.
    Monotonicity implies that any $n \geq \lceil n_0\rceil = \lceil 2^{64K}\rceil$ suffices.
\end{proof}

\end{document}